\documentclass[fleqn,usenatbib]{mnras}

\usepackage{newtxtext,newtxmath}

\usepackage[T1]{fontenc}
\usepackage{ae,aecompl}

\usepackage{graphicx}	
\usepackage{amsmath}	
\usepackage{amssymb}	

\title[Twinning: equal-mass binaries at wide separations]{Discovery of an equal-mass ``twin'' binary population reaching 1000+ AU separations}

\author[El-Badry et al.]{
Kareem El-Badry,$^{1, 2}$\thanks{E-mail: kelbadry@berkeley.edu}
Hans-Walter Rix,$^{2}$
Haijun Tian,$^{3,2,4}$
Gaspard Duch\^{e}ne,$^{1,5}$
\newauthor
\,\,and Maxwell Moe$^{6}$
\\
$^{1}$Department of Astronomy and Theoretical Astrophysics Center, University of California Berkeley, Berkeley, CA 94720\\
$^{2}$Max Planck Institute for Astronomy, D-69117 Heidelberg, Germany\\
$^{3}$China Three Gorges University, Yichang 443002, China\\
$^{4}$Center of Astronomy and Space Science Research, China Three Gorges University, Yichang 443002, China\\
$^{5}$Univ. Grenoble Alpes, CNRS, Institut de Plan\'etologie et d'Astrophysique de Grenoble (IPAG), F-38000 Grenoble, France\\
$^{6}$Steward Observatory, University of Arizona, Tucson, AZ 85721}

\defcitealias{ElBadry_2018}{ER18}

\date{Accepted to MNRAS, September 3, 2019}

\pubyear{2019}

\begin{document}
\label{firstpage}
\pagerange{\pageref{firstpage}--\pageref{lastpage}}
\maketitle

\begin{abstract}
We use a homogeneous catalog of 42,000 main-sequence wide binaries identified by \textit{Gaia} to measure the mass ratio distribution, $p(q)$, of binaries with primary masses $0.1 < M_1/M_{\odot} < 2.5$, mass ratios $0.1 \lesssim q < 1$, and separations $50 <s/{\rm AU} < 50,000$. 
A well-understood selection function allows us to constrain $p(q)$ in 35 independent bins of primary mass and separation, with hundreds to thousands of binaries in each bin. 
Our investigation reveals a sharp excess of equal-mass ``twin'' binaries that is statistically significant out to separations of 1,000 to 10,000 AU, depending on primary mass. 
The excess is narrow: a steep increase in $p(q)$ at $0.95  \lesssim q < 1$, with no significant excess at $q\lesssim 0.95$. A range of tests confirm the signal is real, not a data artifact or selection effect.
Combining the {\it Gaia} constraints with those from close binaries, we show that the twin excess decreases with increasing separation, but its width ($q\gtrsim 0.95$) is constant over $0.01 < a/{\rm AU} < 10,000$.
The wide twin population would be difficult to explain if the components of all wide binaries formed via core fragmentation, which is not expected to produce strongly correlated component masses.
We conjecture that wide twins formed at closer separations ($a \lesssim 100$\,AU), likely via accretion from circumbinary disks, and were subsequently widened by dynamical interactions in their birth environments. The separation-dependence of the twin excess then constrains the efficiency of dynamical widening and disruption of binaries in young clusters.
We also constrain $p(q)$ across $0.1 \lesssim q < 1$. Besides changes in the twin fraction, $p(q)$ is independent of separation at fixed primary mass over $100 \lesssim s/{\rm AU} < 50,000$. It is flatter than expected for random pairings from the IMF but more bottom-heavy for wide binaries than for binaries with $a\lesssim$100\,AU.  
\end{abstract}




\section{Introduction}
Binary stars are ubiquitous: roughly half of all field stars have binary companions \citep{Duquennoy_1991, Duchene_2013, Moe_2017}, and the binary fraction is even higher in star forming regions \citep[e.g.][]{Ghez_1993, Leinert_1993, Mathieu_1994, Connelley_2008, Sadavoy_2017, Duchene_2018}.
Significant progress has been made in developing theoretical models to explain the population statistics of observed binaries \citep[e.g.][]{Kroupa_1995, Marks_2011, Bate_2012, Lomax_2015}, but fundamental aspects of the binary formation process remain imperfectly understood. 

The distribution of binary mass ratios has been a subject of interest for at least a century \citep[e.g.][]{Biesbroeck_1916, Opik_1924,  Kuiper_1935}. As a final outcome of the binary formation process, the mass ratio distribution provides useful constraints on theoretical models of star formation. Unlike the distributions of orbital separation and eccentricity, the mass ratio distribution has been suggested to be insensitive to dynamical evolution after formation (such that binaries of different mass ratios are disrupted at similar rates; e.g. \citealt{Parker_2013}). Mapping the mass ratio distribution over a range of binary masses and separations has thus been the focus of many studies \citep[e.g.][]{Trimble_1974, Trimble_1987, Trimble_1990, Eggleton_1989, Mazeh_1992, Hogeveen_1992, Shatsky_2002, Burgasser_2007, Soderhjelm_2007, Raghavan_2010, Tokovinin_2014, Gullikson_2016}.

Observational studies of the mass ratio distribution are complicated by incompleteness. All binary detection methods are biased against low-mass ratio companions, which produce weaker radial-velocity shifts at fixed separation \citep[e.g.][]{Shahaf_2019}, contribute less light to the observed spectra of unresolved binaries \citep[e.g.][]{ElBadry_2018_mock, Elbadry_2018a}, cause weaker eclipses \citep[e.g.][]{Moe_2013}, and are less likely to be detected as visual companions \citep[e.g.][]{Tokovinin_2011, ElBadry_2018}. The detection efficiency also varies with primary mass and orbital separation. This complicates measurement of the mass ratio distribution, because the distributions of separation, primary mass, and mass ratio are not independent \citep[e.g.][]{Moe_2017}. It is thus important for demographic studies of binaries that the selection function of observed samples is well understood. If possible, binaries of different masses and physical separations should be considered independently.

A puzzling feature of the mass ratio distribution identified by previous works is the so-called ``twin'' phenomenon, which refers to a purported statistical excess of nearly equal-mass binaries with mass ratios $0.95 \lesssim q < 1$. Most studies that find an excess of equal-mass twins have focused on spectroscopic binaries with close separations \citep[$a \ll 1$\,AU, e.g. ][]{Lucy_1979, Hogeveen_1992, Tokovinin_2000, Halbwachs_2003, Lucy_2006, Pinsonneault_2006, Simon_2009, Kounkel_2019}. Indeed, several studies have reported a sharp drop-off in the fractional excess of twin binaries beyond periods of $40$\, days \citep[$a \lesssim 0.2$\,AU; e.g.][]{Lucy_1979, Tokovinin_2000, Simon_2009}. Recent studies have not confirmed such a sharp drop-off in the twin excess for solar-type stars \citep{Tokovinin_2014, Moe_2017}, but have still found it to decline monotonically with increasing separation. At least for $M_1 \gtrsim 1 M_{\odot}$ (the mass range on which most previous studies have focused), the twin excess has been found to decrease with increasing primary mass at fixed separation and to extend to wider separations for lower-mass primaries \citep{Moe_2017}. 

Some works have also argued that the twin phenomenon may be a selection effect \citep{Mazeh_2003, Cantrell_2014}, since equal-mass binaries are brighter, can be detected at larger distances, and are preferentially selected by several binary detection methods \citep[e.g.][]{Branch_1976}. Such biases are minimized for binary samples that are nearly volume-complete  \citep[e.g.][]{Raghavan_2010, Tokovinin_2014} and/or have well-understood completeness. 

If it is a real effect, the physical origin of the twin phenomenon is not fully understood. Several mechanisms have been proposed that could lead to preferential formation of equal-mass binaries, including fragmentation during the late stages of protostellar collapse, mass transfer between pre-main sequence stars, and competitive accretion (see \citealt{Tokovinin_2000} for discussion of different formation mechanisms). Several simulations have predicted that accretion of high-angular momentum gas, particularly from a circumbinary disk, tends to drive binary mass ratios toward unity \citep[e.g.][]{Bate_2000, Farris_2014, Young_2015}. 
However, it is not obvious why, when averaged over a large population of binaries, accretion from a circumbinary disk would produce a sharp peak in the mass ratio distribution at $q\gtrsim 0.95$ as opposed to a gradual increase. 

There have also been hints of an excess of twins among spatially resolved wide binaries with separations ranging from tens to thousands of AU \citep{Trimble_1987, Soderhjelm_2007}. The selection functions of the wide binary samples studied in these works were poorly understood, causing investigators to remain agnostic of whether the excess of equal-brightness pairs in their catalogs was of astrophysical origin or rooted in selection biases. Recently, \citet{Moe_2017} measured the twin excess at different separations in a small but relatively complete sample of solar-type binaries within 25\,pc of the Sun. They found the twin excess to decline with separation, but found it inconsistent with 0 out to separations of 200\,AU. At even wider separations, they set an upper limit of $\sim$5\% on the excess twin fraction.

High-quality parallaxes and proper motions from the recent \textit{Gaia} data releases \citep{Gaia_2016, Gaia_2018} have simplified the process of constructing samples of wide binaries with (a) little contamination from chance alignments and (b) a well-understood selection function. Using data from {\it Gaia} DR2, \citet[][hereafter ER18]{ElBadry_2018} constructed a high-purity catalog of wide binaries within 200 pc of the Sun consisting mainly of AFGKM dwarfs. In this paper, we use a subset of that catalog to constrain the mass ratio distribution over a wide range of primary masses ($0.1 \lesssim M/M_{\odot} \lesssim 2.5$), mass ratios ($0.1 \lesssim q < 1$) and separations ($50 \lesssim s/{\rm AU} < 50,000$). The large size of the catalog allows us to constrain $p(q)$ in narrow bins of primary mass and separation independently.  This approach make it possible to measure variation in $p(q)$ with mass and separation, and it minimizes the sensitivity of our results to imperfectly known inputs such as the initial mass function (IMF) and separation distribution.  

A striking result of our investigation is the unambiguous evidence that twins are not purely a close-binary phenomenon: a significant excess of equal-mass ($q \gtrsim 0.95$) binaries persists out to separations as wide as 10,000\,AU. We derive  constraints on the excess fraction of twins and the width of the twin excess as a function of mass and separation. We also provide constraints on the full mass ratio distribution over  $0.1 \lesssim q \leq 1$ in all bins.

The remainder of this paper is organized as follows. Section~\ref{sec:data} describes the binary catalog and tests we have done to verify that the twin excess is real. In Section~\ref{sec:modeling}, we describe how we forward-model synthetic binary populations to fit for the intrinsic mass ratio distribution. Results of this fitting are presented in Section~\ref{sec:results}. In Section~\ref{sec:discussion}, we compare to previous work and discuss implications of our results for models of binary star formation and dynamical evolution. The appendices provide additional details about several aspects of the data and model. There we discuss sensitivity to the adopted parametric form of $p(q)$, (Appendix~\ref{sec:func_form}), systematic uncertainties in our model (Appendix~\ref{sec:systmatics}), evidence for a twin excess in archival wide binary catalogs (Appendix~\ref{sec:other_catalogs}), empirical determination of the selection function (Appendix~\ref{sec:selection_function_details}), and validation of the adopted Galactic model and selection function (Appendix~\ref{sec:model_validation}). Constraints on fitting parameters are tabulated in Appendix~\ref{sec:full_constraints}.

\section{Data}
\label{sec:data}

\begin{figure*}
    \includegraphics[width=\textwidth]{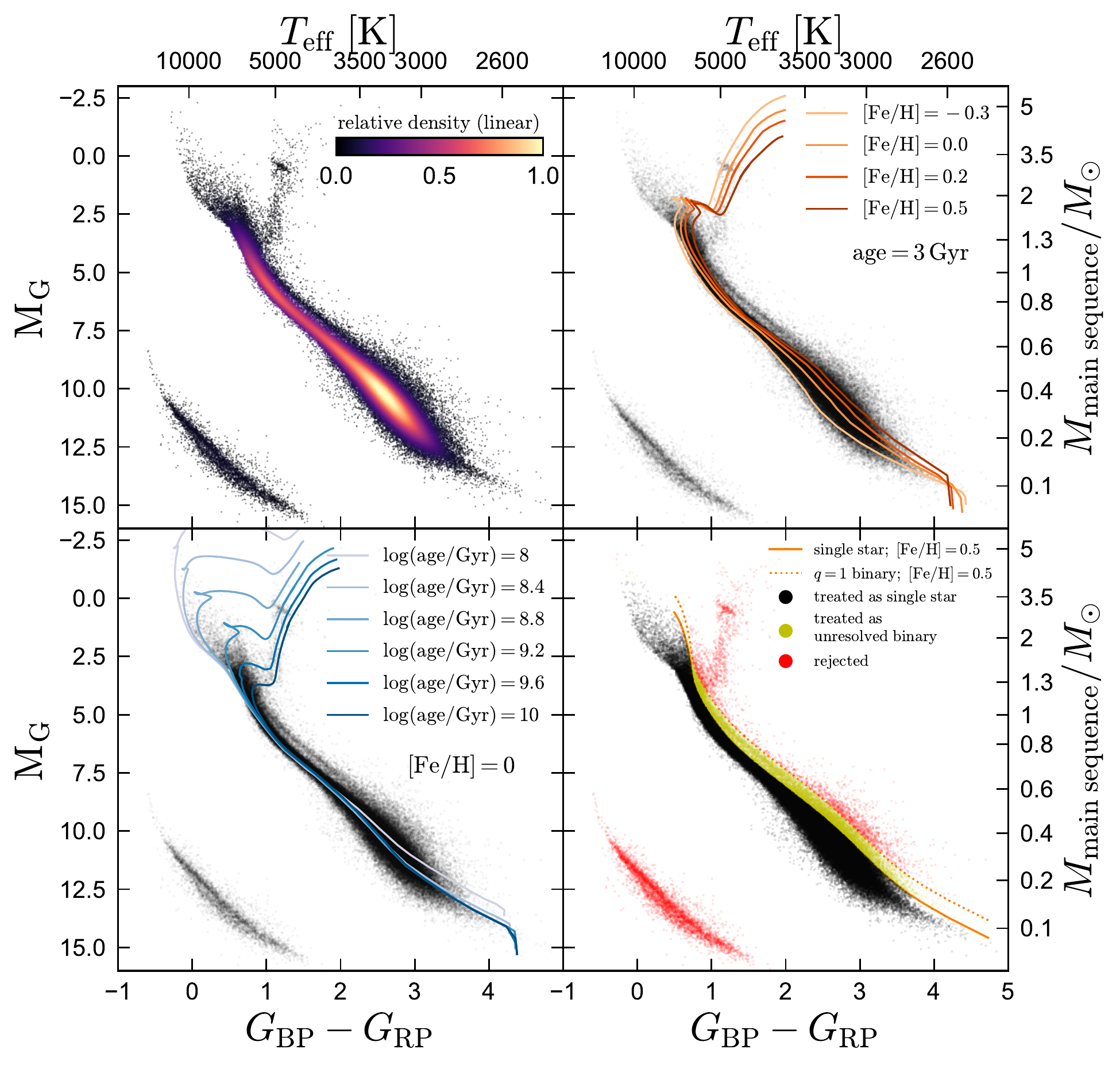}
    \caption{Color-magnitude diagram of all stars (both primaries and secondaries) in the \citetalias{ElBadry_2018} catalog of resolved wide binaries. Right y-axes and upper x-axes show the mass and effective temperature corresponding to a given magnitude and color for main-sequence stars with $\rm [Fe/H]=0$.
    In the upper left panel, points are colored by a Gaussian kernel density estimate of the local density. 
    We compare PARSEC isochrones with a range of metallicity (upper right) and age (lower left) to the data. 
    The bottom-right panel divides the CMD into three regions.
    Black points (main-sequence components with no bright unresolved sub-components) comprise the bulk of our sample, and their masses are estimated using single-star isochrones. 
    Yellow points (components of wide binaries with a bright unresolved companion) are kept, but their masses are estimated using unresolved binary isochrones. 
    Red points (pre-main sequences stars, unresolved triples, and evolved stars) are discarded.}
    \label{fig:cmds}
\end{figure*}

Our primary analysis uses the binary catalog assembled in \citetalias{ElBadry_2018}. This catalog contains $\sim$55,000 spatially resolved wide binaries with main sequence (MS), giant, and white dwarf components, and projected physical separations of $50 \lesssim s/{\rm AU} < 50,000$. We refer to \citetalias{ElBadry_2018} for a full description of the catalog's contents. In brief, it was constructed by searching {\it Gaia} DR2 for nearby ($d < 200$\,pc) pairs of stars whose positions, proper motions, and parallaxes are consistent with being gravitationally bound. Resolved higher-order multiples and suspected members of bound and dissolving clusters were removed. The catalog is designed to be pure but not complete: cuts on photometric and astrometric precision ensure that the contamination rate from chance alignments is low ($\ll1$\%), but they also reduce the number of faint binaries.

We do not use the full catalog from \citetalias{ElBadry_2018}, but impose the following additional cuts:
\begin{itemize}
    \item We only consider MS/MS binaries, removing binaries in which either component is suspected to be a white dwarf, subgiant, giant, or pre-main sequence star. We identify non-MS components from the color-magnitude diagram (CMD; see below). 
    \item We require {\it both} components to have \texttt{parallax} > 5,  \texttt{parallax\_over\_error} > 20,  \texttt{phot\_bp\_mean\_flux\_over\_error} > 20, and \texttt{phot\_rp\_mean\_flux\_over\_error} > 20. \citetalias{ElBadry_2018} used these same cuts for the primary, but used less stringent cuts for the secondary. Here we apply the same cuts to both components in order to symmetrize the selection function.
    \item We reject binaries in which the CMD-inferred mass of the primary falls outside $0.1 < M_1/M_{\odot} < 2.5$. The \citetalias{ElBadry_2018} catalog contains fewer than 100 binaries with estimated primary masses above 2.5 $M_{\odot}$, and none with estimated primary masses below $0.1 M_{\odot}$.
\end{itemize}

These additional cuts remove 23\% of the \citetalias{ElBadry_2018} catalog, leaving us with a sample of 42,338 MS/MS binaries. 

Figure~\ref{fig:cmds} shows the CMD of all stars in the \citetalias{ElBadry_2018} catalog, with primaries and secondaries included on the same axes. Overplotted PARSEC isochrones show that the spread on the lower main sequence is primarily attributable to metallicity (upper right panel), while that on the upper main sequence and red giant branch is primarily due to age (lower left panel). A secondary sequence consisting mainly of unresolved binaries is visible above the main sequence. Wide ``binaries'' with one component in this sequence are primarily hierarchical triples. We do not remove these from our sample but account for them in our model when fitting for the mass ratio distribution in Section~\ref{sec:modeling}.\footnote{We also experimented with removing binaries with suspected unresolved components from the sample; doing so does not qualitatively change any of our results.}

\begin{figure*}
    \includegraphics[width=\textwidth]{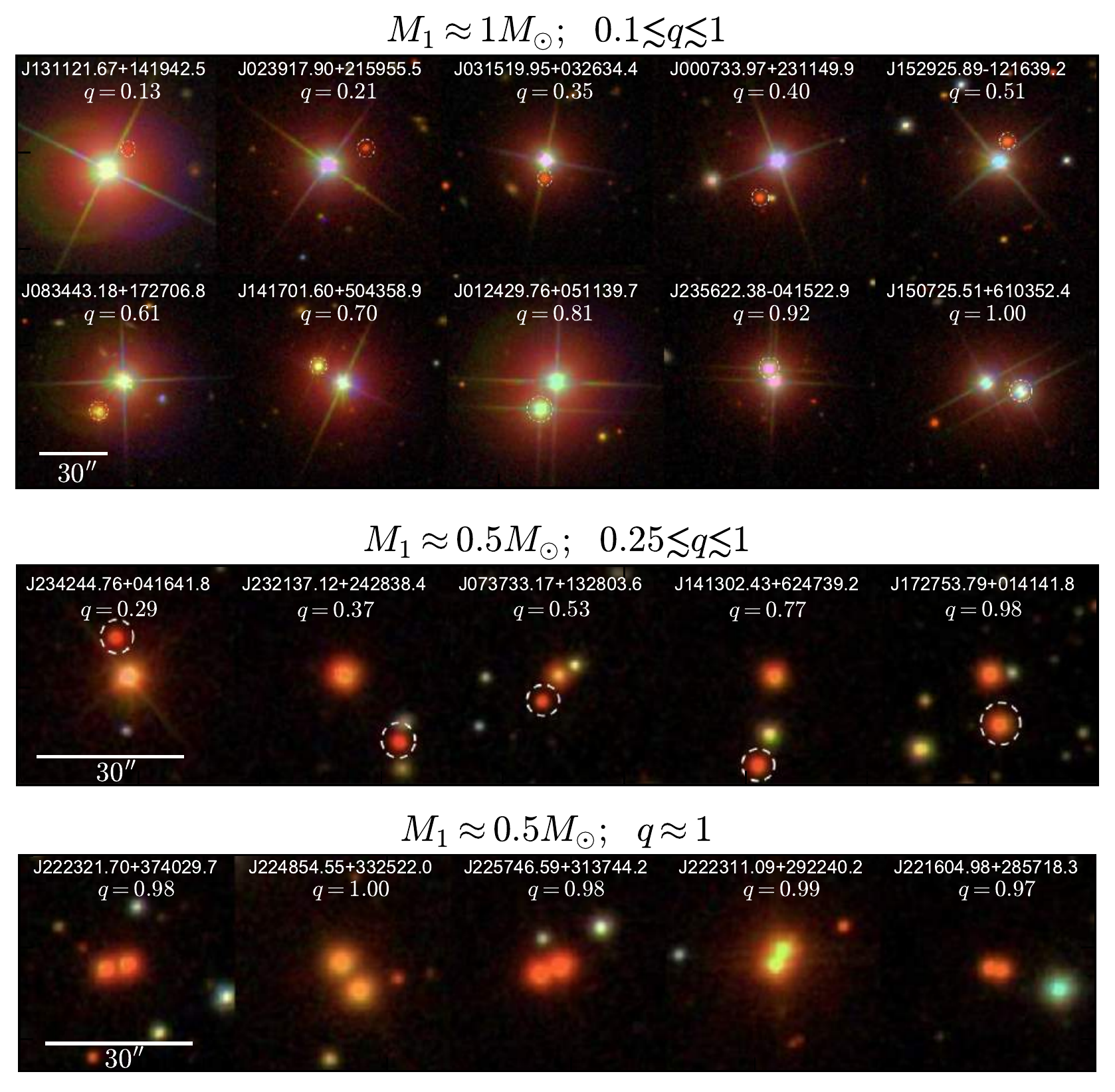}
    \caption{SDSS images of a selection of binaries from our catalog. Top panel shows binaries in which the primary is a solar-type star ($0.9 < M_1/M_{\odot}<1.1$) and the mass ratio varies from $\approx 0.1$ to 1. The primary is at the center of each image, and the secondary is circled. Middle panel shows binaries in which the primary is a late K dwarf ($0.4 < M_1/M_{\odot}<0.6$), again for a range of mass ratios.
    Bottom panels shows examples of ``twin'' binaries with near-identical components, each with $0.4 < M_1/M_{\odot}<0.6$.
    Each image is 100 (top panel) or 45 (bottom panels) arcsec on a side. 
    Our analysis uses photometry from {\it Gaia}, not SDSS. Because the SDSS photometry is ground-based, blending and source contamination affect it more severely. }
    \label{fig:sdss_pictures}
\end{figure*}

MS components that fall below a main sequence PARSEC isochrone with $[\rm Fe/H]=0.5$ (black points in the lower right panel of Figure~\ref{fig:cmds}) are treated as single stars. Those that fall above this isochrone but below an unresolved binary isochrone for two equal-mass stars with $[\rm Fe/H]=0.5$ (yellow points) are treated as unresolved binaries. Finally, sources that fall above this binary isochrone (red points; likely a mix of unresolved triples, pre-MS stars, and giants/subgiants) are rejected, as are white dwarfs. Of the 42,382 wide binaries in our catalog, 35,087 have two components consistent with having no unresolved sub-components (black points), and 7,295 have at least one component suspected to be an unresolved binary (yellow points).

We estimate masses for both components of each binary based on their location in the CMD. The ratios of these masses are not used explicitly in fitting the mass ratio distribution, but they are used to assign the primary vs. secondary components and to assign binaries to bins of primary mass. For MS stars that are suspected to be single, we estimate masses by interpolating from a grid of single-star PARSEC isochrones. This method is reasonably effective for single components, but it would yield biased results for the unresolved sub-components. 

To estimate masses for components suspected to be unresolved binaries, we construct a population of synthetic unresolved binaries (see Section~\ref{sec:modeling}) and, for the subset of this population that falls in the region of the CMD colored in yellow in the bottom right panel of Figure~\ref{fig:cmds}, we calculate the median {\it primary} mass as a function of $\rm M_{G}$ of the unresolved binary. For observed sources in that region of the CMD, we estimate the primary mass by interpolating from $\rm M_{G}$ on this median relation. That is, the mass assigned to unresolved components represents the mass of the primary of the unresolved component, not the total mass. 

This method of assigning masses is not without drawbacks: masses assigned to unresolved binaries are imprecise, because the mass ratio is not known. In addition, some pre-MS stars may be mistaken for unresolved binaries, and some low-mass ratio unresolved binaries may be mistaken for higher-metallicity single stars. However, we expect the typical accuracy to be $\lesssim 0.1 M_{\odot}$, which is good enough for our purpose of assigning binaries to different bins of primary mass prior to fitting. In modeling the mass ratio distribution, it is not critical that the mass ratio of any one binary be measured accurately, but rather that the {\it distribution} of magnitude difference be predicted self-consistently.

Figure~\ref{fig:sdss_pictures} shows ``postage-stamp'' images\footnote{\href{http://skyserver.sdss.org/dr14/en/tools/chart/listinfo.aspx}{skyserver.sdss.org/dr14/en/tools/chart/listinfo.aspx}} of some of the binaries in our catalog from the Sloan Digital Sky Survey (SDSS; \citealt{York_2000}). We note that SDSS photometry is not used in our analysis or in fitting the mass ratio distribution; we show it here because raw images from {\it Gaia} are not publicly available. Source contamination in the SDSS images is generally expected to be more severe than in {\it Gaia} photometry due to atmospheric seeing. To showcase the diversity of binaries in the catalog, we choose a selection of binaries with roughly solar-mass primaries and a range of mass ratios (top panel), binaries with primary masses of $\approx 0.5\,M_{\odot}$ and a range of mass ratios (middle panel), and some examples of equal-mass ``twin'' binaries with component masses of $\approx 0.5 M_{\odot}$ (bottom panel). The twins are easily recognized because their magnitudes and colors are very similar. 

\begin{figure*}
    \centering
    \includegraphics[width=\textwidth]{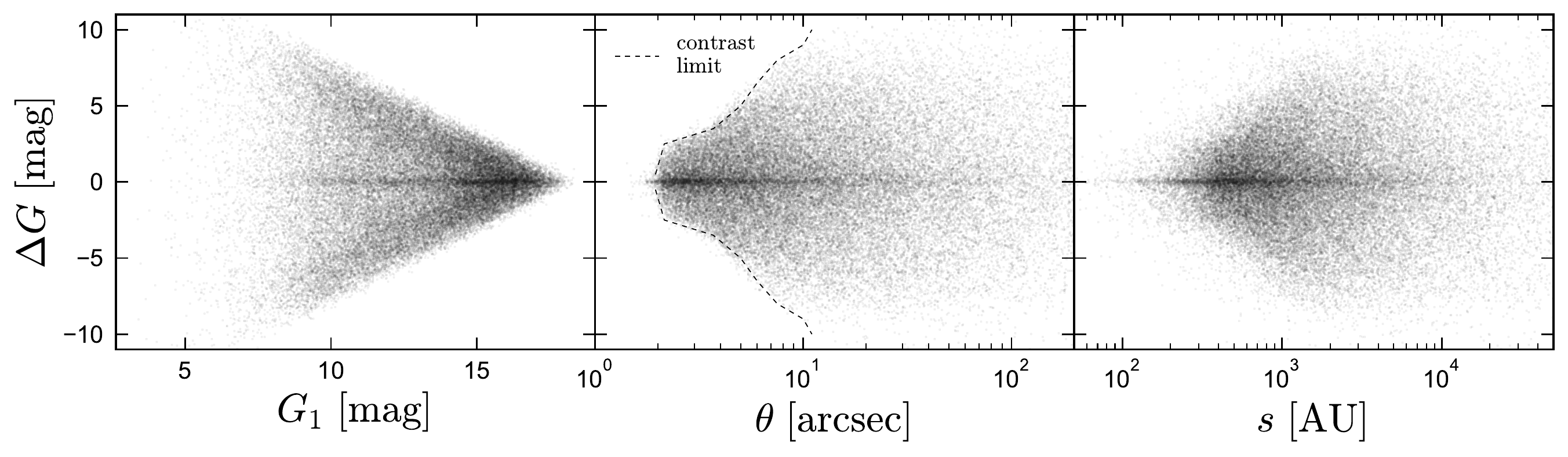}
    \caption{Apparent $G$-band magnitude difference between the two components of binaries in our catalog as a function of the apparent magnitude of the primary (left), angular separation (middle), and projected physical separation (right). The sign of $\Delta G$ is randomized for easier visualization. There is a clear excess of equal-brightness binaries with $\Delta G \approx 0$. These ``twin'' binaries are found over a large range of apparent magnitudes, preferentially at closer physical and angular separations. The middle panel shows the contrast sensitivity limit for our sample; at fixed angular separation, the probability of a companion passing our photometric quality cuts drops rapidly outside this limit due to source contamination (Appendix~\ref{sec:sensitivy}). The lack of binaries with large $\Delta G$ at small separations is a selection effect; the narrow excess at $\Delta G\approx 0$ is not.}
    \label{fig:dG_points}
\end{figure*}

An observable quantity that is closely related to the mass ratio is the difference in apparent magnitude of the two components, $\Delta G = |G_1 - G_2|$.
Figure~\ref{fig:dG_points} shows the distribution of $\Delta G$ as a function of apparent magnitude of the primary, angular separation, and projected physical separation. The sign of $\Delta G$ is randomized, such that the distribution about $\Delta G=0$ is symmetric. This is helpful in making the equal-brightness population stand out, since it would otherwise be squished against the x-axis. An excess of equal-brightness binaries is evident over a wide range of apparent magnitudes; it is strongest at close separations. As we will show, the feature is quite narrow: the density of binaries is enhanced primarily at magnitude differences of $\Delta G < 0.25$\,mag. This is much narrower than the range over which the selection function varies significantly: at $\theta > 2.5$\,arcsec, the contrast sensitivity is basically constant over $0< \Delta G < 2$ (see \citetalias{ElBadry_2018}; their Appendix A). 

The middle panel of Figure~\ref{fig:dG_points} shows that there are no binaries with small separations and large $\Delta G$ in the catalog. This is a consequence of photometric contamination at small angular separations. The dashed line shows the contrast sensitivity limit derived in Appendix~\ref{sec:sensitivy}; this is the value of $\Delta G$ at which the sensitivity is 50\% of its value at asymptotically large separations for a given $\theta$. The contrast limit is derived from the correlation function of chance alignment sources subject to similar quality cuts as real binaries. The fall-off in sensitivity is quite steep beyond the contrast limit, leading to an envelope in $\Delta G (\theta)$ beyond which no binaries are found. The drop-off toward larger $\Delta G$ is less steep as a function of physical separation $s$, since binaries with similar $s$ have different angular separations at different distances.

\begin{figure*}
    \includegraphics[width=\textwidth]{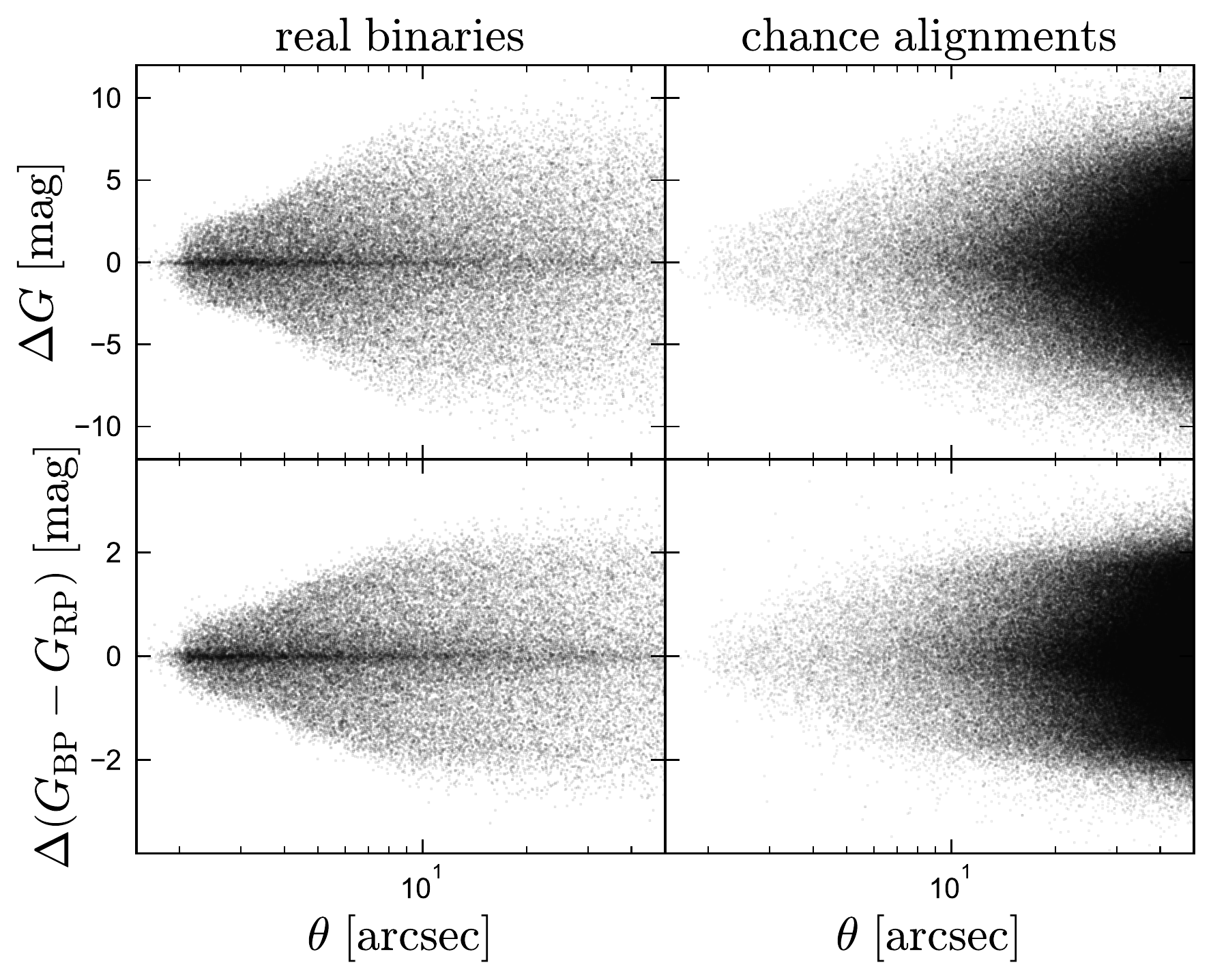}
    \caption{Difference in $G$-band magnitude (top) and color (bottom) between the two components of pairs with a range of angular separations. We compare genuine binaries (left) to chance-alignments (right). The chance alignments are required to pass the same quality cuts as the true binaries. A clear excess of pairs with $\Delta G\approx 0$ and $\Delta (G_{\rm BP} - G_{\rm RP})\approx 0$ is evident for the true binaries, but not for the chance alignments. The absence of the excess among chance alignments bolsters our confidence that the feature found in the real binaries is due to a true excess of equal-mass twins and is not a selection effect or data artifact. }
    \label{fig:chance_align}
\end{figure*}

\subsection{Is the twin excess real?}
\label{sec:sanity_checks}

To test whether the narrow excess of binaries with $\Delta G \approx 0$ is a real astrophysical effect (as opposed to a data artifact), we produced a control sample of chance alignments with similar observable properties to the real binaries. This sample was produced by repeating the procedure used to produce the real binary catalog (applying the same quality cuts and limits on photometric and astrometric precision), but requiring that the two stars have parallaxes and proper motions that are {\it inconsistent} rather than consistent. This selects pairs of stars that are close together on the sky (and thus are affected by contamination and blending in the same way as real binaries) but are not physically associated. We applied the same procedure for removing resolved higher-order multiples and potential members of bound and dissolving clusters that was used for real binaries. Because chance alignments are intrinsically rare at close angular separations, we searched out to 400 pc (rather than 200 pc for the fiducial binary catalog) to obtain better statistics. We verified that our conclusions are unchanged when only the sample within 200 pc is considered. 

In Figure~\ref{fig:chance_align}, we compare the distributions of magnitude and color difference for real binaries in our catalog (left panels) and chance alignments (right panels). Chance alignments are more common at large angular separations. The broad distribution and outer envelope of $\Delta G$ and $\Delta(G_{\rm BP}-G_{\rm RP})$ at a given $\theta$ is similar for binaries and chance alignments, reflecting {\it Gaia's} contrast sensitivity. For the chance alignments, there is no sharp excess of pairs with nearly-equal magnitude and color. Because chance alignments are subject to same cuts on astrometric and photometric quality and signal-to-noise as the real binaries, any aspects of the {\it Gaia} source detection algorithm that might be expected to produce a bias toward equal-brightness pairs should affect real binaries and chance alignments very similarly. We therefore interpret the lack of a thin excess at $\Delta G \approx 0$ and $\Delta(G_{\rm BP}-G_{\rm RP}) \approx 0 $ for chance alignments as strong evidence that the twin excess among real binaries is astrophysical. 

An initial worry was that apparent twins might be duplicate {\it Gaia} sources that were not properly removed: if a source were observed twice in two different scans without being identified as a duplicate, it could manifest in our catalog as an apparent binary pair in which the two components had essentially identical astrometry and photometry. We verified that unrecognized duplicate sources are {\it not} the source of the twin signal using the SDSS images: we visually inspected the postage stamps of several hundred equal-brightness binaries in the SDSS footprint, and all the systems indeed contain two stars. 

\begin{figure*}
    \includegraphics[width=\textwidth]{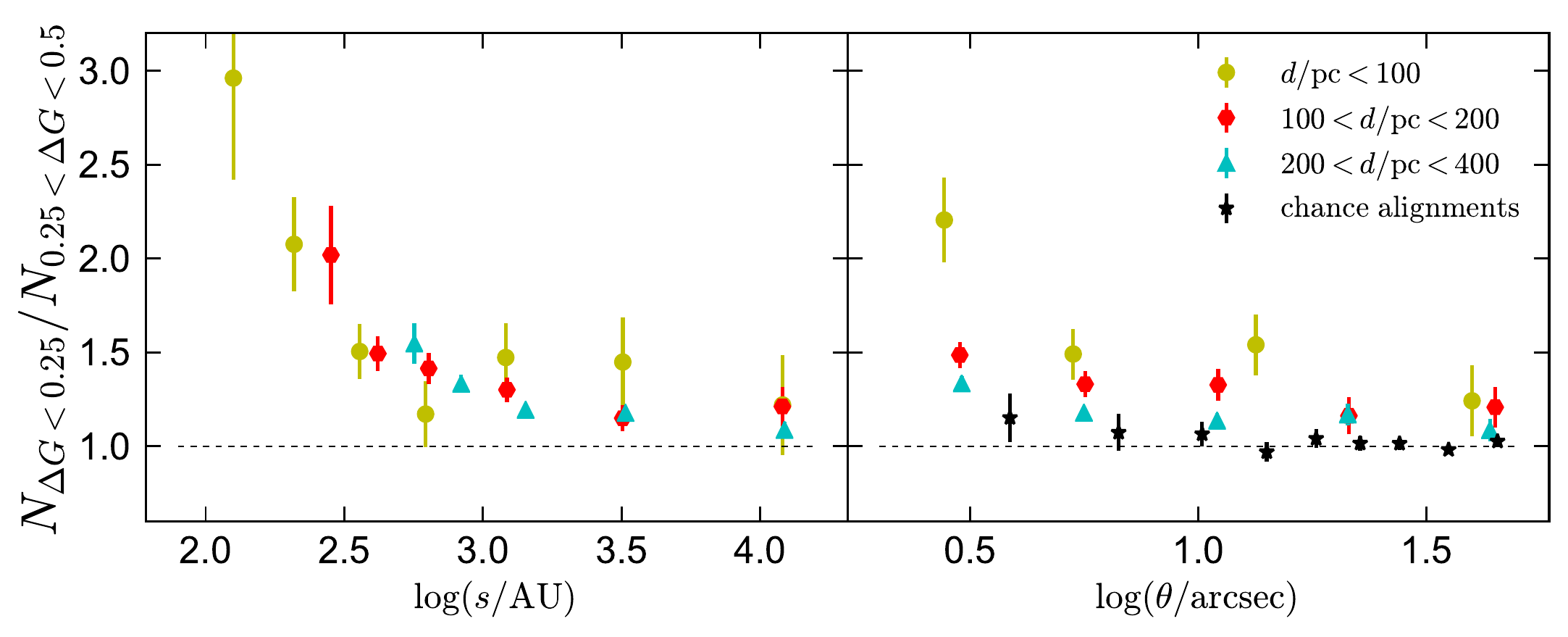}
    \caption{Ratio of the number of binaries with nearly equal magnitudes ($\Delta G < 0.25$) to the number with slightly different magnitude ($0.25 < \Delta G < 0.5$) as a function of physical separation (left) and angular separation (right). This ratio quantifies the excess of equal-brightness ``twins''.  Error bars are 1 sigma. We compare binaries at different distances, as well as a control sample of chance-alignments. 
    At fixed \textit{physical} separation, the twin excess is consistent with being independent of distance. At fixed \textit{angular} separation, it decreases with distance. 
    Together, these trends strongly suggest that the $\Delta G\approx 0$ excess (Figures~\ref{fig:dG_points} and~\ref{fig:chance_align}) reflects a real excess of equal-mass binaries, not a selection effect or photometric issue at close angular separations.}
    \label{fig:three_distances}
\end{figure*}

Another useful test to verify that the $\Delta G\approx 0$ excess is physical is to determine whether its strength depends primarily on physical or angular separation. This can be accomplished by comparing the $\Delta G$ distributions of binaries at different distances. If the excess of equal-brightness pairs were due to an observational bias against binaries that are not nearly equal-mass (or if there were an issue with the {\it Gaia} photometry causing close pairs to erroneously have the same reported magnitude), one would expect the twin excess to depend on angular separation, manifesting itself at different {\it physical} separations for binaries at different distances. If the excess is due to an astrophysical preference for equal-mass binaries, then it should be primarily a function of $s$, manifesting itself at the same physical separation but different $\theta$ for binary samples at different distances.\footnote{An implicit assumption here is that the fractional excess of twins is independent of the intrinsic properties of a binary (e.g. mass), since binaries that pass our quality cuts at larger distances will be more massive on average than those that are nearby. We show in Section~\ref{sec:results} that this assumption does not hold up in detail, which could lead to modest variation with distance in the observed twin excess at fixed physical separation.}

In Figure~\ref{fig:three_distances}, we compare the excess of equal-brightness pairs for binaries in three different distance bins as a function of physical separation (left) and angular separation (right). We measure the excess as the ratio of the number of binaries with $0 < \Delta G < 0.25$ (nearly-equal magnitudes) to the number with $0.25 < \Delta G < 0.5$ (slightly different, but still similar magnitudes). In order to include binaries at distance larger than 200\,pc, we repeated the binary search from \citetalias{ElBadry_2018}, this time searching out to 400\,pc. The additional binaries with $200 < d/{\rm pc} < 400$ are considered for this test only and are not included in the sample used to fit for the mass ratio distribution.

The left panel of Figure~\ref{fig:three_distances} shows that at fixed {\it physical} separation, the fractional excess of twins increases toward closer separations in a manner that is consistent across the three different distance samples. The right panel shows that as expected, the twin excess at fixed {\it angular} separation varies with distance. At fixed $\theta$, larger distances correspond to wider physical separation. Because the twin excess decreases with physical separation, it decreases with distance at fixed $\theta$. The  excess of equal-brightness chance alignments is also shown as a function of angular separation in the right panel of Figure~\ref{fig:three_distances}. As expected, this is nearly consistent with 0 at all angular separations, meaning that there is no strong bias toward equal-brightness pairs compared to pairs with slightly different brightness. At the closest angular separations ($\theta \lesssim 5\,\rm arcsec$), there is a slight excess of equal-brightness chance alignments, suggesting that contrast sensitivity begins to play a role. The excess for the chance alignments is less than that found for the true binaries at all distances and is self-consistently accounted for in the selection function (Appendix~\ref{sec:sensitivy}).

\begin{figure}
    \includegraphics[width=\columnwidth]{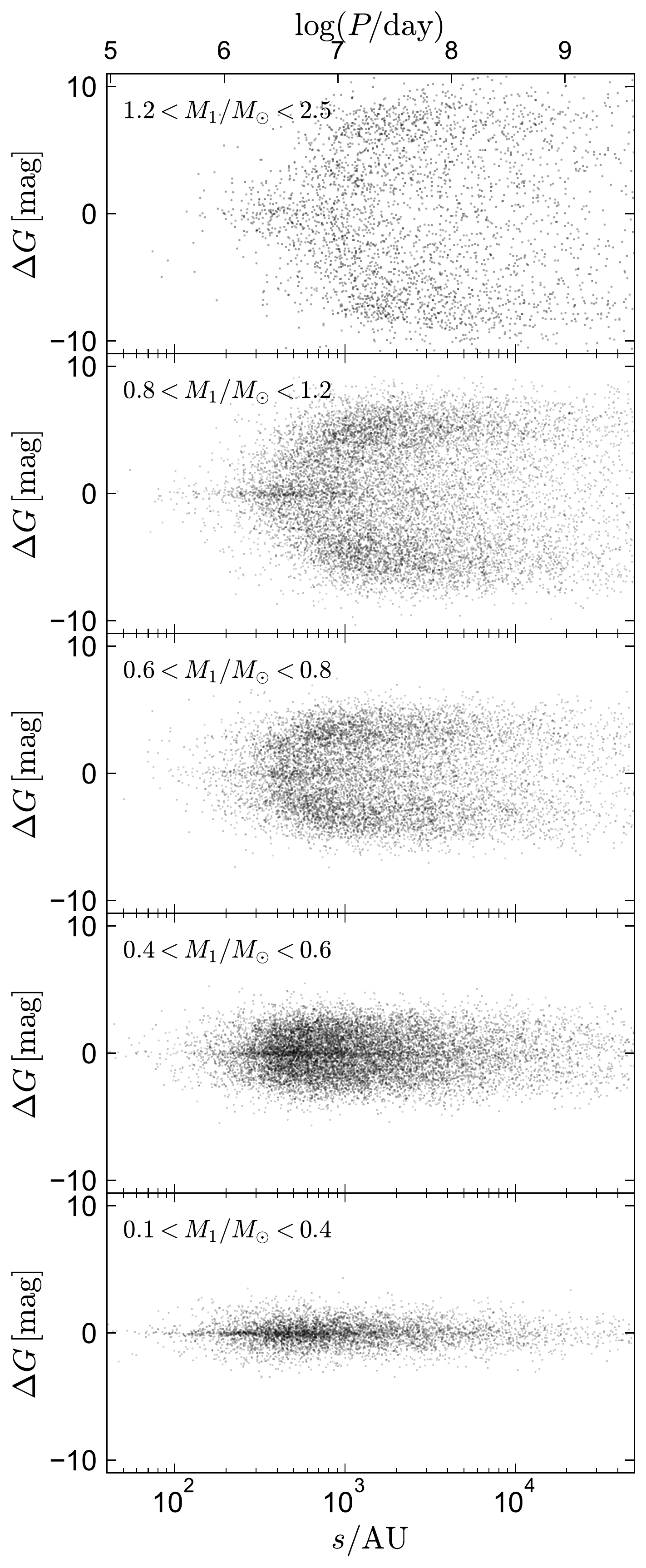}
    \caption{Magnitude difference as a function of projected physical separation for binaries with different primary masses. The sign of $\Delta G$ is randomized for easier visualization, and masses are computed from the CMD. The thin band of ``twin'' binaries with $\Delta G \approx 0$ is clearly visible in all but the highest-mass bin. In the top three bins, there is a clear preference for unequal-mass (large $\Delta G$, low-$q$) binaries. Incompleteness to faint companions prevents the detection of large $\Delta G$ companions to low-mass secondaries. }
    \label{fig:delta_G_mass_bins}
\end{figure}

Figure~\ref{fig:delta_G_mass_bins} shows distributions of $\Delta G$ for binaries in 5 different bins of primary mass. Both the overall shape of the distribution and the strength of the excess at $\Delta G\approx 0$ vary substantially between mass bins. For the highest-mass bin, $1.2 < M_1/M_{\odot} < 2.5$, the excess at $\Delta G \approx 0$ is weak, and the observed distribution (without accounting for incompleteness) peaks at $\Delta G\approx 8$, corresponding to $q\approx 0.3$ (see Figure~\ref{fig:delta_G_vs_q}). For $0.8 < M_1/M_{\odot} < 1.2$ and $0.6 < M_1/M_{\odot} < 0.8$, there is a clear excess of twins out to $s\approx 1000$\,AU, with the observed peaks in the $\Delta G$ distribution corresponding to $q\approx 0.45$ and $q\approx 0.6$, respectively. In the two lowest-mass bins, the visible twin excess appears to extend to larger separations, and there is no secondary peak in the distributions of $\Delta G$. For all mass bins, the binaries with the largest $\Delta G$ have secondaries with $M \approx 0.1 M_{\odot}$. The maximum $\Delta G$ in each panel is probably set largely by observational incompleteness, since at moderately old ages ($\gg 100\,\rm Myr$), objects below the hydrogen burning limit are several magnitudes fainter than those just above it. Incompleteness at small separations due to the angular resolution and contrast sensitivity of our catalog becomes significant at different projected physical separations for different mass bins, because the low-mass binaries that pass our quality cuts are on average at closer distances than the higher-mass binaries. 

\section{Modeling}
\label{sec:modeling}

We now turn to the intrinsic mass ratio distribution, $p(q)$. Because we expect that $p(q)$ may vary with primary mass and/or separation, we split the observed sample into 5 bins of primary mass and 7 bins of projected physical separation, fitting each of the resulting 35 subsamples independently. Our choice of bins is designed to balance the number of binaries in each bin while still covering a large dynamic range of mass and separation. 
We constrain $p(q)$ for each subsample by forward-modeling a simulated population of binaries with a given distribution of primary mass, age, metallicity, distance, physical separation, and mass ratio, passing them through the selection function, and comparing to the data. The ``data'' we consider is the distribution of the observed binaries in the 3-dimensional space of angular separation $\theta$, magnitude difference $\Delta G$, and parallax $\varpi$. The best-fitting $p(q)$ is then the one that best matches the observed data, and uncertainties in $p(q)$ are estimated via Markov chain Monte Carlo from the range of $p(q)$ that adequately reproduce the observed data. This approach requires a parameterized form of $p(q)$ (Section~\ref{sec:functional_form}), knowledge of the selection function (Section~\ref{sec:selection_function}), and a Galactic model from which simulated binaries are drawn (Section~\ref{sec:model_inputs}).

To generate a model prediction for a given set of fitting parameters, we generate a population of $N=10^6$ synthetic binaries and forward-model their distribution into the space of observables. Masses, ages, metallicities, and distances are Monte Carlo sampled from their respective distribution functions. For mass ratios and separations, we use a regular $1000 \times 1000$ grid, weighting the synthetic binary at each gridpoint by the mass ratio and separation distributions. This approach is chosen because in the fiducial model, we fit for the mass ratio and separations distributions but leave the distributions of age, mass, distance, and metallicity fixed. The number of synthetic binaries generated must be large enough that Poisson noise is negligible. We verified that $N=10^6$ binaries in each mass and separation bin is large enough that our constraints are converged and insensitive to the random seed. 

Synthetic photometry is calculated for both components from isochrones (including the effects of unresolved companions in hierarchies; see Section~\ref{sec:model_inputs}), and the observable properties of each binary are passed through the selection function. In constructing the distribution of mock-observables to be compared with the data, each synthetic binary is weighted by the selection function evaluated for its observables. Finally, the observed and simulated distributions are binned on a regular 3D grid. We default to using 100 bins in $\Delta G$, 100 bins in $\log \theta$, and 5 bins in $\varpi$ (because typical errors in $\varpi$ are larger than those in $\Delta G$ or $\theta$). The maximum value of the $\Delta G$ grid for a given primary mass bin is chosen to include the largest $\Delta G$ value for the data in that bin, and the 5 bins in $\varpi$ are chosen so that roughly the same number of observed binaries fall in each bin. Our constraints are not sensitive to the choice of bins, since they are small compared to the scale on which the data exhibit substructure. 

We re-scale the binned model prediction such that the total counts match the observed data. We calculate the likelihood for a particular set of model parameters by summing over all cells in the 3D distribution, assuming that the distribution of counts in each cell is set by a Poisson process. The log-likelihood function is 

\begin{equation}
\ln L=\sum_{m_{i}\neq0} \left[d_{i}\ln m_{i}-m_{i}-\ln(d_{i}!)\right],
\label{eq:lnL}
\end{equation}
where $m_i$ and $d_i$ are the counts in the $i$th cell of the binned model and data (``!'' denotes a factorial). We sample from the posterior using \texttt{emcee} \citep{ForemanMackey_2013}, using priors described in Section~\ref{sec:functional_form}. We use 200 walkers and draw 20,000 samples for each bin of mass and separation after a burn-in period of 200 steps per walker. Inspecting the chains, we find this to be sufficient for convergence in all cases. We carried out tests with mock data that was drawn from a known mass ratio distribution assigned with realistic observational uncertainties in order to verify that our approach yields unbiased constraints on $p(q)$.

This fitting procedure is qualitatively very similar to the method commonly used to constrain population properties such as the IMF, star formation history, unresolved binary fraction, or initial-final mass relation from CMDs \citep[e.g.][]{Dolphin_2002,Bonatto_2012, Geha_2013, ElBadry_2018_IFMR}. The difference between our approach and these studies is that we are forward-modeling the distribution of binaries in the space of angular separation, magnitude difference, and parallax rather than color and magnitude. The approach can in principle be generalized to include other observables, such as color difference or apparent magnitude of the primary, but the computational expense increases rapidly with the dimension of the space in which the likelihood function is calculated. 

We note that the data uncertainties do not enter Equation~(\ref{eq:lnL}). The implicit assumption (which does hold for our problem setup) is that the uncertainties are small compared to the scale of the bins in all quantities. We also note that for a fine grid, a majority of grid cells will contain 0 or 1 real binaries. This is not a problem; Equation~(\ref{eq:lnL}) does not make any assumptions about the magnitude of $d_i$.

\subsection{Parameterization}
\label{sec:functional_form}

\begin{figure}
    \includegraphics[width=\columnwidth]{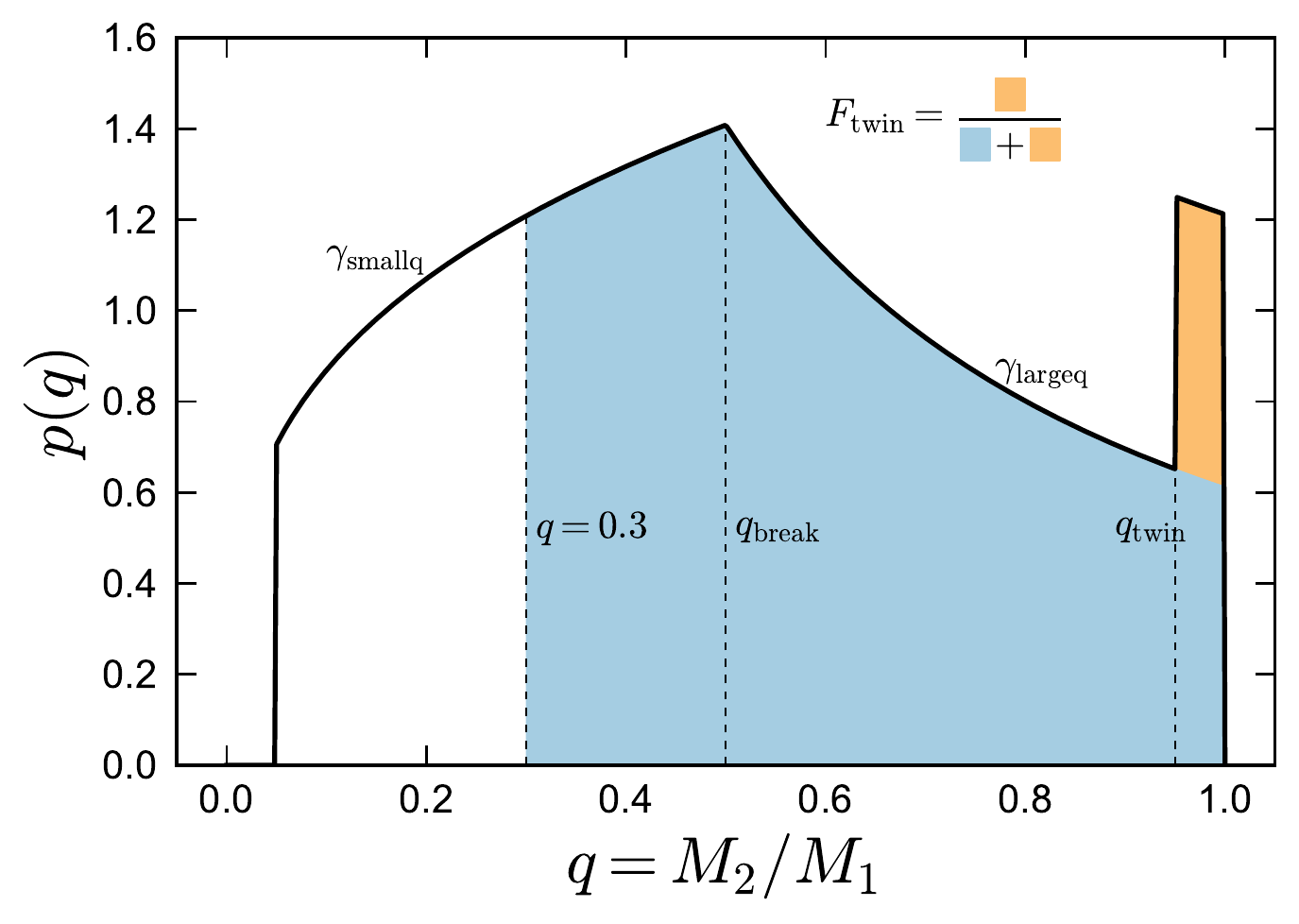}
    \caption{Parameterized mass ratio distribution. The broad part of the distribution is modeled with a broken power law of logarithmic slope $\gamma_{\rm smallq}$ at $q<q_{\rm break}$ and  $\gamma_{\rm largeq}$ at $q>q_{\rm break}$. $F_{\rm twin}$ is the \textit{excess} fraction of nearly equal-mass binaries with $q > q_{\rm twin}$, relative to the underlying power-law distribution for $q>0.3$. For this particular example, $F_{\rm twin} = 0.04$, $q_{\rm twin} = 0.95$, $\gamma_{\rm smallq} = 0.3$, $\gamma_{\rm largeq}=-1.3$, and $q_{\rm break} = 0.5$.}
    \label{fig:schematic_q_dist}
\end{figure}

We fit the mass ratio distribution by assuming a parameterized form of $p(q)$ and then obtaining constraints on the parameters. Our fiducial parameterization is motivated by the one used by \citet{Moe_2017} and is shown in Figure~\ref{fig:schematic_q_dist}. $p(q)$ is parameterized as a broken power law with logarithmic slope $\gamma_{\rm smallq}$ at $q < q_{\rm break}$ and $\gamma_{\rm largeq}$ at $q > q_{\rm break}$. An possible excess (or deficit) of twins is added on top of the power law component at $q>q_{\rm twin}$. This excess is modeled as a step function, with the magnitude such that the integrated excess of twins divided by the total number of binaries with $q > 0.3$ is $F_{\rm twin}$. The reason for this choice (as opposed to normalizing relative to {\it all} binaries) is that, compared to the constraints at $q > 0.3$, the constraints on $p(q)$ at small $q$ are often weak due to incompleteness. $p(q)$ is set to 0 at $q < 0.05$ to prevent divergence when $\gamma_{\rm smallq}<-1$. Because we are not sensitive to companions with $q < 0.05$, this has little effect on our results.  

There is no a priori motivation for this particular parameterization: because the physics that set $p(q)$ are imperfectly understood, we simply require a functional form that is sufficiently flexible to reproduce the observed distributions of $\Delta G$. We have experimented with other forms of $p(q)$, including adding an additional break point to the power law and modeling the twin excess as ramping up linearly instead of increasing stepwise. However, we find the form shown in Figure~\ref{fig:schematic_q_dist} to provide a good fit in all mass and separation bins, with more complicated models providing only marginal improvements. We thus use this functional form for our main analysis. In Appendix~\ref{sec:func_form}, we show results of using alternative parameterizations, including one that smoothly transitions between the two power laws instead of including a sharp break and one that leaves the shape of the twin excess flexible. 

There are two differences between our parameterization of $p(q)$ and the one used by \citet{Moe_2017}, who focused primarily on binaries with $M_1 \gtrsim M_{\odot}$. First, they fixed $q_{\rm break} = 0.3$, a value that was appropriate for their analysis because several studies have found the mass ratio distribution to peak at $q \approx 0.2-0.3$ for binaries with $M_1 \gtrsim M_{\odot}$ \citep{Duquennoy_1991, Gullikson_2016, Murphy_2018}. We also use $q_{\rm break} = 0.3$ for our highest-mass bin, but we find $q_{\rm break} = 0.5$ to provide a better fit at lower masses (see Appendix~\ref{sec:qbreak}). Second, they fixed $q_{\rm twin} = 0.95$, roughly the value found for spectroscopic binaries \citep[e.g.][]{Tokovinin_2000}. In order to identify or rule out trends with mass and separation, we leave $q_{\rm twin}$ as a free parameter.

\begin{table*}
\centering
\caption{Summary of priors adopted in each primary mass bin. We use the same priors for all separation bins. $\mathcal{U}(a,b)$ represents a uniform distribution over $[a,b]$, and $\mathcal{N}(\mu,\sigma)$ represents a normal distribution with mean $\mu$ and dispersion $\sigma$. $\gamma_s$ is the logarithmic slope of the local separation distribution; i.e., $p(s) \propto s^{\gamma_s}$. Other parameters are described in Figure~\ref{fig:schematic_q_dist}.}
\label{tab:priors}

\begin{tabular}{ |c|c|c|c|c|c|} 
 \hline
      & $0.1 < M_1/M_{\odot} < 0.4$ & $0.4 < M_1/M_{\odot} < 0.6$ & $0.6 < M_1/M_{\odot} < 0.8$ & $0.8 < M_1/M_{\odot} < 1.2$ & $1.2 < M_1/M_{\odot} < 2.5$ \\
      \hline
 $F_{\rm twin}$ & $\mathcal{U}(-1, 1)$ & $\mathcal{U}(-1, 1)$ & $\mathcal{U}(-1, 1)$ & $\mathcal{U}(-1, 1)$ & $\mathcal{U}(-1, 1)$ \\ 
 $q_{\rm twin}$ & $\mathcal{U}(0.93, 1)$ & $\mathcal{U}(0.93, 1)$ & $\mathcal{U}(0.93, 1)$ & $\mathcal{U}(0.93, 1)$ & $\mathcal{U}(0.93, 1)$ \\ 
 $\gamma_{\rm largeq}$ & $\mathcal{N}(0.5, 1)$ & $\mathcal{N}(-1,1)$ & $\mathcal{N}(-1,1)$ & $\mathcal{N}(-1,1)$ & $\mathcal{N}(-1,1)$ \\ 
 $\gamma_{\rm smallq}$ & --- & $\mathcal{N}(0.5, 0.5)$ &  $\mathcal{N}(0, 0.5)$ &  $\mathcal{N}(0, 0.5)$ &  $\mathcal{N}(0, 0.5)$ \\ 
 $\gamma_{\rm s}$ & $\mathcal{N}(-1.5, 1)$ & $\mathcal{N}(-1.5, 1)$ & $\mathcal{N}(-1.5, 1)$ & $\mathcal{N}(-1.5, 1)$ & $\mathcal{N}(-1.5, 1)$ \\ 
 $q_{\rm break}$ & --- & 0.5 & 0.5 & 0.5 & 0.3 \\ 
 \hline
\end{tabular}
\end{table*}

Along with the parameters of the mass ratio distribution, we also fit for $\gamma_s$, the local logarithmic slope of the separation distribution in each bin of primary mass and separation. Opik's law (a uniform distribution of $\log(s)$) corresponds to $\gamma_s=0$.

Our adopted priors are listed in Table~\ref{tab:priors}. 
For the lowest-mass bin, the smooth component of $p(q)$ is modeled as a single power law, because no useful constraints can be obtained at $q \lesssim 0.5$. Our priors on $\gamma_{\rm smallq}$ and $\gamma_{\rm largeq}$ are loosely motivated by constraints from the literature (particularly \citealt{Tokovinin_2014} and \citealt{Moe_2017}) but are fairly weak. We use a uniform flat prior on $F_{\rm twin}$. We require $q_{\rm twin}>0.93$ in order to prevent cases where a large $F_{\rm twin}$ is combined with $q_{\rm twin}\ll 1$ such that a broad ``twin excess'' simply modifies the overall shape of $p(q)$ but does not actually correspond to a sharp increase near $q=1$. We show in Appendix~\ref{sec:how_sharp} that, where a twin excess is significant, it is always narrow ($q_{\rm twin} \gtrsim 0.94$).    

\subsection{Selection Function}
\label{sec:selection_function}
Because both components of a binary must pass astrometric and photometric quality cuts, the binaries in our catalog are relatively bright. The median apparent magnitude of all stars in our fitting sample (considering primaries and secondaries together) is $\left\langle G\right\rangle =14.5$, and 90\% (99\%) of stars fall in the range $9.1 < G < 17.4$ ($6.6 < G < 18.1$). For stars in this magnitude range, {\it Gaia} DR2 is nearly complete outside of crowded fields \citep{Arenou_2018, Sollima_2019}. The completeness is not quite 100\% due to a variety of issues,  but it is primarily a function of position on the sky, not color or magnitude. The selection function for our sample is thus determined by the cuts imposed on astrometric and photometric precision. 

In order for a binary to appear in the catalog, (a) both components must be bright enough that they individually pass the cuts we impose on \texttt{parallax\_over\_error} and photometric signal to noise, and (b) they must not be so close on the sky that the photometry of either component is significantly contaminated. The selection function for binaries can thus be expressed as a product of the two components' single-star detection probabilities and a contrast sensitivity cross term: 
\begin{align}
    \label{eq:s_12}
    s_{\rm binary}=s_{1}\times s_{2}\times s_{\Delta G}(\theta).
\end{align}
Here $s_1$ and $s_2$ represent the independent probabilities of detecting an isolated star with the observable properties of star 1 and star 2; they depend primarily on apparent magnitude and color. $s_{\Delta G}(\theta)$ quantifies the reduction in the probability of detecting star 1 and star 2 together, {\it relative to the probability of detecting them at asymptotically large separation}. It is primarily a function of the angular separation of the two stars and their flux ratio. For example, $s_{\Delta G}(\theta)\approx 0$ at $\theta < 2\,{\rm arcsec}$ and 1 at $\theta > 10\,{\rm arcsec}$; at intermediate separations, it depends strongly on $\Delta G$. We calculate the single-star selection function given our quality cuts in Appendix~\ref{sec:single_star_term} and  $s_{\Delta G}(\theta)$ in Appendix~\ref{sec:sensitivy}. The derived selection functions are then validated in Appendix~\ref{sec:model_validation}.

\subsection{Model inputs}
\label{sec:model_inputs}
We draw primary masses and system ages assuming a \citet{Kroupa_2001_IMF} IMF and a constant star formation history over the last 10\,Gyr. Because suspected members of bound and dissolving clusters are removed from our binary catalog, we remove synthetic binaries with $\rm age< 100$\,Myr.  We assume that the wide binary fraction, $f_{\rm wb}$, scales with mass as $f_{\rm wb}\propto M_1^{\alpha_{\rm wb}}$, where $\alpha_{\rm wb} = 0.4$ is a constant (i.e., higher-mass primaries are more likely to have wide binary companions). The effect of this assumption is that primary masses are drawn from a distribution with logarithmic slope $\alpha_{\rm IMF} + \alpha_{\rm wb}$, where $\alpha_{\rm IMF}$ is the local logarithmic slope of the IMF ($\alpha_{\rm IMF} = -2.3$ for $M_1/M_{\odot} > 0.5$ and $\alpha_{\rm IMF} = -1.3$ for $M_1/M_{\odot} < 0.5$). We find that with this choice of $\alpha_{\rm wb}$, our model predicts a distribution of primary magnitudes in reasonably good agreement with that of the binary catalog when the selection function is taken into account. 

We model the intrinsic spatial distribution of all stars as a plane-parallel exponential with the Sun at the midplane. The exponential scale height increases with stellar age \citep[e.g.][]{Nordstrom_2004, Seabroke_2007}, because older stars (a) were born from kinematically hotter gas and (b) have been dynamically heated more since their formation \citep[e.g][]{Ting_2018}. We use a fit to the empirical age-scale height relation recently measured by \citet{Sollima_2019} using \textit{Gaia} star counts:
\begin{align}
    \log\left(h_{z}/{\rm pc}\right)=0.53\log\left({\rm age}/{\rm yr}\right)-2.65.
    \label{eq:hz}
\end{align}
I.e., the scale height increases from 40\,pc for stars of age 100 Myr, to 130 pc at age 1 Gyr, to 450 pc at age 10 Gyr. We show in Appendix~\ref{sec:single_star_validation} that this leads to a predicted distance distribution in good agreement the data. 

We use the tabulated empirical metallicity distribution function (MDF) for our binary catalog that was measured in \citet{ElBadry_2019} by considering a subset of $\sim$8000 binaries in the catalog for which $\rm [Fe/H]$ was measured spectroscopically for at least one component. 
Most of the binaries in our catalog are disk stars, with a median metallicity of $\left\langle \left[{\rm Fe/H}\right]\right\rangle \approx-0.1$ and tails extending to $[\rm Fe/H] \approx -1.0$ and $[\rm Fe/H] \approx +0.4$. We do not include any variation in the MDF with age or distance.

We generate synthetic photometry in the {\it Gaia} DR2 bands from \citet{Evans_2018} using PARSEC\footnote{\href{http://stev.oapd.inaf.it/cgi-bin/cmd}{http://stev.oapd.inaf.it/cgi-bin/cmd}} isochrones \citep{Bressan_2012, Chen_2014}. Just as for the real data, we remove synthetic binaries in which either component has evolved off the main sequence. 
For companions with $M_2 < 0.1\,M_{\odot}$, we supplement the PARSEC models with \texttt{BT-Settl} models for very low-mass stars and brown dwarfs \citep{Allard_2012, Allard_2014}.\footnote{Synthetic photometry in {\it Gaia} DR2 bands is computed for the \texttt{BT-Settl} models using the Pheonix web simulator, available at \href{https://phoenix.ens-lyon.fr/simulator-jsf22-26/index.faces}{phoenix.ens-lyon.fr/simulator-jsf22-26/index.faces}.} We include companions with masses as low as $0.01\,M_{\odot}$ in our model for completeness. However, these have very little effect on our results, because substellar companions are too faint and red to pass our quality cuts except at very young ages.

\begin{figure}
    \includegraphics[width=\columnwidth]{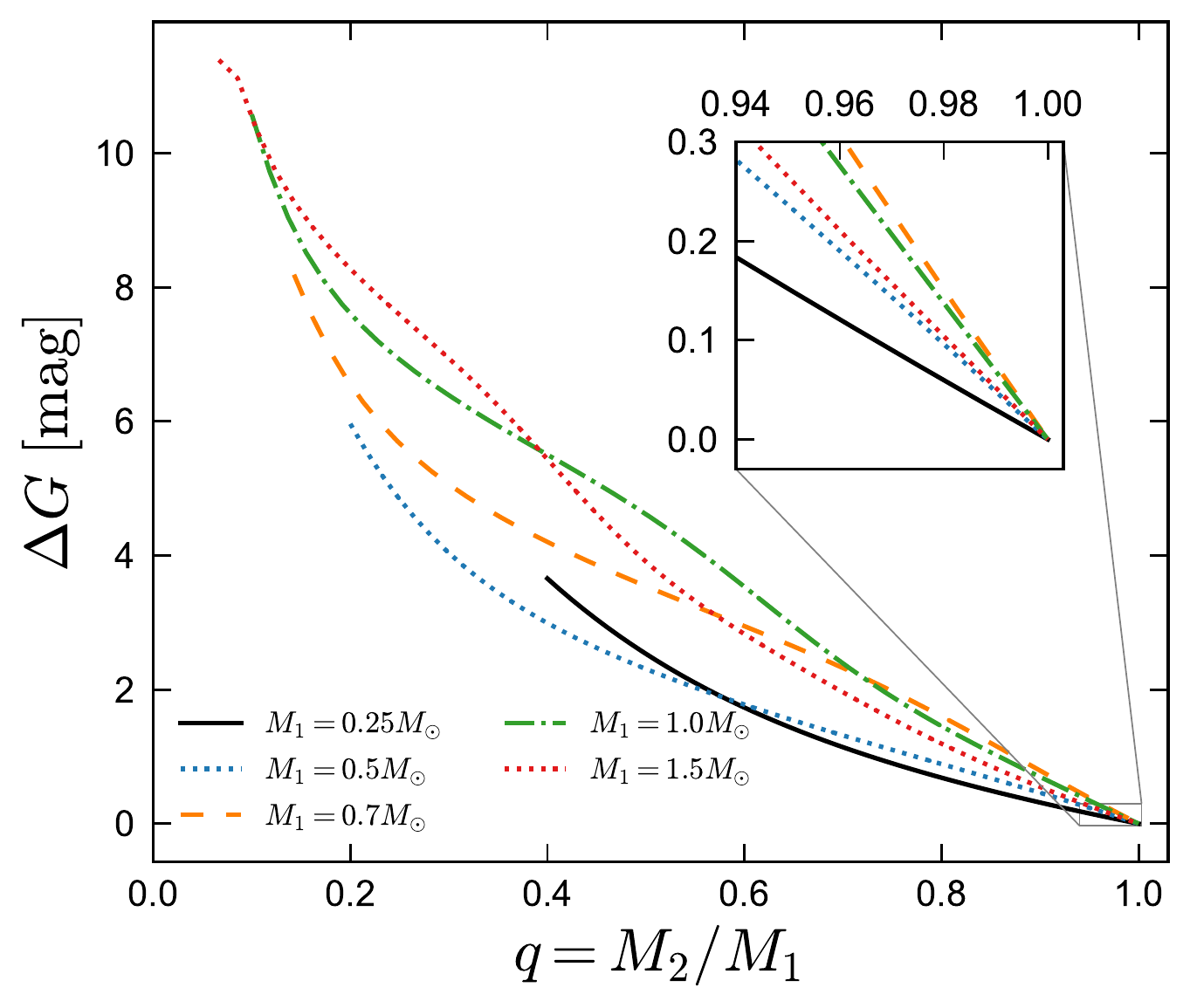}
    \caption{Magnitude difference vs. mass ratio for binaries of different primary mass, computed from PARSEC isochrones. We assume both components have $[\rm Fe/H]=0$, adopting an age of 1 Gyr for the highest primary mass and 5 Gyr for the others. Because the mass-luminosity relation varies with mass, the relation between $\Delta G$ and $q$ does as well. A magnitude difference of 0.25 mag (roughly where the twin excess becomes significant in our catalog) corresponds to a mass ratio of 0.93 to 0.97, depending on primary mass.}
    \label{fig:delta_G_vs_q}
\end{figure}

Figure~\ref{fig:delta_G_vs_q} shows the relation between $q$ and $\Delta G$ predicted for PARSEC isochrones. We show relations for 5 different primary masses, corresponding approximately to the median masses of primaries in each of our five mass bins, and use a minimum companion mass of $0.1M_{\odot}$ in all cases.
Comparing to Figure~\ref{fig:delta_G_mass_bins}, it is evident that the lowest-mass secondaries have $M_2\approx 0.1M_{\odot}$ in all bins of primary mass. As for the excess of equal-brightness binaries, it is primarily manifest over $0 < \Delta G < 0.25$ (though there is some variation with mass; see Figure~\ref{fig:posterior_predictive}), which corresponds to $0.93 \lesssim q < 1$.

Some components of wide binaries have their own spatially-unresolved close companions. We incorporate these in our models following observed binary statistics. The probability that a star has a close companion is taken to be a function of its mass, increasing from 20\% at $M < 0.2 M_{\odot}$, to 30\% at $0.2 < M/M_{\odot} < 0.5$, to 35\% at $0.5 < M/M_{\odot} < 0.8$, to 45\% at $M/M_{\odot} > 0.8$ \citep{Duchene_2013}. For components that are assigned a binary companion, we draw the unresolved companion mass from a mass ratio distribution that is uniform between $q_{\rm min} = 0.1 M_{\odot}/ M_1$ and 1. We assume that the two components' probabilities of having a close companion are independent and neglect dynamical stability constraints. This prescription reproduces the morphology of the observed CMD reasonably well. In particular, the fraction of stars within 200\,pc that fall in the ``suspected unresolved binary'' region of the CMD (yellow points in the bottom right panel of Figure~\ref{fig:cmds}) at a given $\rm M_G$ is reproduced within $\sim$25\% over $0.1 < M/M_{\odot} < 1.0$.

We do not include extinction or reddening due to dust in our model and do not attempt to correct the data for it. Because the stars in our catalog are nearby $(d < 200\,\rm pc)$, the effects of extinction are expected to be modest. The morphology of the CMD (e.g. the compactness of the red clump and main sequence; see Figure~\ref{fig:cmds}) validates this assumption. Moreover, because the two components of a binary have similar position on the sky (within an arcminute in most cases) and similar distance, the extinction toward both components is expected to be similar.

We discuss the sensitivity of our constraints on the mass ratio distribution to various model ingredients in Appendix~\ref{sec:systmatics}. The largest systematic uncertainties come from the choice of stellar models. All the systematics we consider primarily affect constraints at low $q$. This is also true for uncertainties in the completeness function: because the two components of binaries with $q\approx 1$ have similar magnitudes and colors, incompleteness affects them similarly. The translation between the distribution of $\Delta G$ and $p(q)$ is thus more straightforward at $q\approx 1$ than at $q \ll 1$.

\section{Results}
\label{sec:results}

Our fitting produces samples from the posterior distribution of free model parameters for each bin of mass and separation. These samples translate to marginalized constraints on each parameter of the mass ratio distribution and the covariances between them. An example for a single mass and separation bin is shown in Figure~\ref{fig:corner}. Marginalized constraints for all bins are listed in Appendix~\ref{sec:full_constraints}. 

\begin{figure*}
    \includegraphics[width=\textwidth]{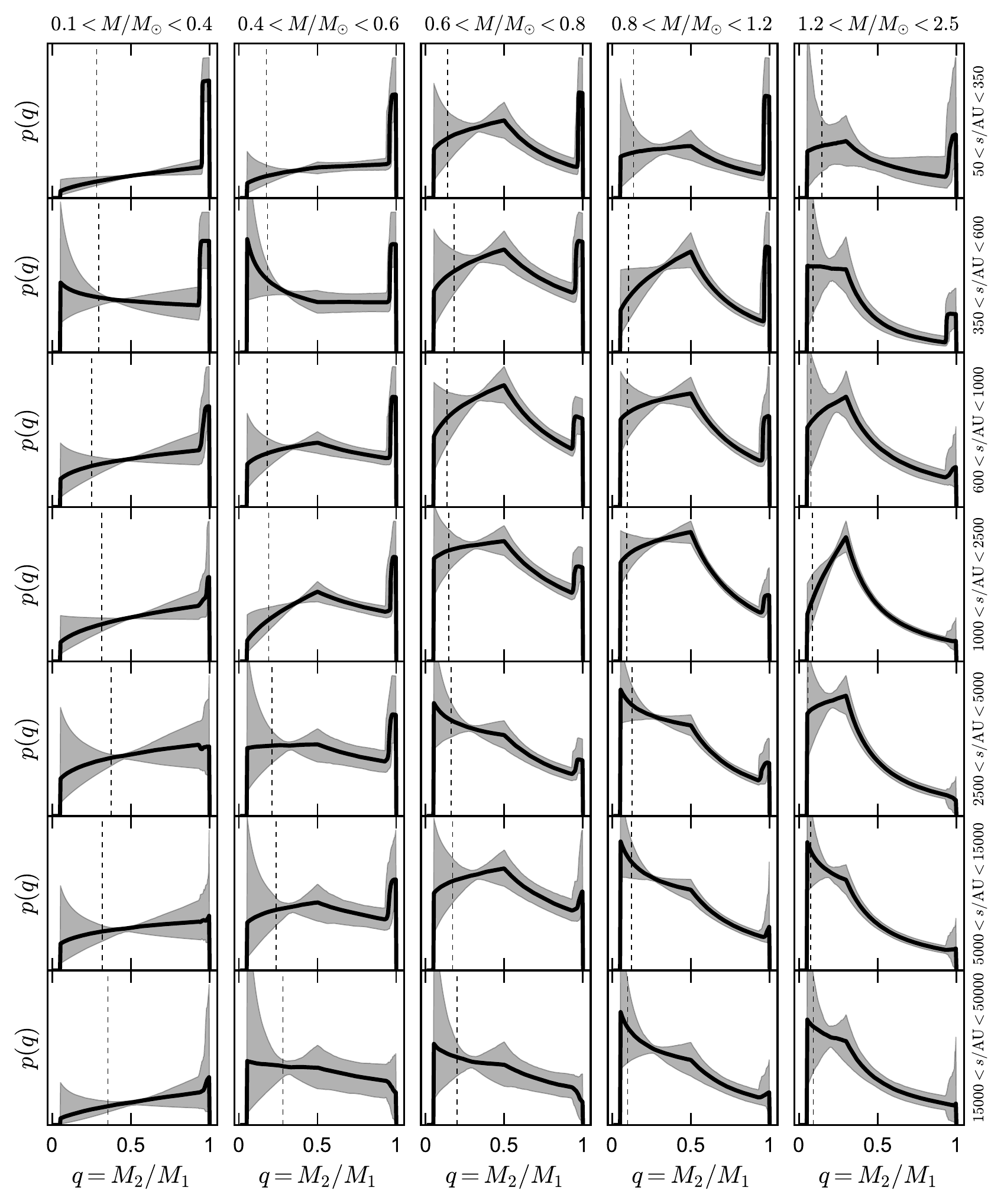}
    \caption{Mass ratio distribution constraints. Each panel corresponds to a single bin of primary mass (increasing left to right) and physical separation (increasing top to bottom). Gray shaded regions show 95.4\% probability. Normalization is arbitrary; the scale of the y-axis is linear and begins at 0. Vertical dashed lines show the lowest mass ratio of observed binaries in each bin. }
    \label{fig:qdist_ranges}
\end{figure*}

The 35 panels of Figure~\ref{fig:qdist_ranges} show median and 2 sigma constraints on the mass ratio distribution for 35 bins of primary mass and separation. We do not show the y-axis ticks to avoid cluttering the figure, as the limits are different in each panel. The uncertainties (shaded regions) are derived by drawing 1000 samples from the posterior, calculating a normalized $p(q)$ for each sample, and then shading the middle 95.4\% range of these samples at each $q$. Solid black lines show the median at each $q$. Because $p(q)$ is normalized, uncertainty in the mass ratio distribution at {\it any} $q$ translates to uncertainty in the normalization of $p(q)$ at {\it all} $q$. This reflects the fact that it is impossible to know the total fraction of all binaries that fall in some mass ratio range if one does not know how many binaries with low mass ratios are missed. However, $\gamma_{\rm largeq}$, the slope of the mass ratio distribution at $q > q_{\rm break}$, is usually well-constrained. Dashed vertical lines in each panel show the lowest-mass ratio observed binary included in that bin. This roughly corresponds to the observational completeness limit and marks the mass ratio below which meaningful constraints cannot be obtained.

The fact that $p(q)$ is modeled as a double power law with a sharp break leads to some unphysical artifacts in Figure~\ref{fig:qdist_ranges}, including a sharp change in slope at $q=q_{\rm break}$ and artificially low uncertainty just below $q_{\rm break}$, which is caused by $p(q)$ ``pivoting'' about this point as $\gamma_{\rm smallq}$ varies and $p(q)$ is renormalized. We show in Appendix~\ref{sec:smoothly_broken} that these features are not present when we fit a more flexible ``smoothly-broken'' power law model. However, doing so introduces parameter covariances that are not present for the fiducial form of $p(q)$. We use the simpler sharply broken power law as our fiducial model to facilitate easier comparison between different mass and separation bins and comparison with the literature. 

\begin{figure*}
    \includegraphics[width=\textwidth]{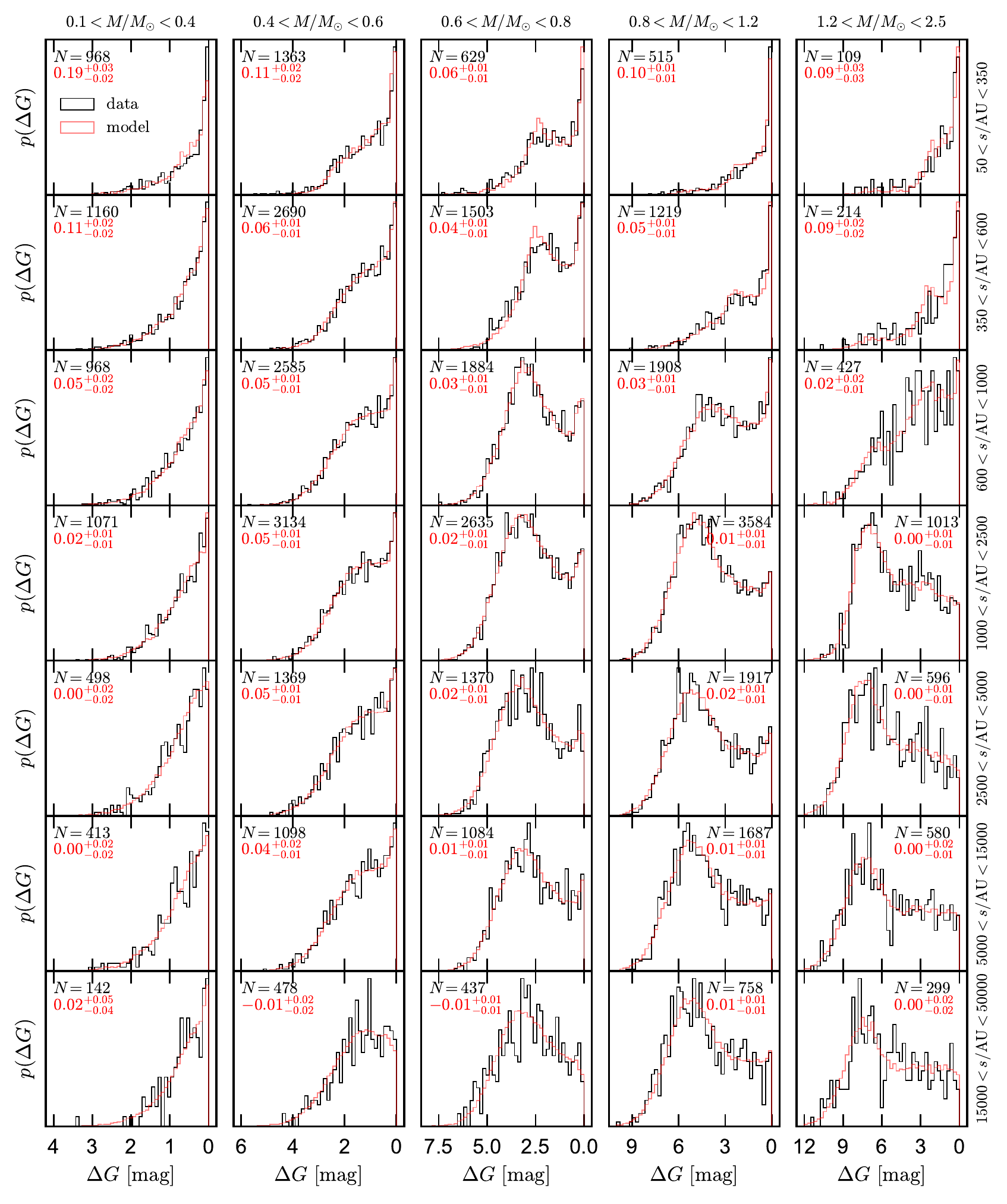}
    \caption{Magnitude difference between the two components of a binary (linear y-axis; normalization is arbitrary). Black histograms show real binaries, split into bins of primary mass (increasing left to right) and physical separation (increasing top to bottom). The twin excess can be seen in many panels as an excess of binaries with $\Delta G \approx 0$. Red histograms show Monte Carlo populations generated from the best-fit model. The number of observed binaries (black) and the marginalized 1 sigma constraints on $F_{\rm twin}$ (red) are listed in each panel. Overall, the model is quite successful in matching the observed distributions. }
    \label{fig:posterior_predictive}
\end{figure*}

Figure~\ref{fig:posterior_predictive} compares the observed 1D distributions of $\Delta G$ to the predictions of the best-fit model; this is useful for assessing the quality of the fits. In each panel, we note the number of observed binaries and the marginalized 1 sigma constraints on $F_{\rm twin}$. In generating the model predictions, we use the median of the 1D marginalized posterior distribution for each free parameter. We then generate a Monte Carlo realization of the binary population, drawing primary masses, mass ratios, separations, ages, metallicities, and distances from the appropriate joint distributions, calculating synthetic photometry, and weighting each binary by the selection function evaluated for its observables. 

Overall, the model predictions are in good agreement with the observed distributions of $\Delta G$. This indicates that our parameterization of $p(q)$ is suitable and sufficiently flexible. In panels where $F_{\rm twin}$ is inconsistent with 0, there is a clear excess of equal-brightness binaries. The distributions of $\Delta G$ near $\Delta G=0$ are also adequately reproduced by the model, indicating that the simple ``step function'' model for the twin excess is consistent with the data. Although it is not shown here, we also find the best-fit models to predict distributions of other observables (angular separation, parallax, and apparent magnitude of the primary) in good agreement with the data.

\begin{figure}
    \includegraphics[width=\columnwidth]{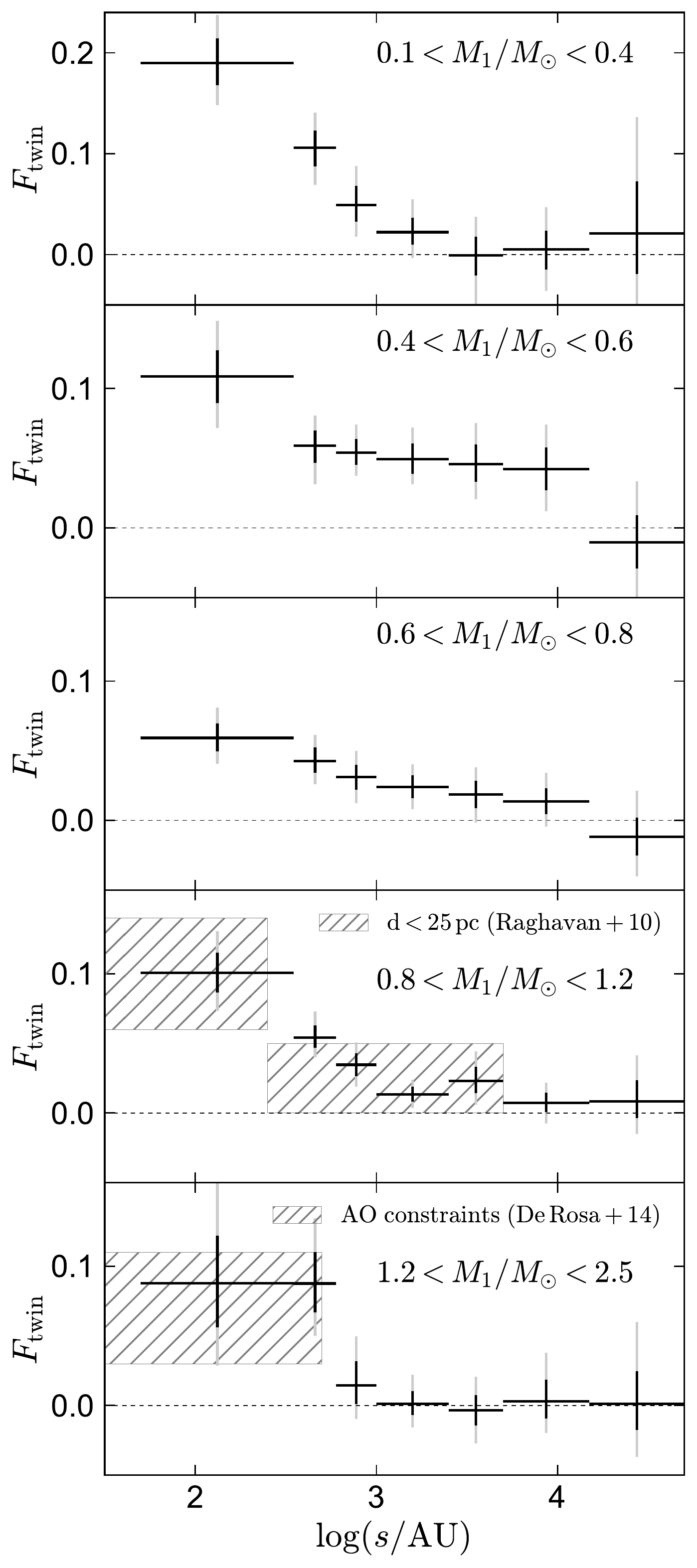}
    \caption{Marginalized constraints on $F_{\rm twin}$, the fractional excess of binaries with nearly equal mass, vs. physical separation. Each panel shows a separate bin of primary mass. Dark and light error bars show 1 and 2 sigma constraints. In all mass bins, $F_{\rm twin}$ declines with increasing separation and is consistent with 0 at the largest separations. The twin excess reaches the widest separations for $0.4 < M_1/M_{\odot} < 0.6$, where $F_{\rm twin}$ is $\sim$5\% out to 15,000 AU. We compare to constraints from the nearby solar-type binary sample of \citet[][panel 4]{Raghavan_2010} and the adaptive optics imaging survey from \citet[][panel 5]{DeRosa_2014}; these were derived by \citet{Moe_2017}. }
    \label{fig:Ftwin_vs_separation}
\end{figure}

Figure~\ref{fig:Ftwin_vs_separation} shows the constraints on $F_{\rm twin}$ as a function of mass and projected physical separation. Light and dark error bars show middle 95.4\% and 68.2\% of the marginalized posterior distributions. Note that the y-axis scale is different in the top panel. In all bins of primary mass, the excess twin fraction falls with increasing separation and is negligible at the largest separations we consider ($s > 15,000$\,AU). At fixed separation, the magnitude of the excess varies with primary mass. For close separations ($50 <s/{\rm AU} < 350$), $F_{\rm twin}$ is largest in the lowest bin of primary mass ($0.1 < M_1/M_{\odot} < 0.4$). This may be in part because low-mass primaries in our sample are at closer distances on average, such that the median separation within the ($50 <s/{\rm AU} < 350$) bin is smaller than for higher-mass primaries. 

The maximum separation out to which there is a significant twin excess also varies with primary mass. For the highest-mass bin, $F_{\rm twin}$ is consistent with 0 at $s>600$\,AU. For solar-type primaries, $F_{\rm twin}$ is only consistent with 0 at $s>5000$\,AU, but it is $<3\%$ for $s > 1000$\,AU. The fall-off in $F_{\rm twin}$ with increasing separation is shallowest for $0.4 < M_1/M_{\odot} < 0.6$ and $0.6 < M_1/M_{\odot} < 0.8$, with a larger normalization for $0.4 < M_1/M_{\odot} < 0.6$. Here $F_{\rm twin}$ is almost independent of separation over $350<s/{\rm AU}<15,000$. Finally, the fall-off steepens again in the lowest primary mass bin, where $F_{\rm twin}\approx 0$ beyond 2500\,AU. We discuss possible interpretations of these trends in Section~\ref{sec:widening}.  

In the bottom two panels of Figure~\ref{fig:Ftwin_vs_separation}, we compare our constraints on $F_{\rm twin}$ to the 1 sigma constraints obtained by \citet[][their Tables 8 and 11]{Moe_2017} for binaries in similar mass and separation ranges. The constraints for solar-type stars ($0.8 <  M_1/M_{\odot} < 1.2$) were obtained from the solar neighborhood sample of \citet{Raghavan_2010}. Those shown in the bottom panel were obtained from the AO-assisted survey of visual binaries with A-type primaries ($1.7 <  M_1/M_{\odot} < 2.3$) described in \citet{DeRosa_2014}. Reassuringly, constraints for both mass bins are consistent with those obtained from our catalog. Because the binary sample we analyze is larger than the \citet{Raghavan_2010} and \citet{DeRosa_2014} samples, we can tighten the uncertainties on $F_{\rm twin}$ at large separations, showing, for instance, that $F_{\rm twin}$ for solar-type primaries is inconsistent with 0 out to $s\approx 5000$\,AU. 

\begin{figure}
    \includegraphics[width=\columnwidth]{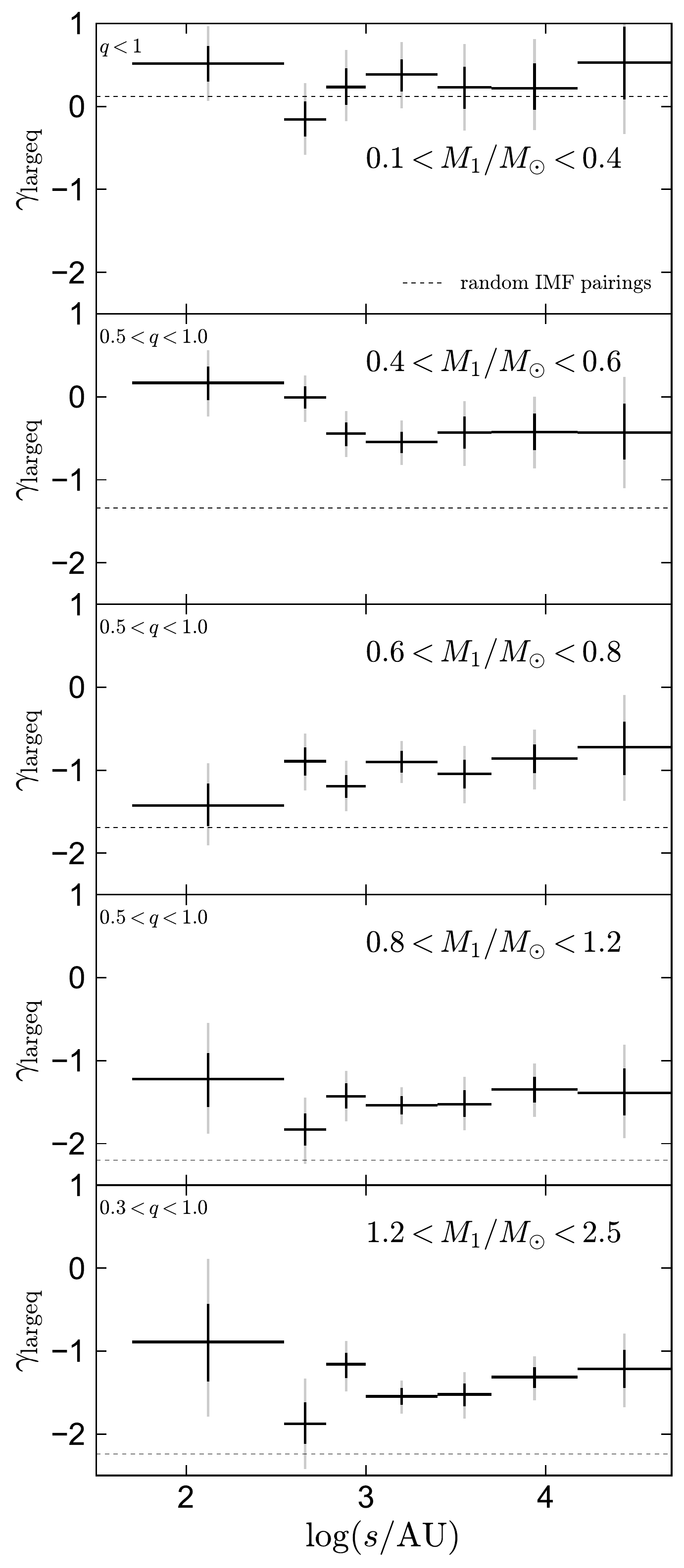}
    \caption{Marginalized constraints on $\gamma_{\rm largeq}$, the logarithmic slope of the power law component of the mass ratio distribution at large $q$, vs. projected separation. Dark and light error bars show 1 and 2 sigma constraints. The range of $q$ over which $\gamma_{\rm largeq}$ is fit in each mass bin is noted in the upper left of each panel.
    Dashed lines show the slope expected if binary component masses were drawn from the IMF and paired randomly. Except in the lowest mass bin, $p(q)$ is more bottom-heavy than uniform, but it is always more top-heavy than expected for random IMF pairings.}
    \label{fig:gamma_vs_separation}
\end{figure}

Figure~\ref{fig:gamma_vs_separation} shows constraints on $\gamma_{\rm largeq}$ as a function of separation, again separating binaries by primary mass. The dashed line in each panel shows the slope that would be expected for random pairings from the IMF. This is obtained by fitting a power law mass ratio distribution over the same range of primary masses and mass ratios to a simulated population of binaries in which the masses of both components are drawn from a \citet{Kroupa_2001_IMF} IMF and paired randomly. 

Consistent with previous work \citep[e.g.][]{Lepine_2007, Reggiani_2011, Duchene_2013, Moe_2017}, we find that the mass ratio distribution is {\it not} consistent with random pairings from the IMF, but is weighted toward higher mass ratios than would be expected in such a scenario. Whether this is an imprint of the binary formation process or in part reflects the fact that binaries with higher mass ratio have higher binding energy and are thus more difficult to disrupt is an open question. Any trends in $\gamma_{\rm largeq}$ with separation are weak over the separation range we probe: at the 2 sigma level, our constraints are consistent with a separation-independent $\gamma_{\rm largeq}$ over $100 \lesssim s/{\rm AU} < 50,000$ in all mass bins. However, they are not consistent with being independent of primary mass: $p(q)$ becomes increasingly bottom-heavy (lower $\gamma_{\rm largeq}$) with increasing $M_1$. 

The fact that $\gamma_{\rm largeq}$ does not vary much with separation can serve as a strong constraint on formation models for wide binaries. 
It has frequently been argued that while binaries with separations of $100 \lesssim s/{\rm AU} \lesssim 5000$ formed primarily by core fragmentation, those with $s \gtrsim 5000$\,AU (the size of typical cloud cores) formed by another process. Candidate processes include cluster dissolution \citep{Kouwenhoven_2010, Moeckel_2010},  unfolding of hierarchical triples \citep{Reipurth_2012}, or pairing of adjacent cores \citep{Tokovinin_2017b}. One might naively expect a change in the mass ratio distribution at $s\sim 5000$\,AU if the binary formation mechanism changes there, but none is observed.

The mass ratio distribution for wide solar-type binaries is not uniform, but is weighted towards low mass ratios. For example, companions with $q\approx 0.5$ are roughly twice as common as those with $q \approx 0.9$. The dominance of low-mass ratio companions can be seen clearly in the data at wide separations (Figure~\ref{fig:posterior_predictive}) and cannot be due to selection effects, which all work {\it against} low-mass ratio binaries. $p(q)$ is thus more bottom-heavy at wide separations than at close separations, where it is basically uniform \citep{Mazeh_1992, Tokovinin_2014}. Analyzing the 25-pc \citet{Raghavan_2010} sample of solar-type binaries, \citet{Moe_2017} found the mass ratio distribution to transition from $\gamma_{\rm largeq} = -0.4 \pm 0.3$ (close to uniform) across $10<s/{\rm AU}<200$ to $\gamma_{\rm largeq} = -1.1 \pm 0.3$ across $200<s/{\rm AU}<5000$.\footnote{Because we use $q_{\rm break}=0.5$ for solar-type primaries and \citet{Moe_2017} used $q_{\rm break} = 0.3$, our measurements should not be directly compared. $p(q)$ flattens at $q < q_{\rm break}$, so at fixed $\gamma_{\rm largeq}$, a lower $q_{\rm break}$ corresponds to a more bottom-heavy mass ratio distribution.} Combined with our constraints at wide separations, this implies that the transition between a uniform mass ratio distribution at close separations and a bottom-heavy distribution at wide separations occurs relatively abruptly at $s\sim 100$\,AU. Several other binary population properties are observed to change at $s\sim 100$\,AU (see \citealt{ElBadry_2019} and references therein), perhaps due to a transition in the dominant binary formation mechanism at this separation. This sharp transition, and the fact that $\gamma_{\rm largeq}$ is nearly constant over $300\lesssim s/{\rm AU} \lesssim 50,000$, provides a useful constraint for star formation models. Because the effects of dynamical processing after formation on the mass ratio distribution are imperfectly understood, similar constraints obtained in star-forming environments will prove useful for disentangling the primordial mass ratio distribution from the effects of dynamical processing.

\begin{figure*}
    \includegraphics[width=\textwidth]{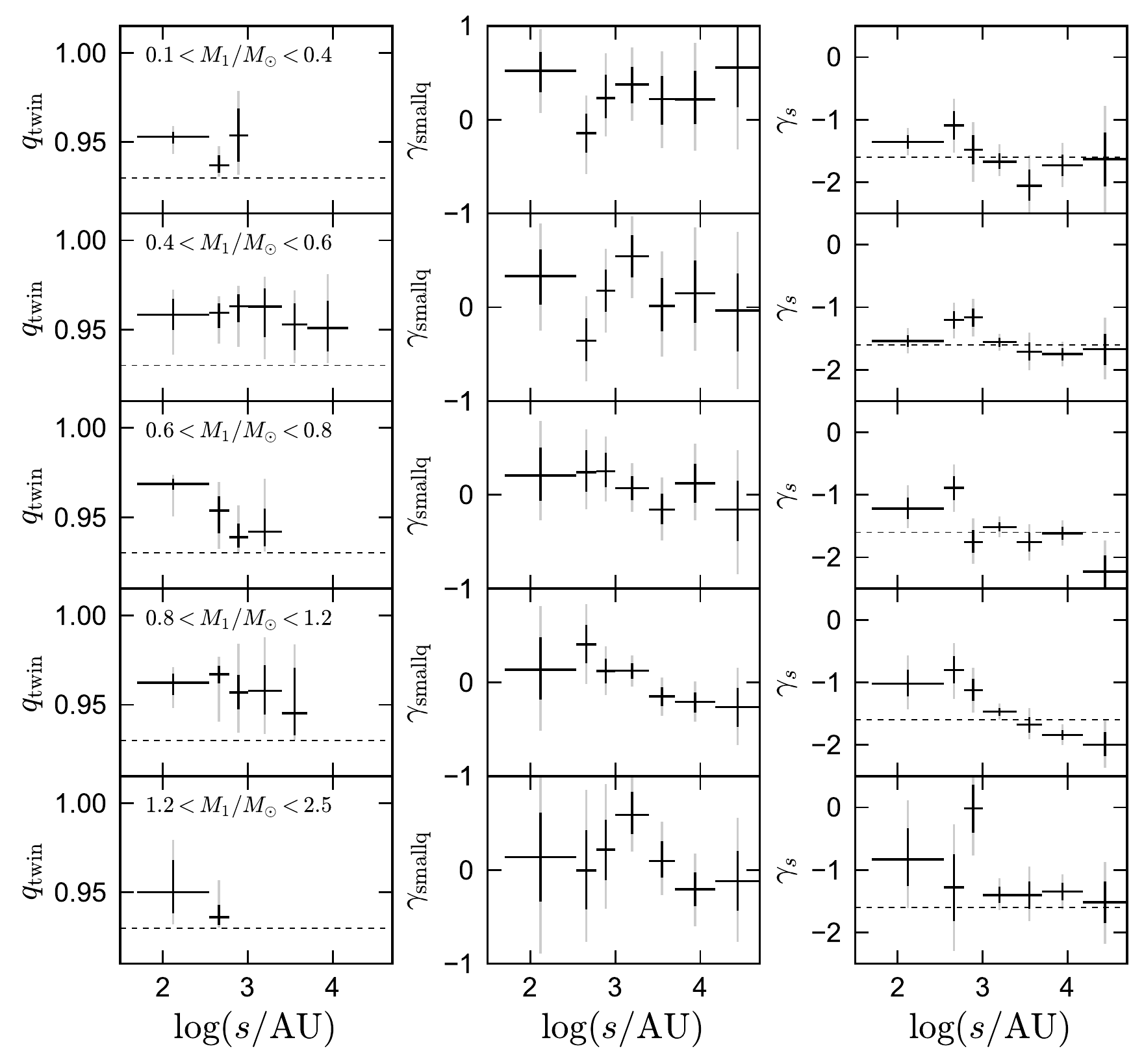}
    \caption{Constraints on $q_{\rm twin}$ (left; the mass ratio above which the twin excess begins), $\gamma_{\rm smallq}$ (middle; the logarithmic slope of the mass ratio distribution at small $q$), and $\gamma_s$ (right; the logarithmic slope of the local separation distribution). Dark and light error bars show 1 and 2 sigma constraints. Primary mass increases from top to bottom. Constraints on $q_{\rm twin}$ are only shown for mass and separation bins where $F_{\rm twin}$ is inconsistent with 0 at the 2 sigma level. Dashed line in the left panels shows $q=0.93$, the lower limit set on $q_{\rm twin}$ by the prior. Dashed lines in the right panels show $\gamma_{s}=-1.6$, the value found by \citetalias{ElBadry_2018} when averaging over all masses and separations. }
    \label{fig:other_parameters}
\end{figure*}

We show constraints on other parameters obtained from our fitting in Figure~\ref{fig:other_parameters}. The left column shows $q_{\rm twin}$. This parameter is only meaningful if $F_{\rm twin}$ is nonzero, so we do not show constraints for bins of mass and separation where $F_{\rm twin}$ is consistent with 0 at the 2 sigma level. There are no strong trends in $q_{\rm twin}$ with mass or separation; the typical best-fit value is $q_{\rm twin}\approx 0.95$. In a few bins (e.g. $1.2 < M_1/M_{\odot} < 2.5$ and $350 < s/{\rm AU} < 600$), the constraint runs up against the prior, implying that a broader excess may be preferred. However, the width of the observed $\Delta G \approx 0$ excess is still reproduced reasonably well in these cases (Figure~\ref{fig:posterior_predictive}).

Constraints on $\gamma_{\rm smallq}$ are shown in the middle column. Most constraints are consistent with $\gamma_{\rm smallq} = 0$ (a flat distribution at small $q$), and trends with separation are weak. For the lowest-mass bin ($M_1 < 0.4 M_{\odot}$), we fit a single power law, so $\gamma_{\rm smallq} = \gamma_{\rm largeq}$. Unlike for $\gamma_{\rm largeq}$, there are no strong trends with primary mass. We note that our data cannot strongly constrain $\gamma_{\rm smallq}$ for low-mass primaries, so the constraints are influenced somewhat by the priors (Section~\ref{sec:functional_form}). 

Finally, the right column of Figure~\ref{fig:other_parameters} shows constraints on the local slope of the separation distribution $\gamma_s$, which is defined such that $p(s)\propto s^{\gamma_s}$.  The dashed line shows $\gamma_s = -1.6$, which is approximately the value that has been found for wide binaries when marginalizing over a larger range of primary masses and separations (\citealt{Andrews_2017}; \citetalias{ElBadry_2018}). The constraints we find here are similar to this value on average but show some evidence for a steepening in $p(s)$ with increasing separation. Any trends with primary mass at fixed separation are weak. We emphasize that these constraints come from the gradient in binary counts as a function of $s$ {\it measured within a narrow bin of $s$}, which is necessarily noisy: trends in $\gamma_s$ with separation represent the second derivative of $p(s)$.

\section{Discussion}
\label{sec:discussion}

\subsection{Comparison to previous work}
\label{sec:literature}
Although most work on twin binaries to date has focused on spectroscopic binaries, hints of a twin excess at wide separation have been reported in several previous works. \citet{Trimble_1987} found a significant excess of equal-brightness pairs among a sample of bright, nearby visual binaries. She suggested that this excess might point toward a formation mechanism that favors equal-mass systems but found significantly different distributions of magnitude difference when comparing different samples of wide binaries and thus did not rule out the possibility that the $q\sim 1$ peak was the result of selection effects. Similar conclusions were reached by \citet{Giannuzzi_1987}.  \citet{Halbwachs_1988} argued that the mass ratio distribution of wide, common proper motion binaries was likely consistent with random pairings from the IMF once selection effects were corrected for. 

Larger and more homogeneous samples of bright wide binaries were identified using astrometry from the Hipparcos satellite for one or both components \citep[e.g.][]{Soderhjelm_2000, Soderhjelm_2007, Eggenberger_2004, Lepine_2007, Shaya_2011}. The mass ratio distribution of Hipparcos binaries with A and F star primaries was modeled in detail by \citet{Soderhjelm_2007}, who found evidence for a $q\approx 1$ peak at $100 \lesssim s/\rm AU \lesssim 1000$. The strength of the peak decreased with primary mass. He argued that the twin excess was not the result of any known selection effect, although he did not reject the possibility that an unknown bias in the Hipparcos input catalog could explain it. The twin feature identified by \citet{Soderhjelm_2007} is likely the same feature apparent in our catalog. We note however, that most of the binaries observed by Hipparcos fall in our highest-primary mass bin (where the twin excess is weaker than at lower masses), as Hipparcos only observed bright stars ($G\lesssim 13$). 

A preference for equal-brightness pairs was also noticed among wide common proper motion disk and halo binaries identified by \citet{Chaname_2004} and \citet{Dhital_2010}. The authors could not rule out the possibility that it was the result of selection effects, which were not well understood for their samples. The twin excess in our catalog is visually quite striking (e.g. Figure~\ref{fig:chance_align}), so a natural question is why it was not as as clear in earlier binary catalogs. In Appendix~\ref{sec:other_catalogs}, we show that a twin excess {\it is} apparent in the large, low-mass wide binary catalog produced by \citet{Dhital_2015} using SDSS photometry, but it only becomes obvious once objects with blended photometry are removed. It is also clearer in the {\it Gaia} photometry than in the ground-based SDSS photometry.

More broadly, the twin excess we identify is visually striking because it is narrow, but it represents only a few percent of the total twin population. This means that it will only become obvious when (a) the photometry is sufficiently precise and uncontaminated that the difference in the components' apparent magnitude can be measured with precision that is good compared to the intrinsic width of the twin excess, and (b) the sample considered is large, containing (at least) hundreds of objects. A twin excess among solar-type visual binaries is also observed in the \citet{Raghavan_2010} 25-pc and \citet{Tokovinin_2014} 67-pc samples, extending out to $s\approx 200$\,AU at a statistically significant level \citet{Moe_2017}. This excess is consistent with our constraints (e.g. Figure~\ref{fig:Ftwin_vs_separation}); at wider separations, these samples did not contain enough objects to detect or rule out a few-percent excess with high significance. 

\subsubsection{Width of the twin excess}
\label{sec:qtwin}
Among spectroscopic binaries, there has been some disagreement in the literature over whether the twin excess is limited to a narrow peak in the mass ratio distribution at $q \gtrsim 0.95$ \citep[e.g.][]{Tokovinin_2000} or is a broader feature, corresponding simply to a positive slope in $p(q)$ at $q\gtrsim 0.8$ (e.g. \citealt{Halbwachs_2003}; see \citealt{Lucy_2006} for further discussion). In our sample, the twin feature is unambiguously narrow, only becoming significant above $q_{\rm twin}\approx 0.95$ (Figure~\ref{fig:other_parameters}). In Appendix~\ref{sec:how_sharp}), we show that $p(q)$ is always consistent with a flat or negative power law at $q< 0.94$; for the majority of the mass and separation bins, the twin excess only becomes strong at $q > 0.97$ (see Figure~\ref{fig:hist_model}). To allow better comparison between close and wide binaries, we now re-examine the twin excess among spectroscopic binaries. 

\begin{figure}
    \includegraphics[width=\columnwidth]{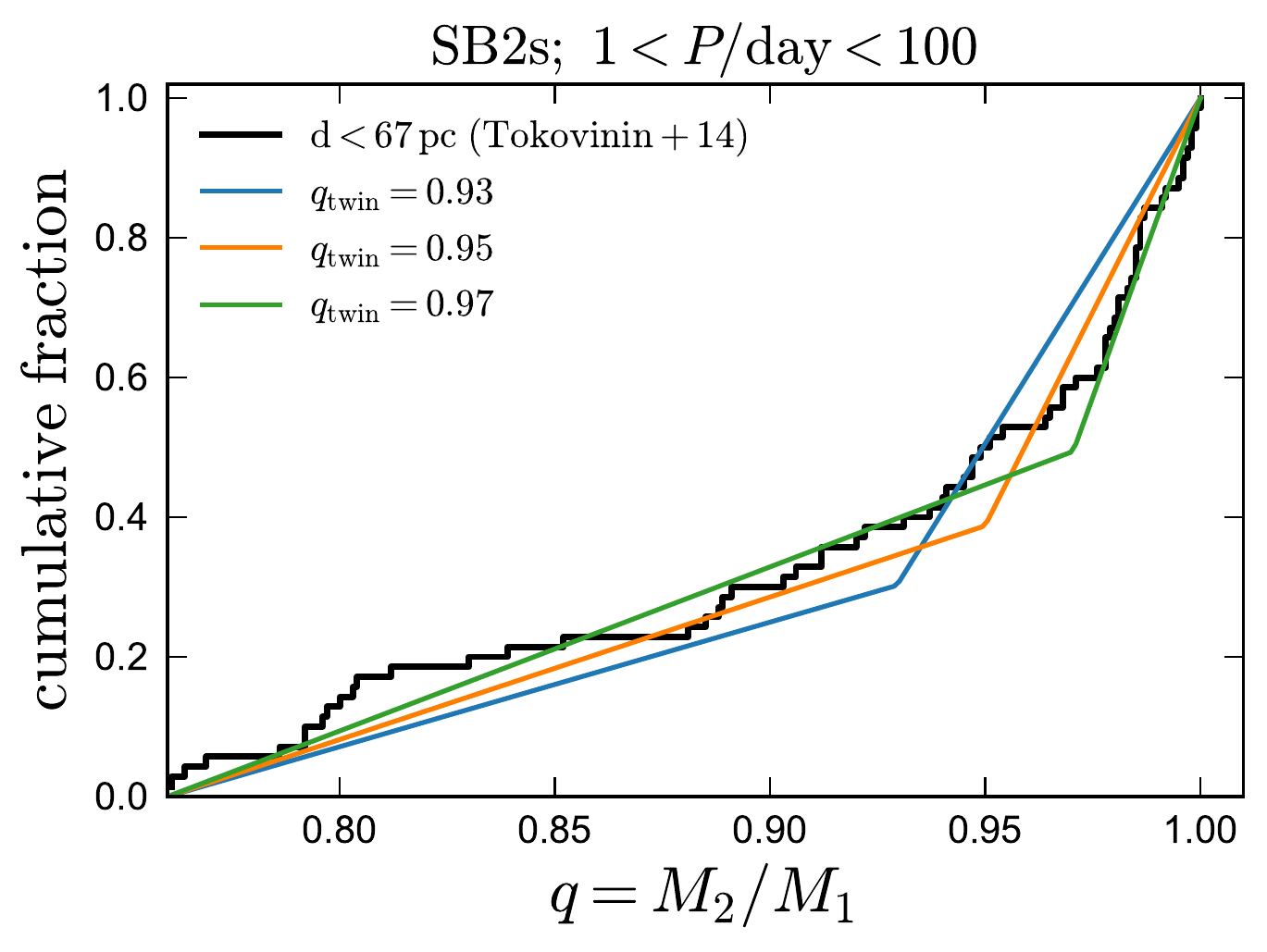}
    \caption{Cumulative distribution function of short-period double-lined spectroscopic binaries with $q > 0.76$ in the volume-limited  67-pc sample of FG dwarfs from \citet{Tokovinin_2014}. We compare predictions for models with three values of $q_{\rm twin}$ (see Section~\ref{sec:qtwin}). For this sample, we find $q_{\rm twin} =  0.964 \pm 0.013$, consistent with the values we find for wide binaries with periods $P=10^{5-9}$\,days.}
    \label{fig:sb2s}
\end{figure}

In the volume-limited 67-pc sample of FG dwarfs, \citet{Tokovinin_2014} identified 98 double-lined spectroscopic binaries (SB2s) with periods $P = 1-100$\,days, 70 of which have dynamical mass ratios $0.76 < q < 1$. Nearly all binaries with $P < 100$\,days and $q > 0.76$ will appear as SB2s, so the 70 observed systems in this parameter space represent a relatively complete subsample. In Figure~\ref{fig:sb2s}, we plot the cumulative mass ratio distribution of the 70 short-period SB2s with $q > 0.76$, about half of which have $q > 0.95$. We model a uniform mass ratio distribution across $0.76 < q < 1$ with an excess twin fraction $F_{\rm twin}$ above $q > q_{\rm twin}$. We use the maximum-likelihood method described in \citet{Moe_2017} to fit the two free parameters $F_{\rm twin}$ and $q_{\rm twin}$ and draw 1,000 bootstrap samples to estimate their uncertainties. We show in Figure~\ref{fig:sb2s} the best-fit models obtained when fixing $q_{\rm twin}$ = 0.93, 0.95, and 0.97. The value $q_{\rm twin}$ = 0.93 is inconsistent with the data ($p = 0.014$), while $q_{\rm twin} = 0.95$ and 0.97 both provide reasonable fits. We formally measure $q_{\rm twin} =  0.964 \pm 0.013$ (1 sigma uncertainties).\footnote{We do not present constraints on $F_{\rm twin}$ for this sample, because $F_{\rm twin}$ depends on $p(q)$ at $0.3 < q < 1$, and many lower-mass ratio binaries will not be double-lined.} By using a larger and more complete sample of short-period SB2s, we thus confirm the conclusions of \citet{Tokovinin_2000} and \citet{Moe_2017} that close solar-type binaries with $a < 0.5$\,AU exhibit a large excess twin fraction and that the twins are narrowly distributed above $q_{\rm twin} \gtrsim 0.95$. 

\begin{figure}
    \includegraphics[width=\columnwidth]{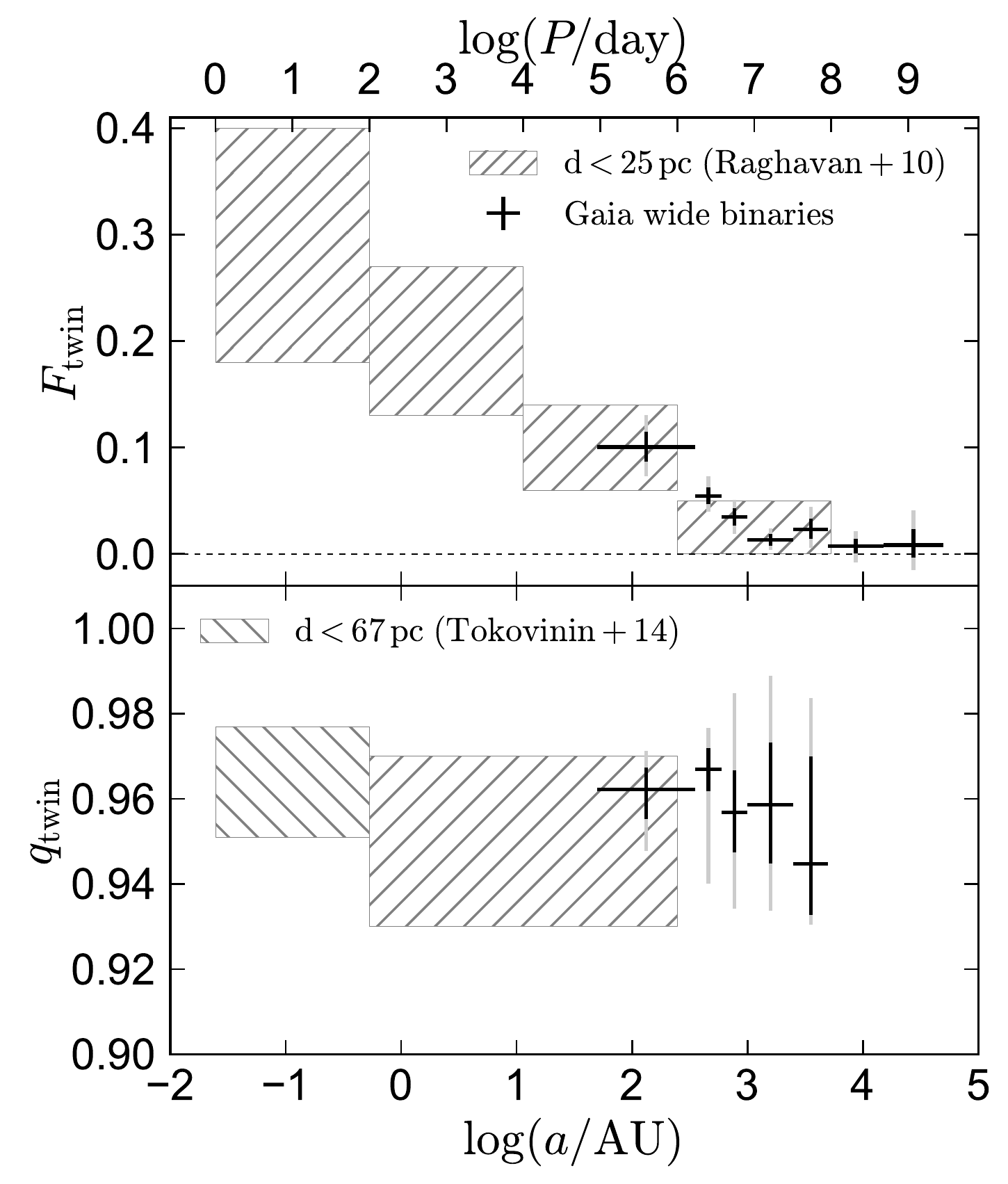}
    \caption{$F_{\rm twin}$ (top) and $q_{\rm twin}$ (bottom) for solar-type binaries ($0.8 \lesssim M_1/M_{\odot} \lesssim 1.2$). We compare results from {\it Gaia} wide binaries (dark and light error bars show 1 and 2 sigma constraints) to 1 sigma constraints at closer separations from the samples of \citet{Raghavan_2010} and \citet[][see Figure~\ref{fig:sb2s}]{Tokovinin_2014}. $F_{\rm twin}$ decreases with separation and is consistent with 0 at $a\gtrsim 5,000$\,AU. However, $q_{\rm twin}\approx 0.95$ is basically constant at all separations.}
    \label{fig:full_period}
\end{figure}

In Figure~\ref{fig:full_period}, we show constraints on $F_{\rm twin}$ and $q_{\rm twin}$ for solar-type binaries across a wide range of periods and separations. At wide separations, the constraints from {\it Gaia} wide binaries are reproduced from Figure~\ref{fig:Ftwin_vs_separation} and~\ref{fig:other_parameters}. At closer separations, we show the constraints on $F_{\rm twin}$ obtained by \citet{Moe_2017} from the \citet{Raghavan_2010} sample, as well as the constraint on $q_{\rm twin}$ at $P=1-100$\,days derived above for the \citet{Tokovinin_2014} sample (Figure~\ref{fig:sb2s}). At intermediate periods ($P=10^{2-6}$\,days), we show $q_{\rm twin}\approx 0.95\pm 0.02$ based on the \citet{Moe_2017} analysis of the \citet{Raghavan_2010} sample. This constraint is not the result of formal fitting but provided a good fit to the data (see Figure 30 of \citealt{Moe_2017}). Figure~\ref{fig:full_period} shows that while $F_{\rm twin}$ decreases with separation, $q_{\rm twin}\approx 0.95$ is consistent with being constant over all separations in this mass range. Similar values of $q_{\rm twin}$ are also found for massive binaries at the short periods where there is a significant twin excess \citep{Moe_2013}. 

\subsection{Origin of twin binaries}
\label{sec:twin_origin}

It is typically assumed that the components of binaries wider than a few hundred AU formed nearly independently of one another \citep[e.g.][]{White_2001, Moe_2017, Tokovinin_2017b,  Moe_2018, ElBadry_2019} during turbulent core fragmentation (for binaries with separation less than a few thousand AU; e.g. \citealt{Offner_2010}) or by becoming bound at slightly later times (for those with the widest separations; e.g. \citealt{Parker_2009}; \citealt{Moeckel_2010}; \citealt{Kouwenhoven_2010}; \citealt{Tokovinin_2017}).

The existence of a narrow twin excess at $q\gtrsim 0.95$ suggests that the components of a fraction of binaries with $s\gg 100$\,AU formed at closer separations in a highly correlated way. We do expect that the dynamical process of becoming and staying bound may lead to a preference for roughly equal-mass binaries (say, $q > 0.5$), because these have higher binding energy. This could quite reasonably explain, at least in part, why the power-law component of the mass ratio distribution is shallower than expected for random pairings from the IMF (Figure~\ref{fig:gamma_vs_separation}). But we do not expect dynamical processes to produce a {\it sharp} twin feature like what is observed: at fixed primary mass, the binding energy at $q=0.9$ is not much less than that at $q=1$. And indeed, simulations of binary formation during cluster dissolution find larger typical mass ratios at wide separations than predicted for random pairings, but they do not predict a narrow excess of twins \citep{Kroupa_1998, Moeckel_2010, Kouwenhoven_2010}.

The excess twin fraction uniformly decreases with separation and eventually goes to 0 at $s> 15,000$\,AU in all mass bins. The shape of the twin excess (i.e. $q_{\rm twin}$ and the slope of $p(q)$ at $q>q_{\rm twin}$) does not vary much between 50 and 15,000 AU in our catalog. Moreover, it is effectively the same for spectroscopic binaries (with separations as close as $0.01$\,AU) and wide binaries (Figure~\ref{fig:full_period}). Invoking Occam's razor, it seems more likely that the wide binary twin phenomenon is an extension of the phenomenon that has previously been observed at $s \lesssim 100$\,AU than that it is produced by a qualitatively different process. 

Even for close binaries, there is not a clear consensus in the literature about the physical origin of the twin phenomenon. Some models for the formation of twins can only apply to very close binaries. In the first paper to highlight the twin phenomenon for spectroscopic binaries, \citet{Lucy_1979} suggested that twins were formed by fragmentation of rapidly rotating pre-main sequence stars during the late stages of dynamical collapse, at scales of $a \ll 1$\,AU. Alternatively, \citet{Krumholz_2007} proposed that twins could be produced by mass transfer between stars of initially different masses during pre-main sequence evolution.\footnote{These authors sought to explain the observed twin excess in massive stars. The specific mechanism they proposed, which relies on deuterium shell burning causing protostars to expand and overflow their Roche lobes, cannot operate in solar-type or lower mass stars. Moreover, the twin excess for massive stars appears to be limited to close separations \citep{Moe_2017}, so mass transfer may adequately explain it. Here we simply suppose, for the sake of argument, that there is some mechanism through which stable mass transfer in lower mass stars could drive the mass ratio to unity.} It seems implausible that such mechanisms can explain the twin phenomenon among wide binaries, because there is no known mechanism to widen the orbits of twins from the separations at which they operate -- a few, or at most a few tens of, solar radii -- to the separations at which they are observed today. Such widening would require a very strong velocity kick, the magnitude of which would have to be fine-tuned in order to not unbind the binaries completely. 

A more plausible formation mechanism for equal-mass twins at wider separations is through competitive accretion from a circumbinary disk. Many studies have found that the accretion rate from a circumbinary disk is usually higher for the secondary than the primary \citep[e.g.][]{Bate_1997, Bate_2000, Farris_2014, Young_2015, Nelson_2016, Matsumoto_2019}. Because the secondary's orbit is larger than that of the primary, it sweeps out a larger radius in the disk and can accrete more rapidly than the primary, unless the material being accreted has low angular momentum.\footnote{It is worth noting that there has  not been a full consensus in the literature whether this mechanism works: some simulations of accretion from a circumbinary disk have actually predicted the opposite trend, with accretion favoring the primary \citep{Ochi_2005, Hanawa_2010, de_val_borro_2011}. These simulations assumed a higher gas temperature than those which have found accretion to favor the secondary; \citet{Young_2015} showed that accretion only favors the secondary when accreted gas is cold, as is appropriate for stellar binaries.} Preferential accretion onto the secondary will necessarily drive the mass ratio towards unity. If such accretion continues for long enough, binaries within circumbinary disks should thus end up with $q\approx 1$. An appeal of this formation mechanism is that it can operate at scales comparable to the size of circumbinary disks, $s \lesssim 100$\,AU. 

It is plausible but not obvious that preferential accretion onto the secondary will give rise to a sharp twin feature like the one found observationally. In order to end up at $q \approx 1$, a binary that initially had an intermediate mass ratio must accrete a large fraction of its mass from a circumbinary disk, such that there is enough time to drive the mass ratio to 1 even while the primary continues to accrete. If twins are formed by accretion from circumbinary disks, then the width of the twin feature (i.e., $q_{\rm twin}$) can tell us about the fraction of the total mass accreted from the disk, as well as the mass ratio above which the accretion rate onto the two components becomes nearly equal.
A sharp twin feature could be expected if a fraction of binaries accrete most of their mass from a circumbinary disk (becoming twins) and the rest either do not develop circumbinary disks or only accrete a subdominant fraction of their total mass from them. 

In accretion-driven explanations of the twin excess, the increase in $F_{\rm twin}$ towards close separations has been interpreted as evidence that gravitational torques within circumbinary disks \citep[e.g.][]{Artymowicz_1991, Shi_2012} shrink the orbits of twins \citep{Young_2015}. Indeed, the observed twin excess is largest at separations $a < 0.2$\,AU (Figure~\ref{fig:full_period}), too close for binaries to have formed at their current separations. This implies that some combination of gravitational torques, viscous dissipation, and dynamical interactions \citep[e.g.][]{Bate_2012, Moe_2018} must have shrunk the orbits of twins at very close separations.

However, several recent simulations of circumbinary disks have found that, contrary to previous results in the literature, accretion can also {\it widen} binaries within circumbinary disks, when the advective torque dominates over the gravitational torque \citep{Miranda_2017, Munoz_2019, Moody_2019}. Whether gravitational or advective torques dominate depends on details such as the sink prescription used for accretion \citep{Tang_2017}. Further work is needed to clarify the effects of circumbinary disks on orbital evolution. However, the fact that the twin excess extends to very wide separations suggests that orbit shrinkage is not a ubiquitous outcome of accretion from circumbinary disks.

High-resolution studies of the dynamics of individual circumbinary disks are generally too idealized, run for too short a time, and are focused on too narrow a range of initial conditions to make ab-initio predictions of the full mass ratio distribution. However, they do typically find that gas is preferentially accreted onto the secondary for realistic accretion geometries once a steady state is reached. On the other hand, global simulations of the fragmentation of molecular clouds \citep[e.g.][]{Bate_2009, Bate_2014, Bate_2019} are reaching the point where they can make realistic predictions of the mass ratio distribution with minimal fine-tuning. Such studies have lower resolution than idealized simulations of individual binaries, so it is not guaranteed that the dynamics within disks are well resolved, but they are able to predict the accretion rate and angular momentum distribution of accreted material, and the mass and size distribution of disks \citep[e.g.][]{Bate_2018}. Such global calculations predict an excess of equal-mass binaries out to separations of order 100 AU (see e.g. \citealt{Bate_2014}, Figure 18). Because they typically only form a few dozen binaries in a cloud, such calculations do not yet have the statistical power to predict or rule out a few-percent effect at wider separations. 

Accretion can plausibly explain an excess of twins out to significantly larger separations than mass transfer or late-stage fragmentation, but it alone cannot explain a signal reaching out to thousands of AU. Observed circumstellar and circumbinary gas disks have typical radii of order $100$\,AU \citep{Ansdell_2018, Eisner_2018}.
The {\it largest} observed circumbinary disks have radii of order 500 AU and host binaries with separations of 50-200 AU \citep[e.g.][]{Hioki_2007, Brinch_2016, Tobin_2016, Takakuwa_2017, Comeron_2018, Czekala_2019}; these preferentially host relatively massive binaries. There are no observed disks with radii exceeding 1000 AU, and simulations also predict the largest circumbinary disks to have radii of several hundred AU \citep{Bate_2018}. It thus seems exceedingly unlikely that twin binaries with $s\gtrsim 200$\,AU formed at their present-day separation by accretion from a circumbinary disk. This implies that either (a) twin binaries formed at closer separations and their orbits were subsequently widened, or (b) some other process is responsible for producing twins at very wide separations. Lacking a good candidate mechanism for (b), we here consider the plausibility of orbit widening.

\subsubsection{Dynamical orbit widening in young clusters}
\label{sec:widening}

In the Galactic field, dynamical interactions have a negligible effect on most binaries with $s\lesssim 10,000$ AU \citep[e.g.][]{Weinberg_1987}. However, dynamical interactions are expected to be more efficient in binaries' birth environments, where the typical stellar density is higher. The dynamical evolution of binaries within their birth clusters has been the subject of considerable study. As a general rule, interactions within birth clusters are expected to widen the orbits of binaries with orbital velocities lower than the cluster velocity dispersion, and to tighten the orbits of those with orbital velocities greater than it \citep{Heggie_1975, Hills_1975}. There are several complicating factors in real clusters. For example, the mass distribution within clusters at early times is not smooth but clumpy \citep{Dorval_2017}, both stars and gas can be dynamically important, and cluster density and velocity dispersion change as clusters age, in part because energy is redistributed among binaries \citep[see e.g.][]{Kroupa_2001, Parker_2009, Goodwin_2010}. Observed binary populations provide constraints on models for binaries' dynamical evolution, but a unified model to explain the diversity of binary populations found in different environments does not exist. We summarize some relevant observational constraints on disruption below. 

\begin{enumerate}
    \item Over a wide range of separations ($10 \lesssim s/{\rm AU} < 3000$), the binary fraction in low-density star forming regions (e.g. Taurus) is higher than in the field by roughly a factor of 2 \citep[e.g.][]{Leinert_1993, Duchene_1999}. This is true especially at wider separations, where the separation distribution of young binaries is roughly log-uniform, but that of field stars declines more steeply \citep[e.g.][]{Connelley_2008, Kraus_2011}.
    \item In dense young clusters (e.g. the ONC), the binary fraction at separations of $s\gtrsim 100$\,AU is similar to the field (lower than in low-density star forming regions) and declines steeply at wide separations \citep{Reipurth_2007}. At closer separations ($10 \lesssim s/{\rm AU} \lesssim 60$) the binary fraction in dense clusters is comparable to that in low-density star forming regions, and higher than that found in the field \citep{Duchene_2018}.
    \item At very close separations ($s\lesssim 5\,\rm AU$), the binary fraction in star forming regions (over a range of densities) is consistent with that in the field \citep[e.g.][]{Kounkel_2019}.
\end{enumerate}
 
Some models \citep[e.g.][]{Kroupa_1995, Marks_2011, Marks_2012} postulate that the initial binary fraction and separation distribution are insensitive to local properties, such that observed variation in binary populations must be due to disruption. These models interpret the higher wide binary fraction in low-density clusters as the primordial population, which is transformed into the field population by dynamical widening and disruption. Because disruption of wide systems is more rapid in dense clusters, these models also predict the binary fraction at $s \gtrsim 100$\,AU to decrease with cluster density, in agreement with observations. However, if such models are correct, it is not clear what happens to the excess of relatively tight binaries  ($10 \lesssim s/{\rm AU} \lesssim 100$), which are over-represented relative to the field in both high- and low-density star forming regions. Such systems have high enough binding energies that they can only be disrupted in very dense clusters. Thus the lower binary fraction in the field at $10 \lesssim s/{\rm AU} \lesssim 100$ would imply that a large fraction of field stars formed in dense environments. 

In any case, the fact that the binary fraction in the field is lower than that in star forming regions down to fairly close separations implies that a significant fraction of young binaries undergo quite energetic interactions, the cumulative effect of which is sufficient to disrupt binaries with initial separations as close as 10-100\,AU. In many cases, binaries will be disrupted by such interactions, but in some cases, they will only be widened \citep[e.g.][]{Kroupa_2001}. It is in this latter context that the twin fraction at wide separations is informative about the fraction of stars at a given present-day separation that formed at significantly closer separations. 

The fact that the dependence of $F_{\rm twin}$ on separation varies with binary mass in a non-monotonic way (Figure~\ref{fig:Ftwin_vs_separation}) suggests that, if the twin excess at $s\gg 100$\,AU is due to dynamical widening of orbits, then the primordial twin statistics (e.g. the twin fraction and the range of separations over which twins are form) must also vary with mass. This may not be unreasonable, since the physical properties of disks do vary with mass \citep{Bate_2018, Eisner_2018}, but it means that disentangling the effects of dynamical widening and the primordial separation distribution of twins is nontrivial. 

\begin{figure}
    \includegraphics[width=\columnwidth]{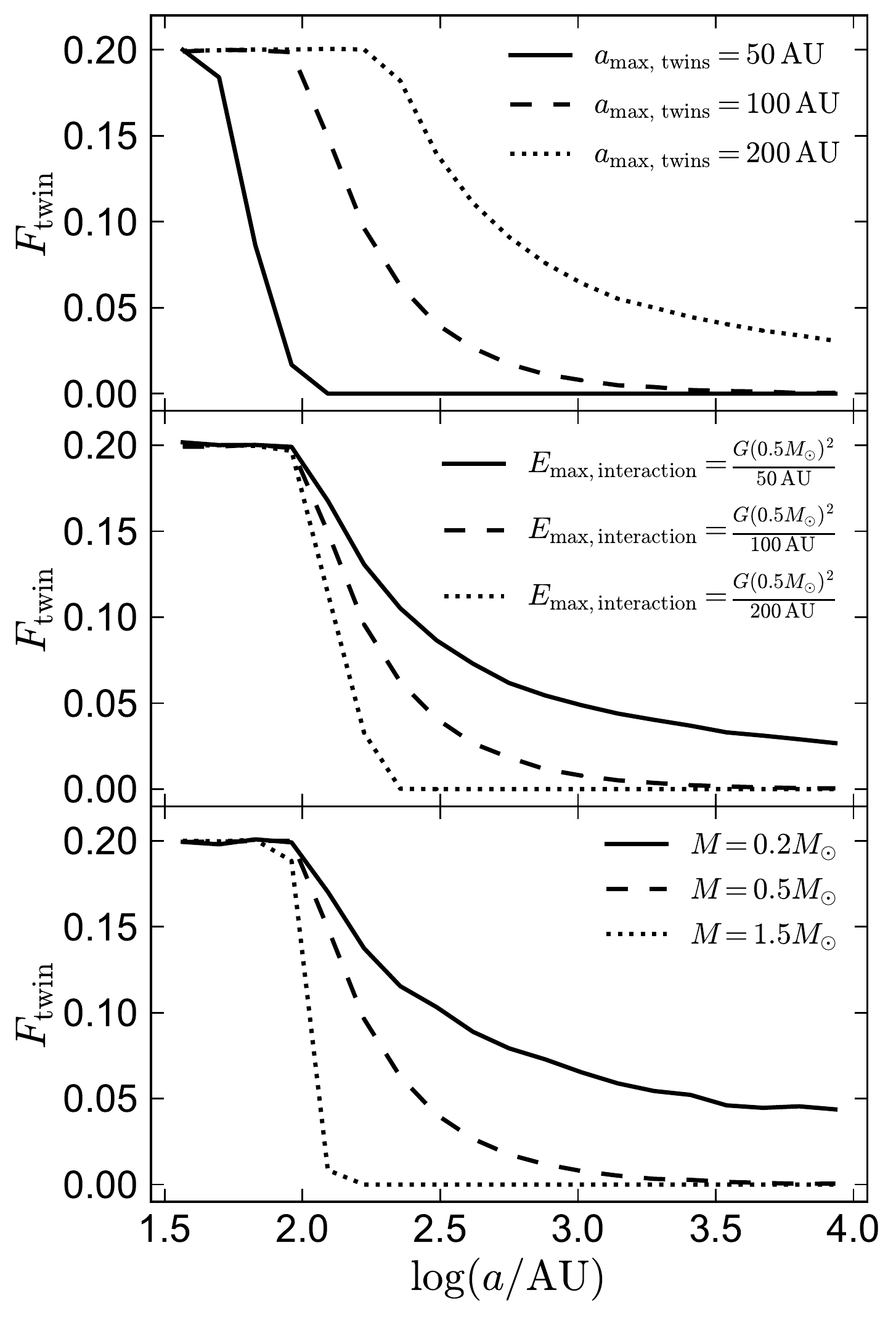}
    \caption{Predictions of the toy model for orbit widening described in Section~\ref{sec:widening}. We assume that the initial twin fraction is 20\% for $a<a_{\rm twin,\,max}$ (varied in the top panel; 100\,AU by default) and 0 at larger separations. We consider an initial separation distribution that is flat in log space. We then widen the orbits of all binaries by adding energy, representing the cumulative effect of gravitational interactions within a birth cluster. The total added energy is drawn from a uniform distribution between 0 and $E_{{\rm max,interaction}}$. Binaries that are disrupted are discarded. We plot the final twin fraction as a function of separation, varying $E_{{\rm max,interaction}}$ (middle panel; default is $G\left(0.5M_{\odot}\right)^{2}/\left(100\,{\rm AU}\right)$) and mass (bottom panel; default is 0.5\,$M_{\odot}$).}
    \label{fig:toy_model_widening}
\end{figure}

We use a simple toy model to explore how the separation-dependence of the twin excess may depend on the initial twin population and the strength of dynamical interactions. We suppose that twin formation is efficient out to a separation of $a_{\rm twin,\,max}\approx 100$\,AU and does not operate at wider initial separations. We consider primordial populations of equal-mass binaries with components of mass $M$ and a uniform distribution of $\log a$ between 10 and $10^4$\,AU. At $a < a_{\rm twin,\,max}$, we assign a random subset of 20\% of the binaries the label of ``twin''. We then assume that dynamical interactions add an energy $E_{\rm int}$ to each orbit (for both twins and non-twins), such that the new orbital energy is $E_{{\rm new}}=E_{{\rm initial}}+E_{\rm int}$, where $E_{{\rm initial}}=-GM^{2}/a_{{\rm initial}}$. Values of $E_{\rm int}$ for each binary are drawn from a uniform distribution between 0 and $E_{\rm max,\,interactions}$. If $E_{\rm new}$ is positive, then the binary is considered unbound and is discarded. The new separation after interactions is $a_{{\rm new}}=-GM^{2}/E_{{\rm new}}$. Finally, we measure what fraction of the surviving binaries bear the twin label as a function of final separation. 
This fraction is proportional to the fraction of binaries at a given present-day separation that formed with $a < a_{\rm twin,\,max}$. We do not model dynamical hardening of close binaries, and thus implicitly assume that all binaries we consider are soft; i.e., that they reside in clusters with velocity dispersion larger than than the highest orbital velocities of binaries being widened. For an initial separation of 100\,AU and binaries with typical component mass of 0.5\,$M_{\odot}$, this corresponds to a dispersion of $\sigma \approx 1.5\,\rm km\,s^{-1}$; for an initial separation of 10\,AU, to $\sigma \approx 5\,\rm km\,s^{-1}$.

We plot the results of this experiment in Figure~\ref{fig:toy_model_widening}, varying $a_{\rm twin,\,max}$ (top), $E_{\rm max,\,interactions}$ (middle), and $M$ (bottom). The final separation-dependence of the twin fraction depends significantly on all of these parameters. The twin excess extends to larger separations, and falls off less steeply with separation, when (a) the initial separation out to which twins form is larger, (b) dynamical interactions are more energetic, or (c) the binding energy of twins is lower. In order to obtain a nonzero twin fraction at very wide separations, it is necessary that a fraction of binaries undergo dynamical interactions energetic enough to unbind binaries at separation at which twins form. Once this is satisfied, twins can contribute significantly to the binary population at wider separations, because the same interactions that widen close binaries will unbind a large fraction of initially wider non-twin binaries. Wide twins are produced most efficiently in clusters where the velocity dispersion is comparable to the orbital velocity at a separation of $a_{\rm twin,\,max}$: significantly denser clusters produce few wide binaries, since most binaries that are not close are disrupted completely. 

The toy model generically predicts that $F_{\rm twin}$ decreases with separation and that low-mass twins can be more efficiently widened than high-mass twins due to their lower binding energy. It thus predicts that the excess twin fraction will fall off less steeply for lower-mass binaries. This trend {\it is} found in the observed binaries over $0.4 < M_1/M_{\odot} < 2.5$ (Figure~\ref{fig:Ftwin_vs_separation}), but it is reversed for the lowest-mass subsample: $F_{\rm twin}$ falls off significantly more steeply for $0.1 < M_1/M_{\odot} < 0.4$ than for $0.4 < M_1/M_{\odot} < 0.6$. This is not easily explained in the context of the toy model. 

It is thought that a fraction of wide binaries (likely at $a \gg 1000$\,AU) form at later times (perhaps during cluster dissolution) than wide binaries at closer separations. Dilution of the twin excess due to these binaries is not accounted for in the toy model. If the fraction of binaries at fixed separation that formed during cluster dissolution were higher at lower masses, this could explain the observed steeper decline in $F_{\rm twin}$ with separation at low masses. 

An addition complication is that the observed population of field binaries is an average over a wide range of formation environments, from low-density regions to dense clusters. The trends in Figure~\ref{fig:toy_model_widening} rely on a number of crude approximations and should not be used to directly interpret the observed trends in $F_{\rm twin}$ (and we have not attempted to tune the model to match observed trends). Here we simply emphasize that the separation distribution of twins is quite sensitive to dynamical processing. We conclude that dynamical orbit widening provides a plausible explanation for the existence of the wide twin excess, but more realistic theoretical modeling is needed to determine whether the observed trends in $F_{\rm twin}$ with primary mass and separation can be reproduced when averaging over a realistic population of star-forming environments.

\subsubsection{Widening of unstable triple systems?}
\label{sec:triples}
Another possible mechanism for dynamically widening twins is through unfolding of unstable triple systems, in which initially close companions can be scattered to much wider orbits \citep[e.g.][]{Reipurth_2012}. Triples are not uncommon: more than a third of all wide binaries contain subsystems \citep[e.g.][]{Tokovinin_2002, Tokovinin_2014}, and a large fraction of binaries are thought to have formed as higher-order multiples that subsequently decayed \citep[e.g.][]{Sterzik_1998}. 

\begin{figure}
    \includegraphics[width=\columnwidth]{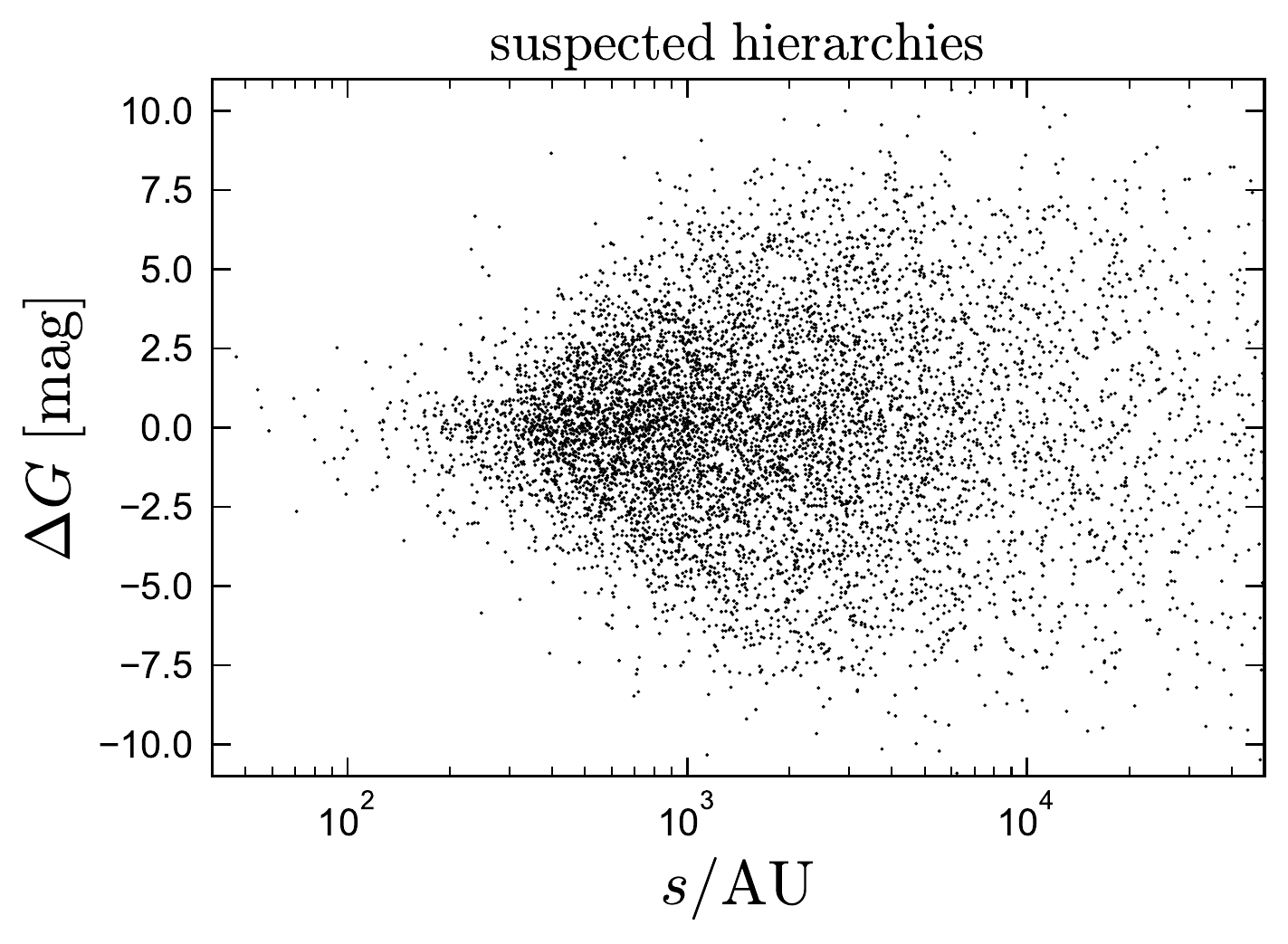}
    \caption{Magnitude difference between the two resolved components of all wide binaries in which one or both components is suspected to have a bright, unresolved companion ($N=7,295$; yellow points in Figure~\ref{fig:cmds}). There is no obvious excess at $\Delta G\approx 0$ (compare to the right panel of Figure~\ref{fig:dG_points}). }
    \label{fig:dG_triples}
\end{figure}

Widening of the outer orbit in triples generally comes at the expense of shrinking of the inner orbit. Since the outer orbit is the more fragile one, a triple-origin of the wide twin excess would imply that one of the component stars in wide twins had, and in most cases still has, a close companion, which in most cases would be spatially unresolved. However, the twin excess at $\Delta G \lesssim 0.25$\,mag cannot be due to systems in which either resolved component has an unresolved companion any less than $\approx 2.5$ magnitudes fainter than it; otherwise the light from the unresolved companion would make it more than 0.25 mag brighter than the other resolved component. Figure~\ref{fig:dG_triples} shows that there is indeed no obvious excess at $\Delta G \approx 0$ for wide binaries in which either component has a bright unresolved companion. There is thus a photometric upper limit on the mass ratio of any unresolved companions to twins with $\Delta G \approx 0$; it ranges from $q < 0.45$ to $q < 0.7$, depending on primary mass (see Figure~\ref{fig:delta_G_vs_q}).

Unresolved companions with lower mass ratios cannot be ruled out based on photometry, but at close separations, they can be detected from radial velocity (RV) variability. In Appendix~\ref{sec:rv_variability}, we show that the {\it Gaia} visit-to-visit radial velocity scatter implies a {\it lower} close binary fraction for the components of wide twins than for components of non-twins with similar separations and masses. This speaks against a triple-origin of wide twins. It is, however, consistent with our expectations if twins are formed through accretion in a circumbinary disk, since the presence of a third close companion in the disk would complicate the mechanism through which accretion drives the mass ratio of two stars in a disk toward unity.

\section{Summary and conclusions}
\label{sec:conclusions}

We have analyzed a pure and homogeneous sample of $\sim$42,000 main-sequence wide binaries selected from {\it Gaia} DR2 to constrain the mass ratio distribution of binaries with projected separations $50\lesssim s/{\rm AU} < 50,000$ and primary masses $0.1 < M_1/M_{\odot} < 2.5$. High-precision photometry allows us to measure mass ratios with unprecedented accuracy, and a well-understood selection function makes it possible to account for biases arising from the magnitude and contrast sensitivity limits of {\it Gaia} DR2 given our quality cuts.  

A striking feature of the catalog is a high-significance excess of ``twin'' binaries with nearly equal brightness (Figure~\ref{fig:dG_points}). The excess is present over a narrow range of magnitude differences, $0 < \Delta G \lesssim 0.25$, corresponding to mass ratios $0.95 \lesssim q \lesssim 1$, and extends over a wide range of masses and separations. The twin excess is reminiscent of the excess of equal-mass binaries historically reported at very close separations ($a < 0.2$\,AU) and recently found to extend to $a\sim 100$\,AU, but it extends to separations of several thousand AU, where binary formation models do not predict strongly correlated component masses.

We have done a variety of tests to confirm that the twin feature is caused by an astrophysical excess of equal-mass binaries, not selection effects or data artifacts. We first repeated the search that produced the binary catalog, but required that the two components of a pair have {\it inconsistent} rather than consistent astrometry. This yields a catalog of physically unassociated ``chance alignments'' that are subject to essentially the same selection function as the real binaries and have similar distributions of most observable properties. A narrow excess of equal-brightness pairs is not found among chance alignments (Figure~\ref{fig:chance_align}).

By considering binaries at different distances, we verified that the strength of the twin feature varies primarily with physical, not angular, separation  (Figure~\ref{fig:three_distances}). This speaks to the physicality of the twin feature, since most observational biases are expected to depend on angular separation, which is the observable quantity. Finally, we verified that the twin feature is not an artifact of the {\it Gaia} photometry; it is visible in photometry from other surveys. We also identified the same twin excess in archival data from another binary catalog after removing objects with contaminated photometry (Appendix~\ref{sec:other_catalogs}).

In order to measure the intrinsic mass ratio distribution, we forward-model the joint distribution of magnitude difference, angular separation, and parallax given a Galactic model, a selection function, and a parameterized mass ratio distribution. We measure the selection function for our catalog empirically, using chance alignments to quantify the contrast sensitivity as a function of angular separation. Our primary results are as follows: 
\begin{enumerate}
    \item {\it Twin fraction}: we quantify the twin excess as $F_{\rm twin}$, the fractional excess of binaries with $q > q_{\rm twin}$ relative to the full population with $q>0.3$ (Figure~\ref{fig:schematic_q_dist}; $q_{\rm twin}\approx 0.95$ quantifies the width of the excess). Typical values of $F_{\rm twin}$ are 10\% at 100\,AU and 3\% at 1000\,AU, with some dependence on mass (Figure~\ref{fig:Ftwin_vs_separation}). These values are lower than the value found for spectroscopic binaries (Figure~\ref{fig:full_period}), which is $F_{\rm twin}\approx 25\%$, but they are clearly inconsistent with 0.
    \item {\it Width of the twin excess}: at all masses and separations where it is statistically significant, the twin excess we find is quite narrow, only becoming significant at $q \gtrsim 0.95$ (Figure~\ref{fig:other_parameters}). We experimented with different functional forms for the enhancement near $q\approx 1$, but we find a step function increase at $q \gtrsim 0.95$ to perform as well as more complicated models (Figure~\ref{fig:hist_model}). We also re-measured $q_{\rm twin}$ at close separations using a volume-limited sample of spectroscopic binaries (Figure~\ref{fig:sb2s}). The width of the twin excess, $q_{\rm twin}$, is basically constant over 6 orders of magnitude in separation, from the closest spectroscopic binaries to wide binaries at $s\sim 10,000$\,AU. 
    \item {\it Mass and separation dependence}: the fractional excess of twins declines with separation and is consistent with 0 at $s > 15,000$\,AU for all mass bins (Figure~\ref{fig:Ftwin_vs_separation}). The twin excess at $s>1,000$\,AU is strongest for primary masses $0.4 < M_1/M_{\odot} < 0.6$. At closer separations ($50 < s/{\rm AU} < 1,000$), it is strongest for low-mass primaries ($M_1 < 0.4 M_{\odot}$) and roughly consistent for other mass bins. The slope of the fall-off in $F_{\rm twin}$ with separation varies non-monotonically with mass; it is shallowest at $0.4 < M_1/M_{\odot} < 0.6$.
    \item {\it Mass ratio distribution at lower q:} we provide broken power law fits to the full mass ratio distribution for all mass and separation bins (Figure~\ref{fig:qdist_ranges}; Appendix~\ref{sec:full_constraints}). These fits reproduce the observed data well (Figure~\ref{fig:posterior_predictive}). For solar-type stars, the power law slope is $\gamma_{\rm largeq}\approx -1.3$ for $q > 0.5$ (i.e., weighted toward lower mass ratios than the uniform distribution found at closer separations) and $\gamma_{\rm smallq}\approx 0$ for $q < 0.5$. $p(q)$ becomes more bottom-heavy with increasing primary mass but is always flatter than expected for random pairings from the IMF. Besides variation in the excess twin fraction, $p(q)$ does not vary much with separation over $100 \lesssim s/{\rm AU} < 50,000$ (Figures~\ref{fig:gamma_vs_separation} and~\ref{fig:other_parameters}).
    \item {\it Origin of the twin excess}: no theoretical models that have been proposed to explain twin binaries at close separations predict them to form at $s \gtrsim 100$\,AU. Dynamical processes may lead to a formation bias against low-$q$ binaries, but they are not expected to produce a sharp peak at $q \approx 1$. Given the monotonic fall-off in $F_{\rm twin}$ with separation, the similar width of the twin feature between close and wide binaries, and the lack of a plausible mechanism for forming twins at very wide separations, we conjecture that the excess twins must have formed at closer separations ($s \lesssim 100$\,AU; likely through accretion from a circumbinary disk) and subsequently been widened by dynamical interactions. 
    
    In this scenario, the separation-dependence of the twin fraction is an imprint of dynamical orbit widening in binaries' birth environments (see Section~\ref{sec:discussion}). The plausibility of this explanation is not straightforward to assess because (a) present-day field binaries formed in a wide range of environments and (b) existing models for widening and disruption of binaries in star forming regions do not fully explain the diversity of observed binary populations in young clusters and in the field. A simple toy model suggests that a separation-dependence in $F_{\rm twin}$ comparable to that which is observed is predicted if typical dynamical interactions are strong enough to disrupt binaries at the separation inside which twins are expected to form ($a \lesssim 100$\,AU). However, the mass-dependence of $F_{\rm twin}$ at wide separations is not fully explained in such models. Further theoretical work is required to (a) predict the primordial separation distribution of twins at different masses and (b) constrain the efficiency of dynamical orbit widening for a realistic ensemble of star forming environments. In future work, we will search for a wide twin excess in observed star forming regions to shed more light on the primordial separation distribution of twins and the density-dependence of orbit widening. 

\end{enumerate}

\section*{Acknowledgements}
We are grateful to the referee, Andrei Tokovinin, for thoughtful comments.
We thank Matthew Bate, Anthony Brown,  Eugene Chiang,  Ian Czekala, Paul Duffel, Morgan Fouesneau, Harshil Kamdar, Tomoaki Matsumoto, Chris Mckee, Eliot Quataert, and Dan Weisz for helpful discussions. KE was supported in part by an NSF graduate research fellowship and by SFB 881. HJT acknowledges the National Natural Science Foundation of China (NSFC) under grants 11873034. MM acknowledges financial support from NASA under Grant No. ATP-170070. This project was developed in part at the 2019 Santa Barbara Gaia Sprint, hosted by the Kavli Institute for Theoretical Physics at the University of California, Santa Barbara. This research was supported in part at KITP by the Heising-Simons Foundation and the National Science Foundation under Grant No. NSF PHY-1748958. This work has made use of data from the European Space Agency (ESA) mission {\it Gaia} (\url{https://www.cosmos.esa.int/gaia}), processed by the {\it Gaia} Data Processing and Analysis Consortium (DPAC, \url{https://www.cosmos.esa.int/web/gaia/dpac/consortium}). Funding for the DPAC has been provided by national institutions, in particular the institutions  participating in the {\it Gaia} Multilateral Agreement.



\bibliographystyle{mnras}

\begin{thebibliography}{}
\makeatletter
\relax
\def\mn@urlcharsother{\let\do\@makeother \do\$\do\&\do\#\do\^\do\_\do\%\do\~}
\def\mn@doi{\begingroup\mn@urlcharsother \@ifnextchar [ {\mn@doi@}
  {\mn@doi@[]}}
\def\mn@doi@[#1]#2{\def\@tempa{#1}\ifx\@tempa\@empty \href
  {http://dx.doi.org/#2} {doi:#2}\else \href {http://dx.doi.org/#2} {#1}\fi
  \endgroup}
\def\mn@eprint#1#2{\mn@eprint@#1:#2::\@nil}
\def\mn@eprint@arXiv#1{\href {http://arxiv.org/abs/#1} {{\tt arXiv:#1}}}
\def\mn@eprint@dblp#1{\href {http://dblp.uni-trier.de/rec/bibtex/#1.xml}
  {dblp:#1}}
\def\mn@eprint@#1:#2:#3:#4\@nil{\def\@tempa {#1}\def\@tempb {#2}\def\@tempc
  {#3}\ifx \@tempc \@empty \let \@tempc \@tempb \let \@tempb \@tempa \fi \ifx
  \@tempb \@empty \def\@tempb {arXiv}\fi \@ifundefined
  {mn@eprint@\@tempb}{\@tempb:\@tempc}{\expandafter \expandafter \csname
  mn@eprint@\@tempb\endcsname \expandafter{\@tempc}}}

\bibitem[\protect\citeauthoryear{{Allard}}{{Allard}}{2014}]{Allard_2014}
{Allard} F.,  2014, in {Booth} M.,  {Matthews} B.~C.,   {Graham} J.~R.,  eds,
  IAU Symposium Vol. 299, Exploring the Formation and Evolution of Planetary
  Systems. pp 271--272, \mn@doi{10.1017/S1743921313008545}

\bibitem[\protect\citeauthoryear{{Allard}, {Homeier}  \& {Freytag}}{{Allard}
  et~al.}{2012}]{Allard_2012}
{Allard} F.,  {Homeier} D.,   {Freytag} B.,  2012, \mn@doi [Philosophical
  Transactions of the Royal Society of London Series A]
  {10.1098/rsta.2011.0269}, \href
  {http://adsabs.harvard.edu/abs/2012RSPTA.370.2765A} {370, 2765}

\bibitem[\protect\citeauthoryear{{Andrews}, {Chanam{\'e}}  \&
  {Ag{\"u}eros}}{{Andrews} et~al.}{2017}]{Andrews_2017}
{Andrews} J.~J.,  {Chanam{\'e}} J.,   {Ag{\"u}eros} M.~A.,  2017, \mn@doi
  [\mnras] {10.1093/mnras/stx2000}, \href
  {https://ui.adsabs.harvard.edu/abs/2017MNRAS.472..675A} {472, 675}

\bibitem[\protect\citeauthoryear{{Ansdell} et~al.,}{{Ansdell}
  et~al.}{2018}]{Ansdell_2018}
{Ansdell} M.,  et~al., 2018, \mn@doi [\apj] {10.3847/1538-4357/aab890}, \href
  {https://ui.adsabs.harvard.edu/abs/2018ApJ...859...21A} {859, 21}

\bibitem[\protect\citeauthoryear{{Arenou} et~al.,}{{Arenou}
  et~al.}{2018}]{Arenou_2018}
{Arenou} F.,  et~al., 2018, \mn@doi [\aap] {10.1051/0004-6361/201833234}, \href
  {https://ui.adsabs.harvard.edu/abs/2018A&A...616A..17A} {616, A17}

\bibitem[\protect\citeauthoryear{{Artymowicz}, {Clarke}, {Lubow}  \&
  {Pringle}}{{Artymowicz} et~al.}{1991}]{Artymowicz_1991}
{Artymowicz} P.,  {Clarke} C.~J.,  {Lubow} S.~H.,   {Pringle} J.~E.,  1991,
  \mn@doi [\apj] {10.1086/185971}, \href
  {https://ui.adsabs.harvard.edu/abs/1991ApJ...370L..35A} {370, L35}

\bibitem[\protect\citeauthoryear{{Bate}}{{Bate}}{2000}]{Bate_2000}
{Bate} M.~R.,  2000, \mn@doi [\mnras] {10.1046/j.1365-8711.2000.03333.x}, \href
  {http://adsabs.harvard.edu/abs/2000MNRAS.314...33B} {314, 33}

\bibitem[\protect\citeauthoryear{{Bate}}{{Bate}}{2009}]{Bate_2009}
{Bate} M.~R.,  2009, \mn@doi [\mnras] {10.1111/j.1365-2966.2008.14106.x}, \href
  {http://adsabs.harvard.edu/abs/2009MNRAS.392..590B} {392, 590}

\bibitem[\protect\citeauthoryear{{Bate}}{{Bate}}{2012}]{Bate_2012}
{Bate} M.~R.,  2012, \mn@doi [\mnras] {10.1111/j.1365-2966.2011.19955.x}, \href
  {https://ui.adsabs.harvard.edu/abs/2012MNRAS.419.3115B} {419, 3115}

\bibitem[\protect\citeauthoryear{{Bate}}{{Bate}}{2014}]{Bate_2014}
{Bate} M.~R.,  2014, \mn@doi [\mnras] {10.1093/mnras/stu795}, \href
  {http://adsabs.harvard.edu/abs/2014MNRAS.442..285B} {442, 285}

\bibitem[\protect\citeauthoryear{{Bate}}{{Bate}}{2018}]{Bate_2018}
{Bate} M.~R.,  2018, \mn@doi [\mnras] {10.1093/mnras/sty169}, \href
  {https://ui.adsabs.harvard.edu/abs/2018MNRAS.475.5618B} {475, 5618}

\bibitem[\protect\citeauthoryear{{Bate}}{{Bate}}{2019}]{Bate_2019}
{Bate} M.~R.,  2019, \mn@doi [\mnras] {10.1093/mnras/stz103}, \href
  {http://adsabs.harvard.edu/abs/2019MNRAS.484.2341B} {484, 2341}

\bibitem[\protect\citeauthoryear{{Bate} \& {Bonnell}}{{Bate} \&
  {Bonnell}}{1997}]{Bate_1997}
{Bate} M.~R.,  {Bonnell} I.~A.,  1997, \mn@doi [\mnras]
  {10.1093/mnras/285.1.33}, \href
  {http://adsabs.harvard.edu/abs/1997MNRAS.285...33B} {285, 33}

\bibitem[\protect\citeauthoryear{{Bonatto}, {Lima}  \& {Bica}}{{Bonatto}
  et~al.}{2012}]{Bonatto_2012}
{Bonatto} C.,  {Lima} E.~F.,   {Bica} E.,  2012, \mn@doi [\aap]
  {10.1051/0004-6361/201118576}, \href
  {https://ui.adsabs.harvard.edu/abs/2012A&A...540A.137B} {540, A137}

\bibitem[\protect\citeauthoryear{{Bovy}, {Rix}, {Green}, {Schlafly}  \&
  {Finkbeiner}}{{Bovy} et~al.}{2016}]{Bovy_2016}
{Bovy} J.,  {Rix} H.-W.,  {Green} G.~M.,  {Schlafly} E.~F.,   {Finkbeiner}
  D.~P.,  2016, \mn@doi [\apj] {10.3847/0004-637X/818/2/130}, \href
  {https://ui.adsabs.harvard.edu/abs/2016ApJ...818..130B} {818, 130}

\bibitem[\protect\citeauthoryear{{Branch}}{{Branch}}{1976}]{Branch_1976}
{Branch} D.,  1976, \mn@doi [\apj] {10.1086/154841}, \href
  {https://ui.adsabs.harvard.edu/abs/1976ApJ...210..392B} {210, 392}

\bibitem[\protect\citeauthoryear{{Brandeker} \& {Cataldi}}{{Brandeker} \&
  {Cataldi}}{2019}]{Brandeker_2019}
{Brandeker} A.,  {Cataldi} G.,  2019, \mn@doi [\aap]
  {10.1051/0004-6361/201834321}, \href
  {http://adsabs.harvard.edu/abs/2019A%26A...621A..86B} {621, A86}

\bibitem[\protect\citeauthoryear{{Bressan}, {Marigo}, {Girardi}, {Salasnich},
  {Dal Cero}, {Rubele}  \& {Nanni}}{{Bressan} et~al.}{2012}]{Bressan_2012}
{Bressan} A.,  {Marigo} P.,  {Girardi} L.,  {Salasnich} B.,  {Dal Cero} C.,
  {Rubele} S.,   {Nanni} A.,  2012, \mn@doi [\mnras]
  {10.1111/j.1365-2966.2012.21948.x}, \href
  {https://ui.adsabs.harvard.edu/abs/2012MNRAS.427..127B} {427, 127}

\bibitem[\protect\citeauthoryear{{Brinch}, {J{\o}rgensen}, {Hogerheijde},
  {Nelson}  \& {Gressel}}{{Brinch} et~al.}{2016}]{Brinch_2016}
{Brinch} C.,  {J{\o}rgensen} J.~K.,  {Hogerheijde} M.~R.,  {Nelson} R.~P.,
  {Gressel} O.,  2016, \mn@doi [\apj] {10.3847/2041-8205/830/1/L16}, \href
  {https://ui.adsabs.harvard.edu/abs/2016ApJ...830L..16B} {830, L16}

\bibitem[\protect\citeauthoryear{{Burgasser}, {Reid}, {Siegler}, {Close},
  {Allen}, {Lowrance}  \& {Gizis}}{{Burgasser} et~al.}{2007}]{Burgasser_2007}
{Burgasser} A.~J.,  {Reid} I.~N.,  {Siegler} N.,  {Close} L.,  {Allen} P.,
  {Lowrance} P.,   {Gizis} J.,  2007, in {Reipurth} B.,  {Jewitt} D.,   {Keil}
  K.,  eds, Protostars and Planets V. p.~427 (\mn@eprint {arXiv}
  {astro-ph/0602122})

\bibitem[\protect\citeauthoryear{{Cantrell} \& {Dougan}}{{Cantrell} \&
  {Dougan}}{2014}]{Cantrell_2014}
{Cantrell} A.~G.,  {Dougan} T.~J.,  2014, \mn@doi [\mnras]
  {10.1093/mnras/stu1890}, \href
  {https://ui.adsabs.harvard.edu/abs/2014MNRAS.445.2028C} {445, 2028}

\bibitem[\protect\citeauthoryear{{Chambers} et~al.,}{{Chambers}
  et~al.}{2016}]{Chambers_2016}
{Chambers} K.~C.,  et~al., 2016, arXiv e-prints, \href
  {https://ui.adsabs.harvard.edu/abs/2016arXiv161205560C} {p. arXiv:1612.05560}

\bibitem[\protect\citeauthoryear{{Chanam{\'e}} \& {Gould}}{{Chanam{\'e}} \&
  {Gould}}{2004}]{Chaname_2004}
{Chanam{\'e}} J.,  {Gould} A.,  2004, \mn@doi [\apj] {10.1086/380442}, \href
  {https://ui.adsabs.harvard.edu/abs/2004ApJ...601..289C} {601, 289}

\bibitem[\protect\citeauthoryear{{Chen}, {Girardi}, {Bressan}, {Marigo},
  {Barbieri}  \& {Kong}}{{Chen} et~al.}{2014}]{Chen_2014}
{Chen} Y.,  {Girardi} L.,  {Bressan} A.,  {Marigo} P.,  {Barbieri} M.,   {Kong}
  X.,  2014, \mn@doi [\mnras] {10.1093/mnras/stu1605}, \href
  {https://ui.adsabs.harvard.edu/abs/2014MNRAS.444.2525C} {444, 2525}

\bibitem[\protect\citeauthoryear{{Choi}, {Dotter}, {Conroy}, {Cantiello},
  {Paxton}  \& {Johnson}}{{Choi} et~al.}{2016}]{Choi_2016}
{Choi} J.,  {Dotter} A.,  {Conroy} C.,  {Cantiello} M.,  {Paxton} B.,
  {Johnson} B.~D.,  2016, \mn@doi [\apj] {10.3847/0004-637X/823/2/102}, \href
  {https://ui.adsabs.harvard.edu/abs/2016ApJ...823..102C} {823, 102}

\bibitem[\protect\citeauthoryear{{Choi}, {Conroy}, {Ting}, {Cargile}, {Dotter}
  \& {Johnson}}{{Choi} et~al.}{2018}]{Choi_2018}
{Choi} J.,  {Conroy} C.,  {Ting} Y.-S.,  {Cargile} P.~A.,  {Dotter} A.,
  {Johnson} B.~D.,  2018, \mn@doi [\apj] {10.3847/1538-4357/aad18c}, \href
  {https://ui.adsabs.harvard.edu/abs/2018ApJ...863...65C} {863, 65}

\bibitem[\protect\citeauthoryear{{Comer{\'o}n}, {Reipurth}, {Yen}  \&
  {Connelley}}{{Comer{\'o}n} et~al.}{2018}]{Comeron_2018}
{Comer{\'o}n} F.,  {Reipurth} B.,  {Yen} H.-W.,   {Connelley} M.~S.,  2018,
  \mn@doi [\aap] {10.1051/0004-6361/201730917}, \href
  {https://ui.adsabs.harvard.edu/abs/2018A&A...612A..73C} {612, A73}

\bibitem[\protect\citeauthoryear{{Connelley}, {Reipurth}  \&
  {Tokunaga}}{{Connelley} et~al.}{2008}]{Connelley_2008}
{Connelley} M.~S.,  {Reipurth} B.,   {Tokunaga} A.~T.,  2008, \mn@doi [\aj]
  {10.1088/0004-6256/135/6/2526}, \href
  {https://ui.adsabs.harvard.edu/abs/2008AJ....135.2526C} {135, 2526}

\bibitem[\protect\citeauthoryear{{Czekala}, {Chiang}, {Andrews}, {Jensen},
  {Torres}, {Wilner}, {Stassun}  \& {Macintosh}}{{Czekala}
  et~al.}{2019}]{Czekala_2019}
{Czekala} I.,  {Chiang} E.,  {Andrews} S.~M.,  {Jensen} E. L.~N.,  {Torres} G.,
   {Wilner} D.~J.,  {Stassun} K.~G.,   {Macintosh} B.,  2019, arXiv e-prints,
  \href {https://ui.adsabs.harvard.edu/abs/2019arXiv190603269C} {p.
  arXiv:1906.03269}

\bibitem[\protect\citeauthoryear{{De Rosa} et~al.,}{{De Rosa}
  et~al.}{2014}]{DeRosa_2014}
{De Rosa} R.~J.,  et~al., 2014, \mn@doi [\mnras] {10.1093/mnras/stt1932}, \href
  {https://ui.adsabs.harvard.edu/abs/2014MNRAS.437.1216D} {437, 1216}

\bibitem[\protect\citeauthoryear{{Dhital}, {West}, {Stassun}  \&
  {Bochanski}}{{Dhital} et~al.}{2010}]{Dhital_2010}
{Dhital} S.,  {West} A.~A.,  {Stassun} K.~G.,   {Bochanski} J.~J.,  2010,
  \mn@doi [\aj] {10.1088/0004-6256/139/6/2566}, \href
  {https://ui.adsabs.harvard.edu/abs/2010AJ....139.2566D} {139, 2566}

\bibitem[\protect\citeauthoryear{{Dhital}, {West}, {Stassun}, {Schluns}  \&
  {Massey}}{{Dhital} et~al.}{2015}]{Dhital_2015}
{Dhital} S.,  {West} A.~A.,  {Stassun} K.~G.,  {Schluns} K.~J.,   {Massey}
  A.~P.,  2015, \mn@doi [\aj] {10.1088/0004-6256/150/2/57}, \href
  {http://adsabs.harvard.edu/abs/2015AJ....150...57D} {150, 57}

\bibitem[\protect\citeauthoryear{{Dolphin}}{{Dolphin}}{2002}]{Dolphin_2002}
{Dolphin} A.~E.,  2002, \mn@doi [\mnras] {10.1046/j.1365-8711.2002.05271.x},
  \href {https://ui.adsabs.harvard.edu/abs/2002MNRAS.332...91D} {332, 91}

\bibitem[\protect\citeauthoryear{{Dorval}, {Boily}, {Moraux}  \&
  {Roos}}{{Dorval} et~al.}{2017}]{Dorval_2017}
{Dorval} J.,  {Boily} C.~M.,  {Moraux} E.,   {Roos} O.,  2017, \mn@doi [\mnras]
  {10.1093/mnras/stw2880}, \href
  {https://ui.adsabs.harvard.edu/abs/2017MNRAS.465.2198D} {465, 2198}

\bibitem[\protect\citeauthoryear{{Duch{\^e}ne}}{{Duch{\^e}ne}}{1999}]{Duchene_1999}
{Duch{\^e}ne} G.,  1999, \aap, \href
  {https://ui.adsabs.harvard.edu/abs/1999A&A...341..547D} {341, 547}

\bibitem[\protect\citeauthoryear{{Duch{\^e}ne} \& {Kraus}}{{Duch{\^e}ne} \&
  {Kraus}}{2013}]{Duchene_2013}
{Duch{\^e}ne} G.,  {Kraus} A.,  2013, \mn@doi [\araa]
  {10.1146/annurev-astro-081710-102602}, \href
  {http://adsabs.harvard.edu/abs/2013ARA%26A..51..269D} {51, 269}

\bibitem[\protect\citeauthoryear{{Duch{\^e}ne}, {Lacour}, {Moraux}, {Goodwin}
  \& {Bouvier}}{{Duch{\^e}ne} et~al.}{2018}]{Duchene_2018}
{Duch{\^e}ne} G.,  {Lacour} S.,  {Moraux} E.,  {Goodwin} S.,   {Bouvier} J.,
  2018, \mn@doi [\mnras] {10.1093/mnras/sty1180}, \href
  {https://ui.adsabs.harvard.edu/abs/2018MNRAS.478.1825D} {478, 1825}

\bibitem[\protect\citeauthoryear{{Duquennoy} \& {Mayor}}{{Duquennoy} \&
  {Mayor}}{1991}]{Duquennoy_1991}
{Duquennoy} A.,  {Mayor} M.,  1991, \aap, \href
  {https://ui.adsabs.harvard.edu/abs/1991A&A...248..485D} {500, 337}

\bibitem[\protect\citeauthoryear{{Eggenberger}, {Halbwachs}, {Udry}  \&
  {Mayor}}{{Eggenberger} et~al.}{2004}]{Eggenberger_2004}
{Eggenberger} A.,  {Halbwachs} J.-L.,  {Udry} S.,   {Mayor} M.,  2004, in
  {Allen} C.,  {Scarfe} C.,  eds,  Revista Mexicana de Astronomia y
  Astrofisica, vol.~27 Vol. 21, Revista Mexicana de Astronomia y Astrofisica
  Conference Series. pp 28--32

\bibitem[\protect\citeauthoryear{{Eggleton}, {Fitchett}  \& {Tout}}{{Eggleton}
  et~al.}{1989}]{Eggleton_1989}
{Eggleton} P.~P.,  {Fitchett} M.~J.,   {Tout} C.~A.,  1989, \mn@doi [\apj]
  {10.1086/168190}, \href
  {https://ui.adsabs.harvard.edu/abs/1989ApJ...347..998E} {347, 998}

\bibitem[\protect\citeauthoryear{{Eisner} et~al.,}{{Eisner}
  et~al.}{2018}]{Eisner_2018}
{Eisner} J.~A.,  et~al., 2018, \mn@doi [\apj] {10.3847/1538-4357/aac3e2}, \href
  {https://ui.adsabs.harvard.edu/abs/2018ApJ...860...77E} {860, 77}

\bibitem[\protect\citeauthoryear{{El-Badry} \& {Rix}}{{El-Badry} \&
  {Rix}}{2018}]{ElBadry_2018}
{El-Badry} K.,  {Rix} H.-W.,  2018, \mn@doi [\mnras] {10.1093/mnras/sty2186},
  \href {http://adsabs.harvard.edu/abs/2018MNRAS.480.4884E} {480, 4884}

\bibitem[\protect\citeauthoryear{{El-Badry} \& {Rix}}{{El-Badry} \&
  {Rix}}{2019}]{ElBadry_2019}
{El-Badry} K.,  {Rix} H.-W.,  2019, \mn@doi [\mnras] {10.1093/mnrasl/sly206},
  \href {http://adsabs.harvard.edu/abs/2019MNRAS.482L.139E} {482, L139}

\bibitem[\protect\citeauthoryear{{El-Badry}, {Weisz}  \& {Quataert}}{{El-Badry}
  et~al.}{2017}]{ElBadry_2017}
{El-Badry} K.,  {Weisz} D.~R.,   {Quataert} E.,  2017, \mn@doi [\mnras]
  {10.1093/mnras/stx436}, \href
  {https://ui.adsabs.harvard.edu/abs/2017MNRAS.468..319E} {468, 319}

\bibitem[\protect\citeauthoryear{{El-Badry}, {Rix}, {Ting}, {Weisz},
  {Bergemann}, {Cargile}, {Conroy}  \& {Eilers}}{{El-Badry}
  et~al.}{2018a}]{ElBadry_2018_mock}
{El-Badry} K.,  {Rix} H.-W.,  {Ting} Y.-S.,  {Weisz} D.~R.,  {Bergemann} M.,
  {Cargile} P.,  {Conroy} C.,   {Eilers} A.-C.,  2018a, \mn@doi [\mnras]
  {10.1093/mnras/stx2758}, \href
  {https://ui.adsabs.harvard.edu/abs/2018MNRAS.473.5043E} {473, 5043}

\bibitem[\protect\citeauthoryear{{El-Badry} et~al.,}{{El-Badry}
  et~al.}{2018b}]{Elbadry_2018a}
{El-Badry} K.,  et~al., 2018b, \mn@doi [\mnras] {10.1093/mnras/sty240}, \href
  {https://ui.adsabs.harvard.edu/abs/2018MNRAS.476..528E} {476, 528}

\bibitem[\protect\citeauthoryear{{El-Badry}, {Rix}  \& {Weisz}}{{El-Badry}
  et~al.}{2018c}]{ElBadry_2018_IFMR}
{El-Badry} K.,  {Rix} H.-W.,   {Weisz} D.~R.,  2018c, \mn@doi [\apjl]
  {10.3847/2041-8213/aaca9c}, \href
  {https://ui.adsabs.harvard.edu/abs/2018ApJ...860L..17E} {860, L17}

\bibitem[\protect\citeauthoryear{{Evans} et~al.,}{{Evans}
  et~al.}{2018}]{Evans_2018}
{Evans} D.~W.,  et~al., 2018, \mn@doi [\aap] {10.1051/0004-6361/201832756},
  \href {https://ui.adsabs.harvard.edu/abs/2018A&A...616A...4E} {616, A4}

\bibitem[\protect\citeauthoryear{{Farris}, {Duffell}, {MacFadyen}  \&
  {Haiman}}{{Farris} et~al.}{2014}]{Farris_2014}
{Farris} B.~D.,  {Duffell} P.,  {MacFadyen} A.~I.,   {Haiman} Z.,  2014,
  \mn@doi [\apj] {10.1088/0004-637X/783/2/134}, \href
  {http://adsabs.harvard.edu/abs/2014ApJ...783..134F} {783, 134}

\bibitem[\protect\citeauthoryear{{Flewelling} et~al.,}{{Flewelling}
  et~al.}{2016}]{Flewelling_2016}
{Flewelling} H.~A.,  et~al., 2016, arXiv e-prints, \href
  {https://ui.adsabs.harvard.edu/abs/2016arXiv161205243F} {p. arXiv:1612.05243}

\bibitem[\protect\citeauthoryear{{Foreman-Mackey}, {Hogg}, {Lang}  \&
  {Goodman}}{{Foreman-Mackey} et~al.}{2013}]{ForemanMackey_2013}
{Foreman-Mackey} D.,  {Hogg} D.~W.,  {Lang} D.,   {Goodman} J.,  2013, \mn@doi
  [\pasp] {10.1086/670067}, \href
  {https://ui.adsabs.harvard.edu/abs/2013PASP..125..306F} {125, 306}

\bibitem[\protect\citeauthoryear{{Gaia Collaboration} et~al.,}{{Gaia
  Collaboration} et~al.}{2016}]{Gaia_2016}
{Gaia Collaboration} et~al., 2016, \mn@doi [\aap]
  {10.1051/0004-6361/201629272}, \href
  {https://ui.adsabs.harvard.edu/abs/2016A&A...595A...1G} {595, A1}

\bibitem[\protect\citeauthoryear{{Gaia Collaboration} et~al.,}{{Gaia
  Collaboration} et~al.}{2018}]{Gaia_2018}
{Gaia Collaboration} et~al., 2018, \mn@doi [\aap]
  {10.1051/0004-6361/201833051}, \href
  {https://ui.adsabs.harvard.edu/abs/2018A&A...616A...1G} {616, A1}

\bibitem[\protect\citeauthoryear{{Geha} et~al.,}{{Geha}
  et~al.}{2013}]{Geha_2013}
{Geha} M.,  et~al., 2013, \mn@doi [\apj] {10.1088/0004-637X/771/1/29}, \href
  {https://ui.adsabs.harvard.edu/abs/2013ApJ...771...29G} {771, 29}

\bibitem[\protect\citeauthoryear{{Ghez}, {Neugebauer}  \& {Matthews}}{{Ghez}
  et~al.}{1993}]{Ghez_1993}
{Ghez} A.~M.,  {Neugebauer} G.,   {Matthews} K.,  1993, \mn@doi [\aj]
  {10.1086/116782}, \href
  {https://ui.adsabs.harvard.edu/abs/1993AJ....106.2005G} {106, 2005}

\bibitem[\protect\citeauthoryear{{Giannuzzi}}{{Giannuzzi}}{1987}]{Giannuzzi_1987}
{Giannuzzi} M.~A.,  1987, \mn@doi [\apss] {10.1007/BF00641559}, \href
  {http://adsabs.harvard.edu/abs/1987Ap%26SS.135..245G} {135, 245}

\bibitem[\protect\citeauthoryear{{Goodwin}}{{Goodwin}}{2010}]{Goodwin_2010}
{Goodwin} S.~P.,  2010, \mn@doi [Philosophical Transactions of the Royal
  Society of London Series A] {10.1098/rsta.2009.0254}, \href
  {https://ui.adsabs.harvard.edu/abs/2010RSPTA.368..851G} {368, 851}

\bibitem[\protect\citeauthoryear{{Gullikson}, {Kraus}  \&
  {Dodson-Robinson}}{{Gullikson} et~al.}{2016}]{Gullikson_2016}
{Gullikson} K.,  {Kraus} A.,   {Dodson-Robinson} S.,  2016, \mn@doi [\aj]
  {10.3847/0004-6256/152/2/40}, \href
  {https://ui.adsabs.harvard.edu/abs/2016AJ....152...40G} {152, 40}

\bibitem[\protect\citeauthoryear{{Halbwachs}}{{Halbwachs}}{1988}]{Halbwachs_1988}
{Halbwachs} J.~L.,  1988, \mn@doi [\apss] {10.1007/BF00656194}, \href
  {https://ui.adsabs.harvard.edu/abs/1988Ap&SS.142..139H} {142, 139}

\bibitem[\protect\citeauthoryear{{Halbwachs}, {Mayor}, {Udry}  \&
  {Arenou}}{{Halbwachs} et~al.}{2003}]{Halbwachs_2003}
{Halbwachs} J.~L.,  {Mayor} M.,  {Udry} S.,   {Arenou} F.,  2003, \mn@doi
  [\aap] {10.1051/0004-6361:20021507}, \href
  {http://adsabs.harvard.edu/abs/2003A%26A...397..159H} {397, 159}

\bibitem[\protect\citeauthoryear{{Hanawa}, {Ochi}  \& {Ando}}{{Hanawa}
  et~al.}{2010}]{Hanawa_2010}
{Hanawa} T.,  {Ochi} Y.,   {Ando} K.,  2010, \mn@doi [\apj]
  {10.1088/0004-637X/708/1/485}, \href
  {http://adsabs.harvard.edu/abs/2010ApJ...708..485H} {708, 485}

\bibitem[\protect\citeauthoryear{{Heggie}}{{Heggie}}{1975}]{Heggie_1975}
{Heggie} D.~C.,  1975, \mn@doi [\mnras] {10.1093/mnras/173.3.729}, \href
  {http://adsabs.harvard.edu/abs/1975MNRAS.173..729H} {173, 729}

\bibitem[\protect\citeauthoryear{{Hills}}{{Hills}}{1975}]{Hills_1975}
{Hills} J.~G.,  1975, \mn@doi [\aj] {10.1086/111815}, \href
  {https://ui.adsabs.harvard.edu/abs/1975AJ.....80..809H} {80, 809}

\bibitem[\protect\citeauthoryear{{Hioki} et~al.,}{{Hioki}
  et~al.}{2007}]{Hioki_2007}
{Hioki} T.,  et~al., 2007, \mn@doi [\aj] {10.1086/519737}, \href
  {https://ui.adsabs.harvard.edu/abs/2007AJ....134..880H} {134, 880}

\bibitem[\protect\citeauthoryear{{Hogeveen}}{{Hogeveen}}{1992}]{Hogeveen_1992}
{Hogeveen} S.~J.,  1992, \mn@doi [\apss] {10.1007/BF00692896}, \href
  {http://adsabs.harvard.edu/abs/1992Ap%26SS.196..299H} {196, 299}

\bibitem[\protect\citeauthoryear{{Juri{\'c}} et~al.,}{{Juri{\'c}}
  et~al.}{2008}]{Juric_2008}
{Juri{\'c}} M.,  et~al., 2008, \mn@doi [\apj] {10.1086/523619}, \href
  {http://adsabs.harvard.edu/abs/2008ApJ...673..864J} {673, 864}

\bibitem[\protect\citeauthoryear{{Katz} et~al.,}{{Katz}
  et~al.}{2019}]{Katz_2019}
{Katz} D.,  et~al., 2019, \mn@doi [\aap] {10.1051/0004-6361/201833273}, \href
  {https://ui.adsabs.harvard.edu/abs/2019A&A...622A.205K} {622, A205}

\bibitem[\protect\citeauthoryear{{Kounkel} et~al.,}{{Kounkel}
  et~al.}{2019}]{Kounkel_2019}
{Kounkel} M.,  et~al., 2019, \mn@doi [\aj] {10.3847/1538-3881/ab13b1}, \href
  {https://ui.adsabs.harvard.edu/abs/2019AJ....157..196K} {157, 196}

\bibitem[\protect\citeauthoryear{{Kouwenhoven}, {Goodwin}, {Parker}, {Davies},
  {Malmberg}  \& {Kroupa}}{{Kouwenhoven} et~al.}{2010}]{Kouwenhoven_2010}
{Kouwenhoven} M.~B.~N.,  {Goodwin} S.~P.,  {Parker} R.~J.,  {Davies} M.~B.,
  {Malmberg} D.,   {Kroupa} P.,  2010, \mn@doi [\mnras]
  {10.1111/j.1365-2966.2010.16399.x}, \href
  {https://ui.adsabs.harvard.edu/abs/2010MNRAS.404.1835K} {404, 1835}

\bibitem[\protect\citeauthoryear{{Kraus}, {Ireland}, {Martinache}  \&
  {Hillenbrand}}{{Kraus} et~al.}{2011}]{Kraus_2011}
{Kraus} A.~L.,  {Ireland} M.~J.,  {Martinache} F.,   {Hillenbrand} L.~A.,
  2011, \mn@doi [\apj] {10.1088/0004-637X/731/1/8}, \href
  {https://ui.adsabs.harvard.edu/abs/2011ApJ...731....8K} {731, 8}

\bibitem[\protect\citeauthoryear{{Kroupa}}{{Kroupa}}{1995}]{Kroupa_1995}
{Kroupa} P.,  1995, \mn@doi [\mnras] {10.1093/mnras/277.4.1507}, \href
  {https://ui.adsabs.harvard.edu/abs/1995MNRAS.277.1507K} {277, 1507}

\bibitem[\protect\citeauthoryear{Kroupa}{Kroupa}{1998}]{Kroupa_1998}
Kroupa P.,  1998, \mn@doi [\mnras] {10.1046/j.1365-8711.1998.01603.x}, \href
  {https://ui.adsabs.harvard.edu/abs/1998MNRAS.298..231K} {298, 231}

\bibitem[\protect\citeauthoryear{{Kroupa}}{{Kroupa}}{2001}]{Kroupa_2001_IMF}
{Kroupa} P.,  2001, \mn@doi [\mnras] {10.1046/j.1365-8711.2001.04022.x}, \href
  {https://ui.adsabs.harvard.edu/abs/2001MNRAS.322..231K} {322, 231}

\bibitem[\protect\citeauthoryear{{Kroupa} \& {Burkert}}{{Kroupa} \&
  {Burkert}}{2001}]{Kroupa_2001}
{Kroupa} P.,  {Burkert} A.,  2001, \mn@doi [\apj] {10.1086/321515}, \href
  {http://adsabs.harvard.edu/abs/2001ApJ...555..945K} {555, 945}

\bibitem[\protect\citeauthoryear{{Krumholz} \& {Thompson}}{{Krumholz} \&
  {Thompson}}{2007}]{Krumholz_2007}
{Krumholz} M.~R.,  {Thompson} T.~A.,  2007, \mn@doi [\apj] {10.1086/515566},
  \href {http://adsabs.harvard.edu/abs/2007ApJ...661.1034K} {661, 1034}

\bibitem[\protect\citeauthoryear{{Kuiper}}{{Kuiper}}{1935}]{Kuiper_1935}
{Kuiper} G.~P.,  1935, \mn@doi [\pasp] {10.1086/124531}, \href
  {http://adsabs.harvard.edu/abs/1935PASP...47...15K} {47, 15}

\bibitem[\protect\citeauthoryear{{Leinert}, {Zinnecker}, {Weitzel}, {Christou},
  {Ridgway}, {Jameson}, {Haas}  \& {Lenzen}}{{Leinert}
  et~al.}{1993}]{Leinert_1993}
{Leinert} C.,  {Zinnecker} H.,  {Weitzel} N.,  {Christou} J.,  {Ridgway} S.~T.,
   {Jameson} R.,  {Haas} M.,   {Lenzen} R.,  1993, \aap, \href
  {https://ui.adsabs.harvard.edu/abs/1993A&A...278..129L} {278, 129}

\bibitem[\protect\citeauthoryear{{L{\'e}pine} \& {Bongiorno}}{{L{\'e}pine} \&
  {Bongiorno}}{2007}]{Lepine_2007}
{L{\'e}pine} S.,  {Bongiorno} B.,  2007, \mn@doi [\aj] {10.1086/510333}, \href
  {https://ui.adsabs.harvard.edu/abs/2007AJ....133..889L} {133, 889}

\bibitem[\protect\citeauthoryear{{Lindegren} et~al.,}{{Lindegren}
  et~al.}{2018}]{Lindegren_2018}
{Lindegren} L.,  et~al., 2018, \mn@doi [\aap] {10.1051/0004-6361/201832727},
  \href {https://ui.adsabs.harvard.edu/abs/2018A&A...616A...2L} {616, A2}

\bibitem[\protect\citeauthoryear{{Lomax}, {Whitworth}, {Hubber}, {Stamatellos}
  \& {Walch}}{{Lomax} et~al.}{2015}]{Lomax_2015}
{Lomax} O.,  {Whitworth} A.~P.,  {Hubber} D.~A.,  {Stamatellos} D.,   {Walch}
  S.,  2015, \mn@doi [\mnras] {10.1093/mnras/stu2530}, \href
  {https://ui.adsabs.harvard.edu/abs/2015MNRAS.447.1550L} {447, 1550}

\bibitem[\protect\citeauthoryear{{Lucy}}{{Lucy}}{2006}]{Lucy_2006}
{Lucy} L.~B.,  2006, \mn@doi [\aap] {10.1051/0004-6361:20065746}, \href
  {http://adsabs.harvard.edu/abs/2006A%26A...457..629L} {457, 629}

\bibitem[\protect\citeauthoryear{{Lucy} \& {Ricco}}{{Lucy} \&
  {Ricco}}{1979}]{Lucy_1979}
{Lucy} L.~B.,  {Ricco} E.,  1979, \mn@doi [\aj] {10.1086/112434}, \href
  {http://adsabs.harvard.edu/abs/1979AJ.....84..401L} {84, 401}

\bibitem[\protect\citeauthoryear{{Marks} \& {Kroupa}}{{Marks} \&
  {Kroupa}}{2012}]{Marks_2012}
{Marks} M.,  {Kroupa} P.,  2012, \mn@doi [\aap] {10.1051/0004-6361/201118231},
  \href {https://ui.adsabs.harvard.edu/abs/2012A&A...543A...8M} {543, A8}

\bibitem[\protect\citeauthoryear{{Marks}, {Kroupa}  \& {Oh}}{{Marks}
  et~al.}{2011}]{Marks_2011}
{Marks} M.,  {Kroupa} P.,   {Oh} S.,  2011, \mn@doi [\mnras]
  {10.1111/j.1365-2966.2011.19257.x}, \href
  {https://ui.adsabs.harvard.edu/abs/2011MNRAS.417.1684M} {417, 1684}

\bibitem[\protect\citeauthoryear{{Mathieu}}{{Mathieu}}{1994}]{Mathieu_1994}
{Mathieu} R.~D.,  1994, \mn@doi [\araa] {10.1146/annurev.aa.32.090194.002341},
  \href {https://ui.adsabs.harvard.edu/abs/1994ARA&A..32..465M} {32, 465}

\bibitem[\protect\citeauthoryear{{Matsumoto}, {Saigo}  \&
  {Takakuwa}}{{Matsumoto} et~al.}{2019}]{Matsumoto_2019}
{Matsumoto} T.,  {Saigo} K.,   {Takakuwa} S.,  2019, \mn@doi [\apj]
  {10.3847/1538-4357/aaf6ab}, \href
  {http://adsabs.harvard.edu/abs/2019ApJ...871...36M} {871, 36}

\bibitem[\protect\citeauthoryear{{Mazeh}, {Goldberg}, {Duquennoy}  \&
  {Mayor}}{{Mazeh} et~al.}{1992}]{Mazeh_1992}
{Mazeh} T.,  {Goldberg} D.,  {Duquennoy} A.,   {Mayor} M.,  1992, \mn@doi
  [\apj] {10.1086/172058}, \href
  {https://ui.adsabs.harvard.edu/abs/1992ApJ...401..265M} {401, 265}

\bibitem[\protect\citeauthoryear{{Mazeh}, {Simon}, {Prato}, {Markus}  \&
  {Zucker}}{{Mazeh} et~al.}{2003}]{Mazeh_2003}
{Mazeh} T.,  {Simon} M.,  {Prato} L.,  {Markus} B.,   {Zucker} S.,  2003,
  \mn@doi [\apj] {10.1086/379346}, \href
  {https://ui.adsabs.harvard.edu/abs/2003ApJ...599.1344M} {599, 1344}

\bibitem[\protect\citeauthoryear{{Miranda}, {Mu{\~n}oz}  \& {Lai}}{{Miranda}
  et~al.}{2017}]{Miranda_2017}
{Miranda} R.,  {Mu{\~n}oz} D.~J.,   {Lai} D.,  2017, \mn@doi [\mnras]
  {10.1093/mnras/stw3189}, \href
  {http://adsabs.harvard.edu/abs/2017MNRAS.466.1170M} {466, 1170}

\bibitem[\protect\citeauthoryear{{Moe} \& {Di Stefano}}{{Moe} \& {Di
  Stefano}}{2013}]{Moe_2013}
{Moe} M.,  {Di Stefano} R.,  2013, \mn@doi [\apj] {10.1088/0004-637X/778/2/95},
  \href {https://ui.adsabs.harvard.edu/abs/2013ApJ...778...95M} {778, 95}

\bibitem[\protect\citeauthoryear{{Moe} \& {Di Stefano}}{{Moe} \& {Di
  Stefano}}{2017}]{Moe_2017}
{Moe} M.,  {Di Stefano} R.,  2017, \mn@doi [\apjs] {10.3847/1538-4365/aa6fb6},
  \href {https://ui.adsabs.harvard.edu/abs/2017ApJS..230...15M} {230, 15}

\bibitem[\protect\citeauthoryear{{Moe}, {Kratter}  \& {Badenes}}{{Moe}
  et~al.}{2018}]{Moe_2018}
{Moe} M.,  {Kratter} K.~M.,   {Badenes} C.,  2018, preprint, \href
  {http://adsabs.harvard.edu/abs/2018arXiv180802116M} {} (\mn@eprint {arXiv}
  {1808.02116})

\bibitem[\protect\citeauthoryear{{Moeckel} \& {Bate}}{{Moeckel} \&
  {Bate}}{2010}]{Moeckel_2010}
{Moeckel} N.,  {Bate} M.~R.,  2010, \mn@doi [\mnras]
  {10.1111/j.1365-2966.2010.16347.x}, \href
  {https://ui.adsabs.harvard.edu/abs/2010MNRAS.404..721M} {404, 721}

\bibitem[\protect\citeauthoryear{{Moody}, {Shi}  \& {Stone}}{{Moody}
  et~al.}{2019}]{Moody_2019}
{Moody} M. S.~L.,  {Shi} J.-M.,   {Stone} J.~M.,  2019, \mn@doi [\apj]
  {10.3847/1538-4357/ab09ee}, \href
  {https://ui.adsabs.harvard.edu/abs/2019ApJ...875...66M} {875, 66}

\bibitem[\protect\citeauthoryear{{Mu{\~n}oz}, {Miranda}  \& {Lai}}{{Mu{\~n}oz}
  et~al.}{2019}]{Munoz_2019}
{Mu{\~n}oz} D.~J.,  {Miranda} R.,   {Lai} D.,  2019, \mn@doi [\apj]
  {10.3847/1538-4357/aaf867}, \href
  {https://ui.adsabs.harvard.edu/abs/2019ApJ...871...84M} {871, 84}

\bibitem[\protect\citeauthoryear{{Murphy}, {Moe}, {Kurtz}, {Bedding},
  {Shibahashi}  \& {Boffin}}{{Murphy} et~al.}{2018}]{Murphy_2018}
{Murphy} S.~J.,  {Moe} M.,  {Kurtz} D.~W.,  {Bedding} T.~R.,  {Shibahashi} H.,
   {Boffin} H. M.~J.,  2018, \mn@doi [\mnras] {10.1093/mnras/stx3049}, \href
  {https://ui.adsabs.harvard.edu/abs/2018MNRAS.474.4322M} {474, 4322}

\bibitem[\protect\citeauthoryear{{Nelson} \& {Marzari}}{{Nelson} \&
  {Marzari}}{2016}]{Nelson_2016}
{Nelson} A.~F.,  {Marzari} F.,  2016, \mn@doi [\apj]
  {10.3847/0004-637X/827/2/93}, \href
  {https://ui.adsabs.harvard.edu/abs/2016ApJ...827...93N} {827, 93}

\bibitem[\protect\citeauthoryear{{Nordstr{\"o}m} et~al.,}{{Nordstr{\"o}m}
  et~al.}{2004}]{Nordstrom_2004}
{Nordstr{\"o}m} B.,  et~al., 2004, \mn@doi [\aap] {10.1051/0004-6361:20035959},
  \href {http://adsabs.harvard.edu/abs/2004A%26A...418..989N} {418, 989}

\bibitem[\protect\citeauthoryear{{Ochi}, {Sugimoto}  \& {Hanawa}}{{Ochi}
  et~al.}{2005}]{Ochi_2005}
{Ochi} Y.,  {Sugimoto} K.,   {Hanawa} T.,  2005, \mn@doi [\apj]
  {10.1086/428601}, \href {http://adsabs.harvard.edu/abs/2005ApJ...623..922O}
  {623, 922}

\bibitem[\protect\citeauthoryear{{Offner}, {Kratter}, {Matzner}, {Krumholz}  \&
  {Klein}}{{Offner} et~al.}{2010}]{Offner_2010}
{Offner} S. S.~R.,  {Kratter} K.~M.,  {Matzner} C.~D.,  {Krumholz} M.~R.,
  {Klein} R.~I.,  2010, \mn@doi [\apj] {10.1088/0004-637X/725/2/1485}, \href
  {https://ui.adsabs.harvard.edu/abs/2010ApJ...725.1485O} {725, 1485}

\bibitem[\protect\citeauthoryear{{{\"O}pik}}{{{\"O}pik}}{1924}]{Opik_1924}
{{\"O}pik} E.,  1924, Publications of the Tartu Astrofizica Observatory, \href
  {https://ui.adsabs.harvard.edu/abs/1924PTarO..25f...1O} {25, 1}

\bibitem[\protect\citeauthoryear{{Parker} \& {Reggiani}}{{Parker} \&
  {Reggiani}}{2013}]{Parker_2013}
{Parker} R.~J.,  {Reggiani} M.~M.,  2013, \mn@doi [\mnras]
  {10.1093/mnras/stt600}, \href
  {https://ui.adsabs.harvard.edu/abs/2013MNRAS.432.2378P} {432, 2378}

\bibitem[\protect\citeauthoryear{{Parker}, {Goodwin}, {Kroupa}  \&
  {Kouwenhoven}}{{Parker} et~al.}{2009}]{Parker_2009}
{Parker} R.~J.,  {Goodwin} S.~P.,  {Kroupa} P.,   {Kouwenhoven} M.~B.~N.,
  2009, \mn@doi [\mnras] {10.1111/j.1365-2966.2009.15032.x}, \href
  {https://ui.adsabs.harvard.edu/abs/2009MNRAS.397.1577P} {397, 1577}

\bibitem[\protect\citeauthoryear{{Pinsonneault} \& {Stanek}}{{Pinsonneault} \&
  {Stanek}}{2006}]{Pinsonneault_2006}
{Pinsonneault} M.~H.,  {Stanek} K.~Z.,  2006, \mn@doi [\apj] {10.1086/502799},
  \href {https://ui.adsabs.harvard.edu/abs/2006ApJ...639L..67P} {639, L67}

\bibitem[\protect\citeauthoryear{{Raghavan} et~al.,}{{Raghavan}
  et~al.}{2010}]{Raghavan_2010}
{Raghavan} D.,  et~al., 2010, \mn@doi [\apjs] {10.1088/0067-0049/190/1/1},
  \href {https://ui.adsabs.harvard.edu/abs/2010ApJS..190....1R} {190, 1}

\bibitem[\protect\citeauthoryear{{Reggiani} \& {Meyer}}{{Reggiani} \&
  {Meyer}}{2011}]{Reggiani_2011}
{Reggiani} M.~M.,  {Meyer} M.~R.,  2011, \mn@doi [\apj]
  {10.1088/0004-637X/738/1/60}, \href
  {https://ui.adsabs.harvard.edu/abs/2011ApJ...738...60R} {738, 60}

\bibitem[\protect\citeauthoryear{{Reipurth} \& {Mikkola}}{{Reipurth} \&
  {Mikkola}}{2012}]{Reipurth_2012}
{Reipurth} B.,  {Mikkola} S.,  2012, \mn@doi [\nat] {10.1038/nature11662},
  \href {https://ui.adsabs.harvard.edu/abs/2012Natur.492..221R} {492, 221}

\bibitem[\protect\citeauthoryear{{Reipurth}, {Guimar{\~a}es}, {Connelley}  \&
  {Bally}}{{Reipurth} et~al.}{2007}]{Reipurth_2007}
{Reipurth} B.,  {Guimar{\~a}es} M.~M.,  {Connelley} M.~S.,   {Bally} J.,  2007,
  \mn@doi [\aj] {10.1086/523596}, \href
  {https://ui.adsabs.harvard.edu/abs/2007AJ....134.2272R} {134, 2272}

\bibitem[\protect\citeauthoryear{{Sadavoy} \& {Stahler}}{{Sadavoy} \&
  {Stahler}}{2017}]{Sadavoy_2017}
{Sadavoy} S.~I.,  {Stahler} S.~W.,  2017, \mn@doi [\mnras]
  {10.1093/mnras/stx1061}, \href
  {https://ui.adsabs.harvard.edu/abs/2017MNRAS.469.3881S} {469, 3881}

\bibitem[\protect\citeauthoryear{{Seabroke} \& {Gilmore}}{{Seabroke} \&
  {Gilmore}}{2007}]{Seabroke_2007}
{Seabroke} G.~M.,  {Gilmore} G.,  2007, \mn@doi [\mnras]
  {10.1111/j.1365-2966.2007.12210.x}, \href
  {http://adsabs.harvard.edu/abs/2007MNRAS.380.1348S} {380, 1348}

\bibitem[\protect\citeauthoryear{{Shahaf} \& {Mazeh}}{{Shahaf} \&
  {Mazeh}}{2019}]{Shahaf_2019}
{Shahaf} S.,  {Mazeh} T.,  2019, arXiv e-prints, \href
  {https://ui.adsabs.harvard.edu/abs/2019arXiv190513239S} {p. arXiv:1905.13239}

\bibitem[\protect\citeauthoryear{{Shatsky} \& {Tokovinin}}{{Shatsky} \&
  {Tokovinin}}{2002}]{Shatsky_2002}
{Shatsky} N.,  {Tokovinin} A.,  2002, \mn@doi [\aap]
  {10.1051/0004-6361:20011542}, \href
  {https://ui.adsabs.harvard.edu/abs/2002A&A...382...92S} {382, 92}

\bibitem[\protect\citeauthoryear{{Shaya} \& {Olling}}{{Shaya} \&
  {Olling}}{2011}]{Shaya_2011}
{Shaya} E.~J.,  {Olling} R.~P.,  2011, \mn@doi [\apjs]
  {10.1088/0067-0049/192/1/2}, \href
  {https://ui.adsabs.harvard.edu/abs/2011ApJS..192....2S} {192, 2}

\bibitem[\protect\citeauthoryear{{Shi}, {Krolik}, {Lubow}  \& {Hawley}}{{Shi}
  et~al.}{2012}]{Shi_2012}
{Shi} J.-M.,  {Krolik} J.~H.,  {Lubow} S.~H.,   {Hawley} J.~F.,  2012, \mn@doi
  [\apj] {10.1088/0004-637X/749/2/118}, \href
  {https://ui.adsabs.harvard.edu/abs/2012ApJ...749..118S} {749, 118}

\bibitem[\protect\citeauthoryear{{Simon} \& {Obbie}}{{Simon} \&
  {Obbie}}{2009}]{Simon_2009}
{Simon} M.,  {Obbie} R.~C.,  2009, \mn@doi [\aj]
  {10.1088/0004-6256/137/2/3442}, \href
  {http://adsabs.harvard.edu/abs/2009AJ....137.3442S} {137, 3442}

\bibitem[\protect\citeauthoryear{{S{\"o}derhjelm}}{{S{\"o}derhjelm}}{2000}]{Soderhjelm_2000}
{S{\"o}derhjelm} S.,  2000, \mn@doi [Astronomische Nachrichten]
  {10.1002/1521-3994(200008)321:3<165::AID-ASNA165>3.0.CO;2-Z}, \href
  {https://ui.adsabs.harvard.edu/abs/2000AN....321..165S} {321, 165}

\bibitem[\protect\citeauthoryear{{S{\"o}derhjelm}}{{S{\"o}derhjelm}}{2007}]{Soderhjelm_2007}
{S{\"o}derhjelm} S.,  2007, \mn@doi [\aap] {10.1051/0004-6361:20066024}, \href
  {http://adsabs.harvard.edu/abs/2007A%26A...463..683S} {463, 683}

\bibitem[\protect\citeauthoryear{{Sollima}}{{Sollima}}{2019}]{Sollima_2019}
{Sollima} A.,  2019, arXiv e-prints, \href
  {https://ui.adsabs.harvard.edu/abs/2019arXiv190405646S} {p. arXiv:1904.05646}

\bibitem[\protect\citeauthoryear{{Sterzik} \& {Durisen}}{{Sterzik} \&
  {Durisen}}{1998}]{Sterzik_1998}
{Sterzik} M.~F.,  {Durisen} R.~H.,  1998, \aap, \href
  {https://ui.adsabs.harvard.edu/abs/1998A&A...339...95S} {339, 95}

\bibitem[\protect\citeauthoryear{{Takakuwa}, {Saigo}, {Matsumoto}, {Saito},
  {Lim}, {Hanawa}, {Yen}  \& {Ho}}{{Takakuwa} et~al.}{2017}]{Takakuwa_2017}
{Takakuwa} S.,  {Saigo} K.,  {Matsumoto} T.,  {Saito} M.,  {Lim} J.,  {Hanawa}
  T.,  {Yen} H.-W.,   {Ho} P. T.~P.,  2017, \mn@doi [\apj]
  {10.3847/1538-4357/aa6116}, \href
  {https://ui.adsabs.harvard.edu/abs/2017ApJ...837...86T} {837, 86}

\bibitem[\protect\citeauthoryear{{Tang}, {MacFadyen}  \& {Haiman}}{{Tang}
  et~al.}{2017}]{Tang_2017}
{Tang} Y.,  {MacFadyen} A.,   {Haiman} Z.,  2017, \mn@doi [\mnras]
  {10.1093/mnras/stx1130}, \href
  {https://ui.adsabs.harvard.edu/abs/2017MNRAS.469.4258T} {469, 4258}

\bibitem[\protect\citeauthoryear{{Ting} \& {Rix}}{{Ting} \&
  {Rix}}{2018}]{Ting_2018}
{Ting} Y.-S.,  {Rix} H.-W.,  2018, arXiv e-prints, \href
  {http://adsabs.harvard.edu/abs/2018arXiv180803278T} {}

\bibitem[\protect\citeauthoryear{{Tobin} et~al.,}{{Tobin}
  et~al.}{2016}]{Tobin_2016}
{Tobin} J.~J.,  et~al., 2016, \mn@doi [\nat] {10.1038/nature20094}, \href
  {https://ui.adsabs.harvard.edu/abs/2016Natur.538..483T} {538, 483}

\bibitem[\protect\citeauthoryear{{Tokovinin}}{{Tokovinin}}{2000}]{Tokovinin_2000}
{Tokovinin} A.~A.,  2000, \aap, \href
  {http://adsabs.harvard.edu/abs/2000A%26A...360..997T} {360, 997}

\bibitem[\protect\citeauthoryear{{Tokovinin}}{{Tokovinin}}{2011}]{Tokovinin_2011}
{Tokovinin} A.,  2011, \mn@doi [\aj] {10.1088/0004-6256/141/2/52}, \href
  {https://ui.adsabs.harvard.edu/abs/2011AJ....141...52T} {141, 52}

\bibitem[\protect\citeauthoryear{{Tokovinin}}{{Tokovinin}}{2014}]{Tokovinin_2014}
{Tokovinin} A.,  2014, \mn@doi [\aj] {10.1088/0004-6256/147/4/87}, \href
  {https://ui.adsabs.harvard.edu/abs/2014AJ....147...87T} {147, 87}

\bibitem[\protect\citeauthoryear{{Tokovinin}}{{Tokovinin}}{2017a}]{Tokovinin_2017b}
{Tokovinin} A.,  2017a, \mn@doi [\mnras] {10.1093/mnras/stx707}, \href
  {https://ui.adsabs.harvard.edu/abs/2017MNRAS.468.3461T} {468, 3461}

\bibitem[\protect\citeauthoryear{{Tokovinin}}{{Tokovinin}}{2017b}]{Tokovinin_2017}
{Tokovinin} A.,  2017b, \mn@doi [\apj] {10.3847/1538-4357/aa7746}, \href
  {http://adsabs.harvard.edu/abs/2017ApJ...844..103T} {844, 103}

\bibitem[\protect\citeauthoryear{{Tokovinin} \& {Smekhov}}{{Tokovinin} \&
  {Smekhov}}{2002}]{Tokovinin_2002}
{Tokovinin} A.~A.,  {Smekhov} M.~G.,  2002, \mn@doi [\aap]
  {10.1051/0004-6361:20011586}, \href
  {https://ui.adsabs.harvard.edu/abs/2002A&A...382..118T} {382, 118}

\bibitem[\protect\citeauthoryear{{Trimble}}{{Trimble}}{1974}]{Trimble_1974}
{Trimble} V.,  1974, \mn@doi [\aj] {10.1086/111639}, \href
  {https://ui.adsabs.harvard.edu/abs/1974AJ.....79..967T} {79, 967}

\bibitem[\protect\citeauthoryear{{Trimble}}{{Trimble}}{1987}]{Trimble_1987}
{Trimble} V.,  1987, \mn@doi [Astronomische Nachrichten]
  {10.1002/asna.2113080607}, \href
  {http://adsabs.harvard.edu/abs/1987AN....308..343T} {308, 343}

\bibitem[\protect\citeauthoryear{{Trimble}}{{Trimble}}{1990}]{Trimble_1990}
{Trimble} V.,  1990, \mn@doi [\mnras] {10.1093/mnras/242.1.79}, \href
  {https://ui.adsabs.harvard.edu/abs/1990MNRAS.242...79T} {242, 79}

\bibitem[\protect\citeauthoryear{{Weinberg}, {Shapiro}  \&
  {Wasserman}}{{Weinberg} et~al.}{1987}]{Weinberg_1987}
{Weinberg} M.~D.,  {Shapiro} S.~L.,   {Wasserman} I.,  1987, \mn@doi [\apj]
  {10.1086/164883}, \href
  {https://ui.adsabs.harvard.edu/abs/1987ApJ...312..367W} {312, 367}

\bibitem[\protect\citeauthoryear{{White} \& {Ghez}}{{White} \&
  {Ghez}}{2001}]{White_2001}
{White} R.~J.,  {Ghez} A.~M.,  2001, \mn@doi [\apj] {10.1086/321542}, \href
  {http://adsabs.harvard.edu/abs/2001ApJ...556..265W} {556, 265}

\bibitem[\protect\citeauthoryear{{York} et~al.,}{{York}
  et~al.}{2000}]{York_2000}
{York} D.~G.,  et~al., 2000, \mn@doi [\aj] {10.1086/301513}, \href
  {https://ui.adsabs.harvard.edu/abs/2000AJ....120.1579Y} {120, 1579}

\bibitem[\protect\citeauthoryear{{Young} \& {Clarke}}{{Young} \&
  {Clarke}}{2015}]{Young_2015}
{Young} M.~D.,  {Clarke} C.~J.,  2015, \mn@doi [\mnras]
  {10.1093/mnras/stv1512}, \href
  {http://adsabs.harvard.edu/abs/2015MNRAS.452.3085Y} {452, 3085}

\bibitem[\protect\citeauthoryear{{Ziegler} et~al.,}{{Ziegler}
  et~al.}{2018}]{Ziegler_2018}
{Ziegler} C.,  et~al., 2018, \mn@doi [\aj] {10.3847/1538-3881/aad80a}, \href
  {https://ui.adsabs.harvard.edu/abs/2018AJ....156..259Z} {156, 259}

\bibitem[\protect\citeauthoryear{{de Val-Borro}, {Gahm}, {Stempels}  \&
  {Pepli{\'n}ski}}{{de Val-Borro} et~al.}{2011}]{de_val_borro_2011}
{de Val-Borro} M.,  {Gahm} G.~F.,  {Stempels} H.~C.,   {Pepli{\'n}ski} A.,
  2011, \mn@doi [\mnras] {10.1111/j.1365-2966.2011.18339.x}, \href
  {http://adsabs.harvard.edu/abs/2011MNRAS.413.2679D} {413, 2679}

\bibitem[\protect\citeauthoryear{{van Biesbroeck}}{{van
  Biesbroeck}}{1916}]{Biesbroeck_1916}
{van Biesbroeck} G.,  1916, \mn@doi [\aj] {10.1086/104155}, \href
  {https://ui.adsabs.harvard.edu/abs/1916AJ.....29..173V} {29, 173}

\makeatother
\end{thebibliography}



\appendix

\section{Functional form of \texorpdfstring{$p(q)$}{pq} }
\label{sec:func_form}

\subsection{How sharp is the twin feature?}
\label{sec:how_sharp}
In our default model, the twin feature is characterized by two number: $F_{\rm twin}$, the fractional excess of binaries with near-equal masses, and $q_{\rm twin}$, the mass ratio above which the excess manifests itself. The implicit assumption in this model is that (a) the increase in the mass ratio distribution near $q=1$ occurs abruptly, not gradually, and (b) the amount of twin excess is constant between $q=q_{\rm twin}$ and $q=1$. In this Section, we assess the validity of this assumption by using a more flexible model for the twin excess.

We model the twin excess between $q=0.85$ and $q=1$ as a histogram with 5 equally-spaced bins of width $\Delta q = 0.03$. As in the fiducial model, we assume the smooth underlying distribution in this region is $p(q)\sim q^{\gamma_{\rm  largeq}}$. We then introduce a set of uncorrelated weights, $w_i$ with $i=0\ldots 5$, and multiply $p(q)$ in the $i$-th bin by $w_i$. With this parameterization, $w_i=1$ entails no deviation from a smooth power law in the $i$-th bin, $w_i=2$ corresponds to a factor-of-two enhancement relative to the underlying power law, and $w_i=0$ means that there are no binaries at all with mass ratios in the $i$-th bin. With a sufficiently large number of bins, this parameterization is flexible enough to represent any arbitrary shape of $p(q)$. 5 bins of width 0.03 is a pragmatic choice given the size of our binary catalog, since the shot noise uncertainty increases as the bin size is reduced. 

\begin{figure*}
    \includegraphics[width=\textwidth]{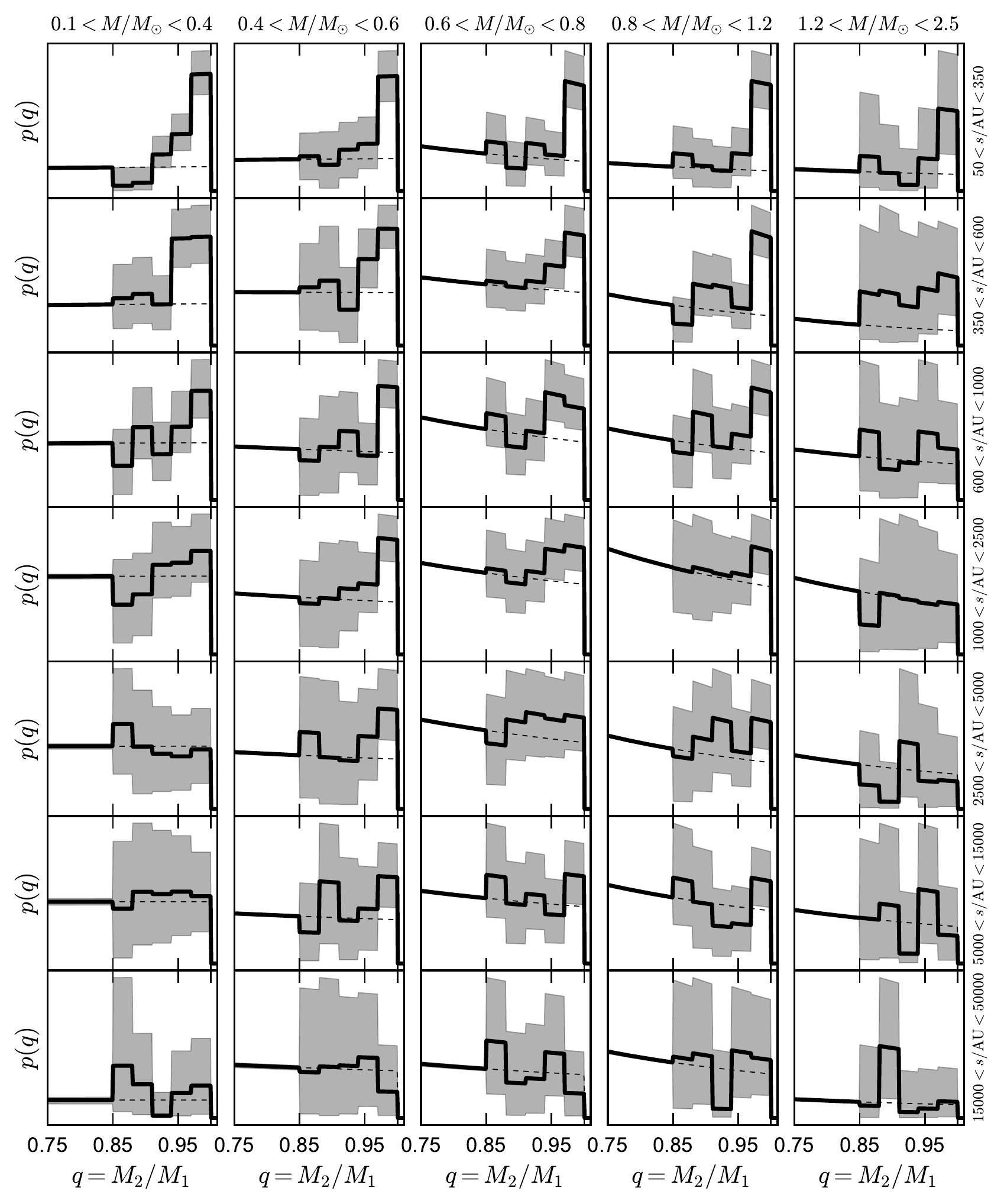}
    \caption{Contraints on $p(q)$ near $q=1$. Solid lines and gray shaded regions represent median and $2\sigma$ (middle 95.4\%) constraints; y-axis scale is linear. We model the mass ratio distribution over $0.85 < q <1$ as a histogram with 5 bins of width $0.03$, fitting for the height of each bin as a free parameter. The twin excess generally becomes statistically significant only at $0.97 < q < 1$, in some cases with a modest enhancement at $0.94 < q < 0.97$. At lower $q$, the mass ratio distribution is generally consistent with a smooth power law (dashed line). }
    \label{fig:hist_model}
\end{figure*}

Figure~\ref{fig:hist_model} shows results of fitting a histogram model. Consistent with Figure~\ref{fig:qdist_ranges}, a statistically significant excess of twins is found out to several thousand AU for $0.1 < M_1/M_{\odot} < 1.2$ and $s \lesssim 2500$\,AU, with the excess reaching $s=15,000$\,AU for $0.4 < M_1/M_{\odot} < 0.6$. In most mass and separation bins, the data is consistent with a smooth power law distribution (with no twin excess) all the way up to $q=0.97$. In a few mass bins, there is also a significant excess relative to the underlying power law at $0.94 < q < 0.97$, but never at $q < 0.94$. That is, in all bins where it is significant, the twin feature is  ``thin'': there is not a broad peak at $q\gtrsim 0.8$, but a narrow excess only at $q \gtrsim 0.95$. It is because of the narrowness of the twin peak that the excess of equal-mass pairs is so clearly apparent in the observed data (e.g. Figure~\ref{fig:dG_points}), even though twins only make up a small fraction ($<5$\%; see Figure~\ref{fig:Ftwin_vs_separation}) of the total binary population at wide separations. 

\subsection{Choice of \texorpdfstring{$q_{\rm break}$}{qbreak}}
\label{sec:qbreak}

\begin{figure*}
    \includegraphics[width=\textwidth]{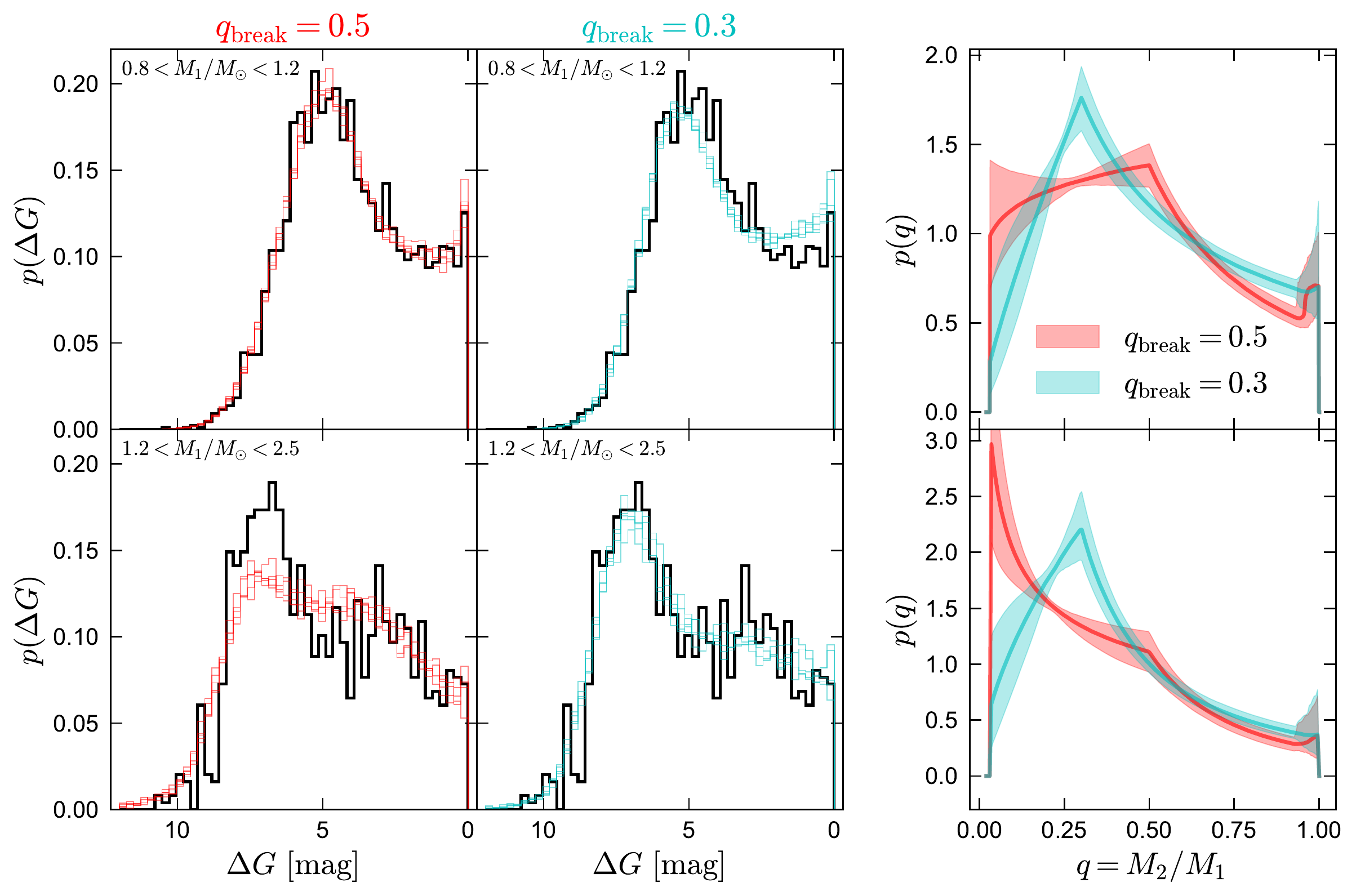}
    \caption{Mass ratio distribution constraints (right) and predicted $\Delta G$ distributions (left) for fits with $q_{\rm break}=0.5$ (red) and $q_{\rm break}=0.3$ (cyan). We show a single separation bin ($1000 < s < 2500$). Top panels show solar-type stars ($0.8 < M_1/M_{\odot} < 1.2$); bottom panels show A and F stars ($1.2 < M_1/M_{\odot} < 2.5$). For each choice of $q_{\rm break}$, we show the best-fit constraints when $q_{\rm break}$ is fixed at that value and other parameters are left free. For solar-type stars, a significantly better fit can be obtained with $q_{\rm break}=0.5$ than with $q_{\rm break=0.3}$. The opposite is true for A and F stars, where the data strongly favor $q_{\rm break} =0.3$.}
    \label{fig:varying_qbreak}
\end{figure*}

As discussed in Section~\ref{sec:modeling}, we model the mass ratio distribution in all but the lowest-mass bin as a broken power law whose slope changes at $q=q_{\rm break}$ (Figure~\ref{fig:schematic_q_dist}). In order to make comparison between different separation bins more straightforward, we do not leave $q_{\rm break}$ free, but instead fix it to a single value in a given mass bin: $q_{\rm break}=0.3$ for $1.2 < M_1/M_{\odot} < 2.5$, and $q_{\rm break}=0.5$ in all other mass bins. These values were chosen by trial and error, but it {\it is} necessary to use different values of $q_{\rm break}$ in different mass bins, as we now show.

Figure~\ref{fig:varying_qbreak} shows constraints on the mass ratio distribution (right) and corresponding predicted $\Delta G$ distributoins (left) for two choices of $q_{\rm break}$ and two bins of primary mass. A single separation bin is shown for illustrative purposes: $1000 < s < 2500$, where the twin excess is weak in both mass bins. In the left panels, we compare the observed distributions of $\Delta G$ to Monte Carlo binary populations produced for 5 draws from the posterior. 

The top panels show that for solar-mass primaries ($0.8 < M_1/M_{\odot} < 1.2$), the best-fit model obtained while assuming $q_{\rm break}=0.5$ provides a significantly better fit to the observed $\Delta G$ distribution than the best-fit model with $q_{\rm break}=0.3$: in the latter case, the observed distribution is poorly reproduced both at $0< \Delta G \lesssim 2$ and at $\Delta G \approx 5$. On the other hand, the bottom panels show that $q_{\rm break}=0.3$ provides a much better fit for A and F star primaries. 

\subsection{Smoothly-broken power law}
\label{sec:smoothly_broken}

\begin{figure}
    \includegraphics[width=\columnwidth]{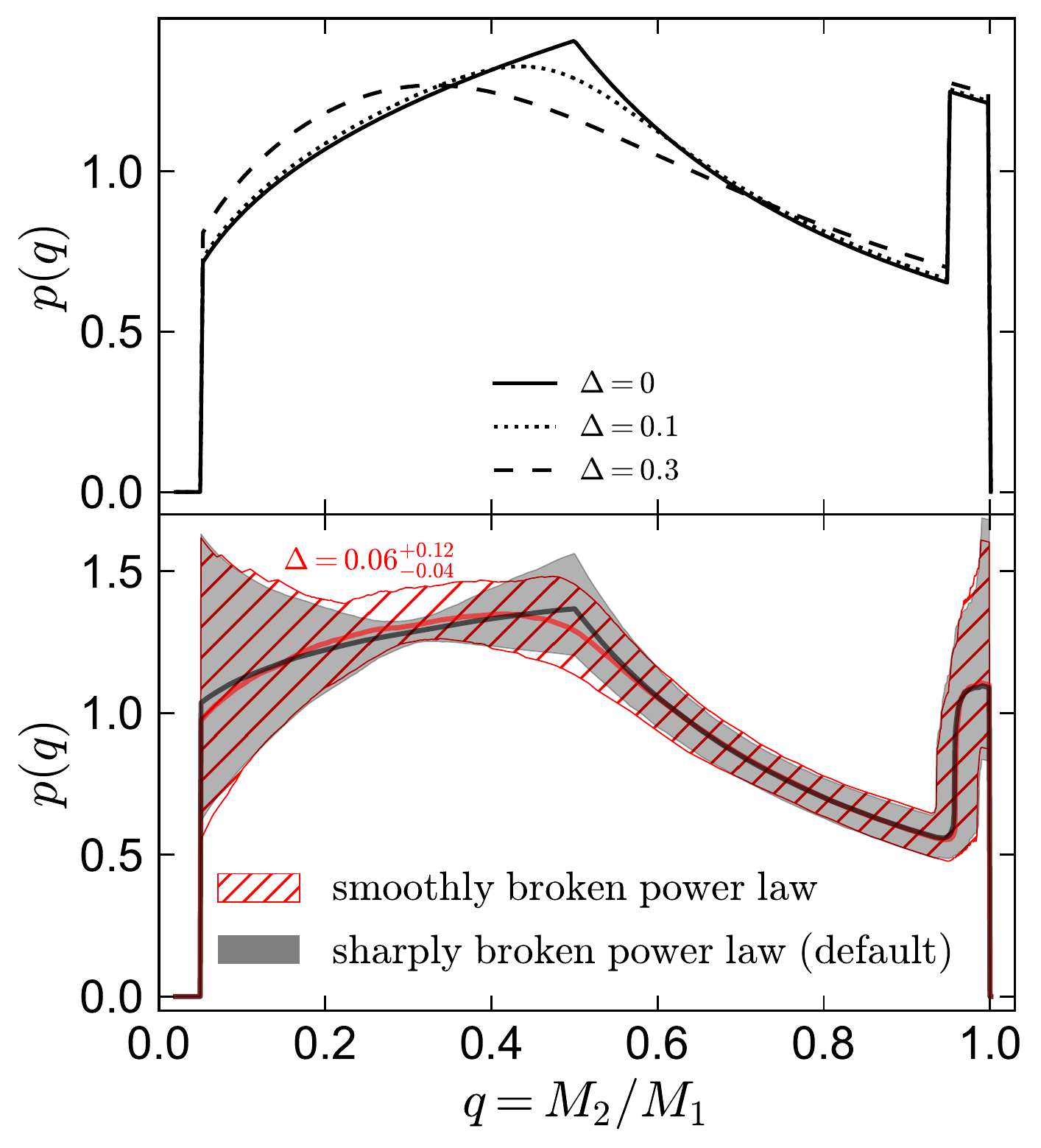}
    \caption{Top panel shows example smoothly broken power laws for different values of the smoothing parameter, $\Delta$. For all cases, $\gamma_{\rm smallq} =0.3$, $\gamma_{\rm largeq} = -1.2$, $F_{\rm twin}=0.04$, and $q_{\rm twin}=0.95$. Increasing $\Delta$ softens the transition between  $p\sim q^{\gamma_{\rm smallq}}$ and $p\sim q^{\gamma_{\rm largeq}}$. Bottom: middle 95.4\% constraints on $p(q)$ for $0.8<M_1/M_{\odot} < 1.2$ and $600 <s/{\rm AU} < 1000$. We compare constraints obtained by fitting a smoothly broken power law (red hatched) to those obtained by fitting our default model, a simple double power law (i.e., forcing $\Delta=0$). The constraints we obtain with the two models are similar, with the most significant difference being that the smoothly broken model lacks (a) the sharp break at $q_{\rm break}=0.5$ and (b) the artificial suppression of uncertainty near a pivot point at $q=0.3$.}
    \label{fig:smoothly_broken}
\end{figure}

As discussed in Section~\ref{sec:results}, our choice to model the mass ratio distribution with a broken power law leads to an unphysical sharp break in the best-fit mass ratio distribution at intermediate $q$. We test the sensitivity of our constraints on $p(q)$ to the assumed functional form below. 
As an alternative to a sharply broken power law, we fit a ``smoothly broken power law'' with the functional form 
\begin{equation}
    p\left(q\right)\propto \left(\frac{q}{q_{\rm break}}\right)^{\gamma_{\rm smallq}}\left[1+\left(\frac{q}{q_{\rm break}}\right)^{1/\Delta}\right]^{\left(\gamma_{\rm largeq}-\gamma_{\rm smallq}\right)\Delta}.
    \label{eq:smoothly_broken}
\end{equation}
This function approaches $p\propto q^{\gamma_{\rm smallq}}$ at $q/q_{\rm break} \ll 1$, and $p\propto q^{\gamma_{\rm largeq}}$ at  $q/q_{\rm break} \gg 1$. The parameter $\Delta$ controls the sharpness of the transition between the two regimes; in the limit of $\Delta\to 0$, it reduces to a simple double power law with a sharp transition between the two slopes.  

In Figure~\ref{fig:smoothly_broken}, we compare the constraints on $p(q)$ obtained for a single mass and separation bin when using the fiducial model (gray) and a smoothly broken power law (hatched red). Overall, the constraints are quite similar, but as expected, fitting a broken power law model smooths the best-fit profile for $p(q)$ and removes the region near $q\sim 0.3$ where the nominal uncertainty is suppressed. We find qualitatively similar results for other bins of mass and separation. However, we note that increasing $\Delta$  shifts the peak of the distribution towards lower $q$ (upper panel of Figure~\ref{fig:smoothly_broken}). Since $\Delta$ itself is often only weakly constrained, this can lead to parameter covariances between $\Delta$, $\gamma_{\rm smallq}$, and $q_{\rm break}$. We therefore use the simpler broken power law as our fidicial model.

\section{Sensitivity to systematics}
\label{sec:systmatics}

\begin{figure*}
    \centering
    \includegraphics[width=\textwidth]{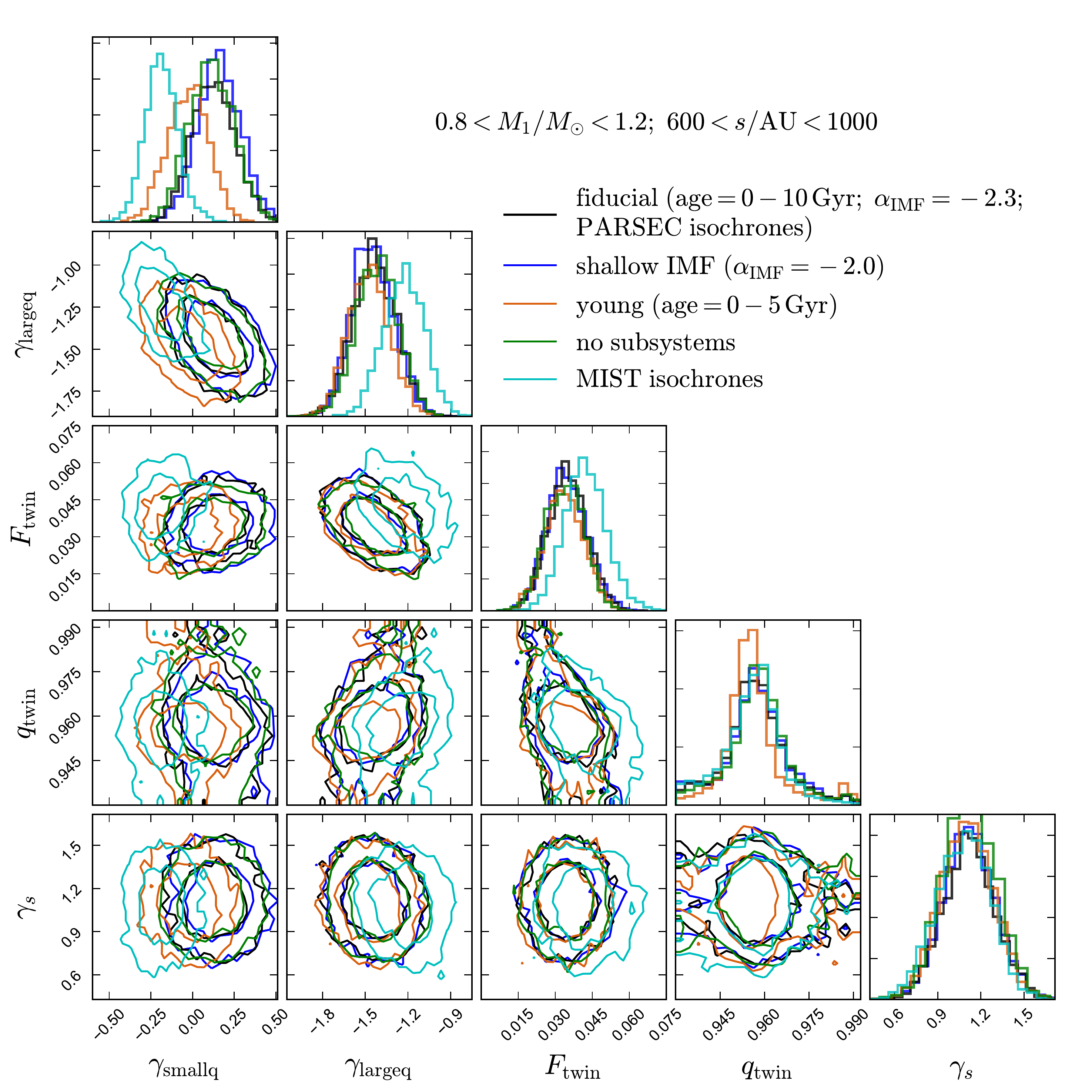}
    \caption{Sixty-eight and ninety-five per cent probability contours for one representative bin of primary mass and physical separation. Panels on the diagonal show marginalized probability distributions. Black contours shows results for fitting the fiducial model. Blue, brown, green, and cyan contours show constraints obtained when the assumed IMF, star formation history, model for unresolved binaries, and stellar models are varied. Overall, constraints are not very sensitive to the assumed IMF, SFH, or unresolved binary model; this is particularly true for $F_{\rm twin}$ and $q_{\rm twin}$.} They are moderately sensitive to the choice of stellar models, particularly at low mass ratios.
    \label{fig:corner}
\end{figure*}

The uncertainties we report on mass ratio distribution parameters represent formal fitting uncertainties due to Poisson errors, but they do not include various systematic uncertainties due to modeling choices we make that are held fixed during fitting. Here we vary several aspects of the model to assess how sensitively our constraints depend on them.  Figure~\ref{fig:corner} shows the effects of varying the assumed IMF, stellar age distribution, and the isochrones used to generate synthetic photometry, on our parameter constraints. We show a single primary mass and separation bin with typical uncertainties and sensitivity to systematics.

Varying the IMF (blue) has very weak effects on our constraints, primarily because we fit narrow bins in primary mass independently. Our constraints are also relatively insensitive to the assumed star formation history (brown). Varying the age distribution has several effects: it changes the number of low-mass companions that are still undergoing Kelvin-Helmholtz contraction and are thus brighter, it changes the mass-luminosity relation at $M\gtrsim 0.7 M_{\odot}$, where age effects are non-negligible; and it changes the distance distribution implied in our model because we adopt an age-dependent scale height (Equation~\ref{eq:hz}). Varying the model for unresolved subsystems has similarly weak effects. The green lines in Figure~\ref{fig:corner} show constraints obtained when unresolved subsystems are ignored entirely in the model (i.e. none of the synthetic wide binary components are assigned an unresolved companion). The thus-obtained constraints are nearly indistinguishable from those obtained when subsystems are included in the model.

Changing the adopted stellar models from PARSEC to MIST \citep{Choi_2016} has the strongest effect on our constraints. Because the two models predict modestly different mass-luminosity relations, a given $\Delta G$ corresponds to different $q$ in the two models. This has weak effects on the constraints on $F_{\rm twin}$ and $q_{\rm twin}$: twins necessarily have similar masses, and the mapping between luminosity and mass varies less over a small range of masses than over a large one. Differences between the models are largest for low-mass stars, where isochrones are known to be more uncertain.

We note that while there clearly {\it are} uncertainties in our results associated with the stellar models, the PARSEC models appear to fit the {\it Gaia} data for low-mass stars significantly better than the MIST models. Particularly on the lower main sequence $\rm M_G \gtrsim 10$, we are unable to reproduce the morphology of the CMD unless we assume a higher-than-expected mean metallicity of $\left\langle \left[{\rm Fe/H}\right]\right\rangle \approx+0.2$. This is why we use the PARSEC synthetic photometry in our fiducial model. We note that the PARSEC models use surface boundary conditions that are empirically calibrated to match the observed mass-radius relation at low masses \citep{Chen_2014}; without this calibration, the tension between their predictions and the observed lower main sequence is similar to that found for MIST models \citep[e.g.][]{Choi_2018}.

\section{Twins in other catalogs}
\label{sec:other_catalogs}

\begin{figure}
    \includegraphics[width=\columnwidth]{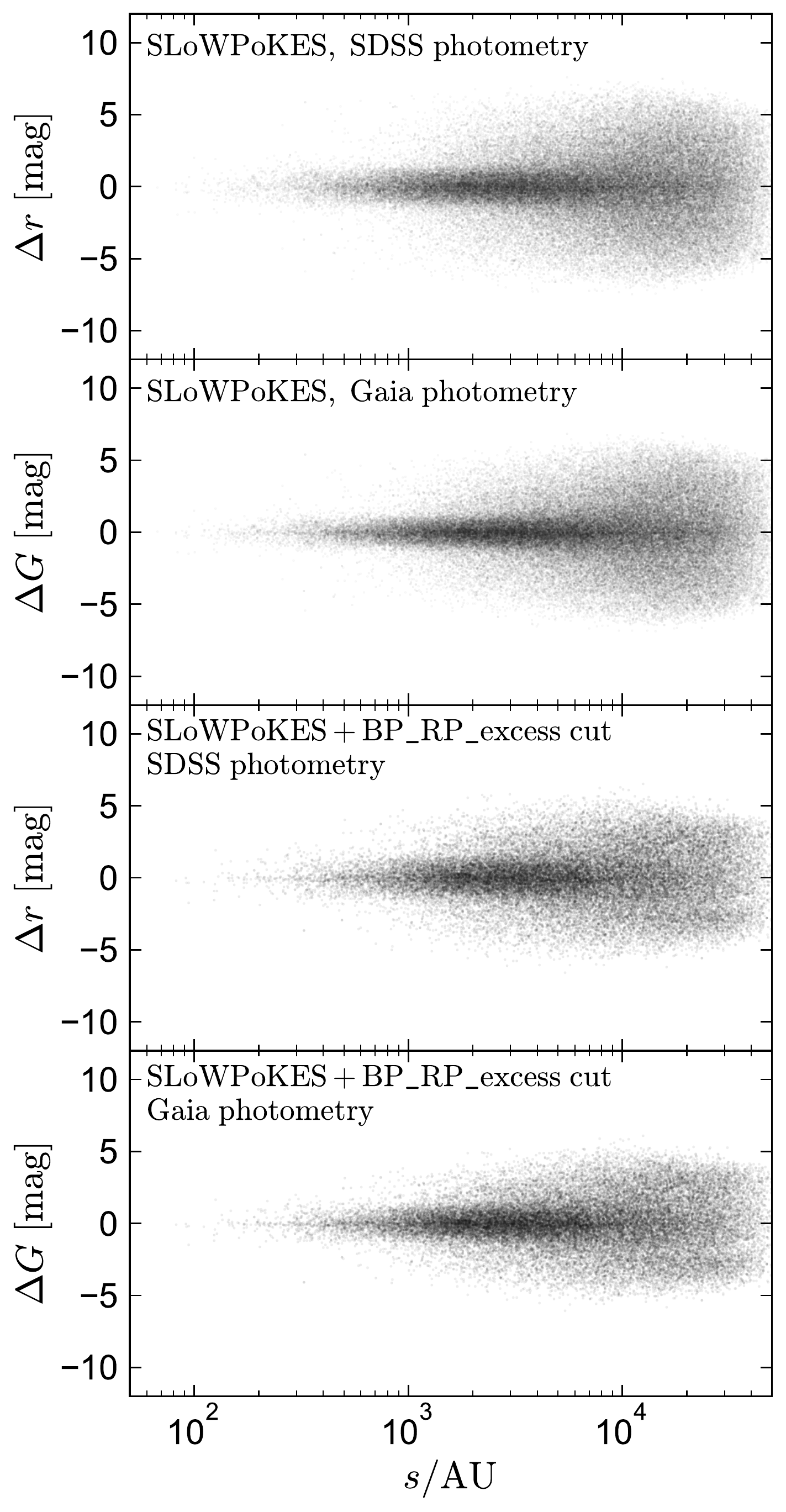}
    \caption{Excess of twins in the SLoWPoKES-II catalog \citep{Dhital_2015}. Top panel shows the full catalog with SDSS photometry. Here there is no obvious excess of equal-brightness twins. Middle panel shows Gaia photometry from the full SLoWPoKES-II catalog. The Gaia photometry is significantly more precise than the ground-based SDSS photometry, so a slight excess of twins is apparent, but the signal is somewhat weaker than that found in our catalog. The bottom two panels show SDSS and Gaia photometry for the subset of the SLoWPoKES-II catalog for which the Gaia photometry passes our cut on \texttt{phot\_bp\_rp\_excess\_factor}. This cut removes binaries in which the photometry for either component is contaminated (either by the binary companion or by a background star). Once sources with contaminated photometry have been removed, an excess of twins is visible in both the SDSS and Gaia photometry.}
    \label{fig:dhital_comparison}
\end{figure}

\begin{figure}
    \includegraphics[width=\columnwidth]{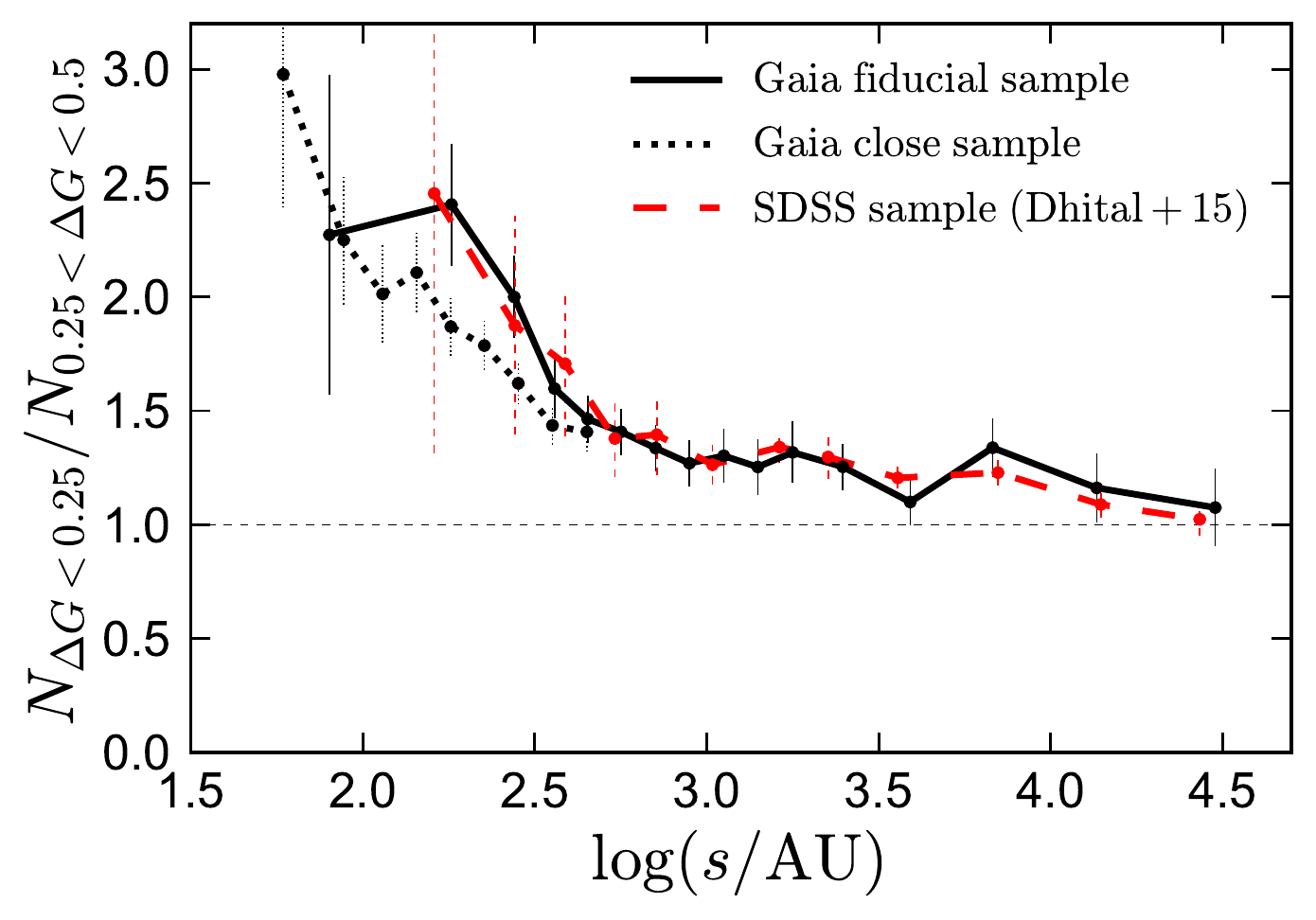}
    \caption{Ratio of the number of binaries with nearly equal magnitudes ($\Delta G < 0.25$) to the number with slightly different magnitude ($0.25 < \Delta G < 0.5$) as a function of projected physical separation (right). Error bars are 1 sigma. We compare the fiducial sample from this work (solid black line) to the subsample of the SLoWPoKES-II catalog \citep{Dhital_2015} with uncontaminated Gaia photometry (dashed red line; bottom panel of Figure~\ref{fig:dhital_comparison}) and the sample of Gaia binaries from \citet{ElBadry_2019}, which  reaches closer angular separations than our fiducial sample but passes less stringent quality cuts. }
    \label{fig:fractional_excess_vs_separation}
\end{figure}

Another large sample of low-mass wide binaries is the SLoWPoKES-II catalog created by \citep{Dhital_2015} using SDSS photometry. This catalog was produced without astrometry and extends to larger distances than our catalog ($d \lesssim 1$\,kpc). Despite the lack of parallaxes and proper motions \citet{Dhital_2015} were able to ensure relatively low contamination by limiting their search to close angular separations ($\theta < 20$\, arcsec) and using an isochrone prior (i.e., requiring both components of a binary to have similar photometric distance). The selection function of their catalog is quite different from ours, so we do not attempt a full probabilistic model of the mass ratio distribution. However, we check whether a twin excess is visible in the magnitude difference distributions of their catalog and whether its strength at fixed physical separation is consistent with that found in our catalog. 

Figure~\ref{fig:dhital_comparison} (top panel) shows the distribution of $\Delta r$, the difference in apparent $r$-band magnitude between the two components, as a function of physical separation for the full SLoWPoKES-II catalog. Projected physical separations are calculated using the mean photometric distance of the two stars estimate by \citet{Dhital_2015}. We verified that for stars that are bright enough to have precise {\it Gaia} parallaxes, the photometric and geometric distances are usually in reasonably good agreement. The SLoWPoKES-II catalog primarily contains binaries with $q\gtrsim 0.5$: there are few pairs with $\Delta r > 5$, and at physical separations $s\lesssim 2000$\,AU, most pairs have $\Delta r < 2$. This is primarily a consequence of the catalog selection function. No clear excess of equal-brightness twins is apparent in the top panel of Figure~\ref{fig:dhital_comparison}: the distribution of $\Delta r$ at fixed separation is reasonably smooth. 

To check whether the lack of obvious twin excess in the full SLoWPoKES-II catalog is a consequence of the SDSS ground-based photometry being poorer, we cross-matched the catalog with {\it Gaia} DR2. A {\it Gaia} source was found with 1 arcsecond for both components for $\sim$90\% of the binaries in the catalog. In the 2nd panel of Figure~\ref{fig:dhital_comparison}, we plot all of these sources (many of which do not pass the photometric and astrometric quality cuts imposed in our catalog), showing the difference in $G$-band magnitude. Here, an excess of equal-brightness binaries {\it is} visible, but it appears somewhat weaker than in our catalog. 

Finally, in the 3rd and 4th panels of Figure~\ref{fig:dhital_comparison}, we show the SDSS and {\it Gaia} photometry for the subset of the SLoWPoKES-II catalog in which the {\it Gaia} photometry for both components passes the quality cut on \texttt{phot\_bp\_rp\_excess\_factor} that is imposed on our catalog. This cut removes objects in which the sum of the flux in the BP and RP bands is not consistent with the flux in the G-band \citep[see][]{Evans_2018}. Because the fluxes in the BP and RP bands are dispersed over several arcsec while the $G$-band flux is obtained by profile fitting a narrower image, this cut efficiently selects objects in which the {\it Gaia} photometry (both BP/RP and $G$-band) is contaminated by a nearby source. Once this cut is applied, a stronger excess of equal-brightness binaries is apparent in both the {\it Gaia} and SDSS photometry. We show below that for the subsample of the SLoWPoKES-II catalog with uncontaminated {\it Gaia} photometry, the twin excess at fixed separation is consistent with that found in our catalog. 

A natural worry is that the cut on \texttt{phot\_bp\_rp\_excess\_factor} could somehow select against pairs that are close on the sky and do not have nearly identical brightness or color, thus erroneously producing an apparent excess of equal-brightness binaries. However, the fact that no excess of equal-brightness pairs is found for chance alignments  subject to the same cut speaks against this possibility. Inspecting the SDSS images of binaries in which the {\it Gaia} photometry for either component does not pass the  \texttt{phot\_bp\_rp\_excess\_factor} cut, we find that in most cases the contamination is quite strong: either the light from the two stars is blended, or another source is blended with one of the components. In such cases, the individual magnitudes of the components cannot be measured with high fidelity, likely leading to underestimated photometric errors. 

In Figure~\ref{fig:fractional_excess_vs_separation}, we compare the strength of the twin excess found in the {\it Gaia} photometry for the SLoWPoKES-II catalog (once the \texttt{phot\_bp\_rp\_excess\_factor} cut is applied) at fixed separation (red) to that found in our catalog (black). The y-axis is similar to that in Figure~\ref{fig:three_distances}. The strength of the twin excess is consistent between the two catalogs at all separations. We note that at close physical separations, binaries are only spatially resolvable if they are nearby, so at $s\lesssim 500$\,AU, there is substantial overlap between the two catalogs. At $s\gtrsim 1000$\,AU, there is little overlap, because most of the SLoWPoKES-II binaries are too distant and faint to enter our catalog. 

We also show in Figure~\ref{fig:fractional_excess_vs_separation} the excess of equal-brightness binaries found in the catalog of {\it Gaia} binaries constructed in \citet{ElBadry_2019}, which targeted closer separations. This catalog only contains binaries with $s < 500$\,AU. Unlike our primary catalog, it did not require the components to have measured $G_{\rm BP}$ and $G_{\rm RP}$ magnitudes, and it did not apply any cuts on \texttt{phot\_bp\_rp\_excess\_factor}. This makes it possible to reach binaries with a factor of $\sim$4 closer angular separations than our primary catalog. Figure~\ref{fig:fractional_excess_vs_separation} shows that in this catalog, the twin excess continues to increase toward smaller separations, at least to $s\approx 50$\.AU. The twin excess is slightly weaker at fixed separation in this catalog, likely because the photometry is more contaminated, but it is strongly inconsistent with 0. This provides further evidence that the twin excess is not the result of the \texttt{phot\_bp\_rp\_excess\_factor} cut creating a bias against binaries with unequal brightness or color. 

In addition to the comparisons to other surveys described above, we have also verified that the twin excess we find is not an artifact of the {\it Gaia} photometry. Cross-matching our full binary catalog with the Pan-STARRS1 survey \citep{Chambers_2016, Flewelling_2016}, we find a twin excess of similar strength in its photometry. The excess is also apparent in SDSS photometry, but it is somewhat less narrow there due to the larger typical photometric errors. 

\section{Selection function}
\label{sec:selection_function_details}

The selection function of {\it Gaia} DR2 is known to exhibit spatial variation on small scales as a result of the scanning law \citep[e.g.][]{Arenou_2018}. Modeling all the small-scale structure in the selection function is beyond the scope of this work, but doing so is not necessary for our purposes. First, spatial variations are almost negligible for sources within the magnitude range of our catalog ($G\lesssim 18$). More generally, since we do not expect the \textit{intrinsic} properties of the binary population to vary much with on-sky position (particularly on small scales), one can construct an effective selection function averaged over the whole sky without introducing biases \citep[e.g.][]{Bovy_2016}.

We model the selection function for binaries as the product of two single-star terms and a cross-term that depends on the angular separation and magnitude difference of the two stars (Equation~\ref{eq:s_12}). Both terms are described below. 

\subsection{Single-star term}
\label{sec:single_star_term}

\begin{figure*}
    \includegraphics[width=\textwidth]{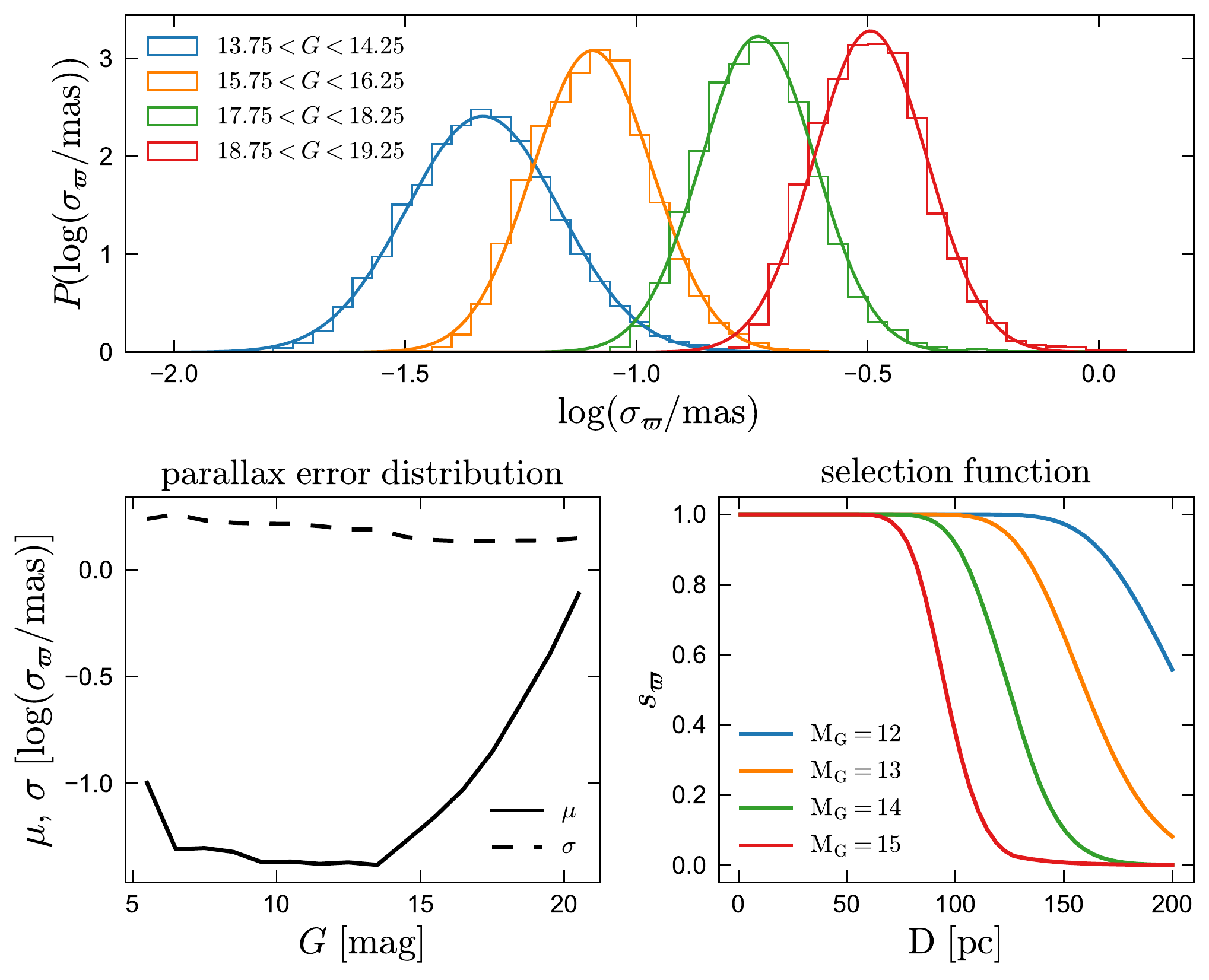}
    \caption{Top panel shows the distributions of log parallax error for nearby stars ($D<100\,\rm pc$) in 0.5-mag wide bins of apparent magnitude. We find that the distribution of $\sigma_{\varpi}$ at fixed apparent magnitude is roughly lognormal; solid curves show Gaussian fits to the observed histograms. Bottom left panel shows the mean and dispersion of the log parallax error distribution as a function of $G$-band magnitude. Modeling the distribution of parallax error as a lognormal with $\mu$ and $\sigma$ following this panel allows us to calculate what fraction of stars with a given absolute magnitude and distance will pass the cut of \texttt{parallax\_over\_error} > 20 (Equation~\ref{eq:s_varpi}; bottom right panel). }
    \label{fig:selection_function_parallax}
\end{figure*}

The single-star term ($s_1$ and $s_2$ in Equation~\ref{eq:s_12}; here we default to $s_1$) is set primarily by the following cuts we imposed in creating the catalog:
\begin{enumerate}
    \item 5\% parallax uncertainty: \texttt{parallax\_over\_error} > 20
    \item precise photometry: \texttt{phot\_bp\_mean\_flux\_over\_error} > 20, \texttt{phot\_rp\_mean\_flux\_over\_error} > 20, and \texttt{phot\_g\_mean\_flux\_over\_error} > 20
    \item good astrometric model fits; $\sqrt{\chi^2/(\nu'-5)}<1.2\times \max(1, \exp(-0.2(G-19.5))$, where $\chi^2$ and $\nu'$ are respectively referred to as \texttt{astrometric\_chi2\_al} and \texttt{astrometric\_n\_good\_obs\_al} in the {\it Gaia} archive. 
    \item uncontaminated photometry; $1.0+0.015({\rm G_{BP}-G_{RP}})^{2} <$ \texttt{phot\_bp\_rp\_excess\_factor} $ < 1.3+0.06({\rm G_{BP}-G_{RP}})^{2}$. 
\end{enumerate}
The motivation for these cuts is described in \citet{Lindegren_2018}. We discuss the effects of each cut below. We note that \citetalias{ElBadry_2018} also required difference in total proper motion of the two stars to be precise, satisfying $\sigma_{\Delta \mu } <1.5$\,mas\,yr$^{-1}$. We find that for the subset of the binaries studied in this work, that cut has a negligible effect on the selection function, as sources that satisfy (i) and (ii) already satisfy it. 

For (i), parallax error is expected to depend mainly on apparent magnitude. Figure~\ref{fig:selection_function_parallax} shows that the parallax error at a given $G-$band magnitude roughly follows a lognormal distribution, the mean value of which increases for fainter stars. This allows us to calculate the fraction of stars at a given distance and magnitude that will have \texttt{parallax\_over\_error} > 20. In particular, the distribution of $\varpi/\sigma_{\varpi}$ at a given distance and magnitude is
\begin{align}
    \label{eq:P_varpi}
    P\left(\varpi/\sigma_{\varpi}\right)=P\left(\log\sigma_{\varpi}\right)\left|\frac{{\rm d}\log\sigma_{\varpi}}{{\rm d}\left(\varpi/\sigma_{\varpi}\right)}\right|,
\end{align}
where $P\left(\log\sigma_{\varpi}\right)$ is the (Gaussian) distribution of log parallax error. The fraction of stars with $\varpi/\sigma_{\varpi}>20$ is then 
\begin{align}
    \label{eq:s_varpi}
    s_{\varpi}(G)	&=\int_{20}^{\infty}P\left(\varpi/\sigma_{\varpi}\right)\,{\rm d}\left(\varpi/\sigma_{\varpi}\right)\\
	&=\frac{1}{2}\left[1-{\rm erf}\left(\frac{\mu_{f}\ln10+\ln\left(20/\varpi\right)}{\sqrt{2}\sigma_{f}\ln10}\right)\right].
\end{align}
Here $\mu_f$ and $\sigma_f$ represent the mean and dispersion of the log parallax error distributions at magnitude $G$; i.e., the quantities plotted in the lower left panel of Figure~\ref{fig:selection_function_parallax}. Given Equation~(\ref{eq:s_varpi}), we can can calculate the probability that a star with given absolute magnitude at a given distance will satisfy \texttt{parallax\_over\_error} > 20; this is shown in the lower right panel of Figure~\ref{fig:selection_function_parallax}.

\begin{figure*}
    \includegraphics[width=\textwidth]{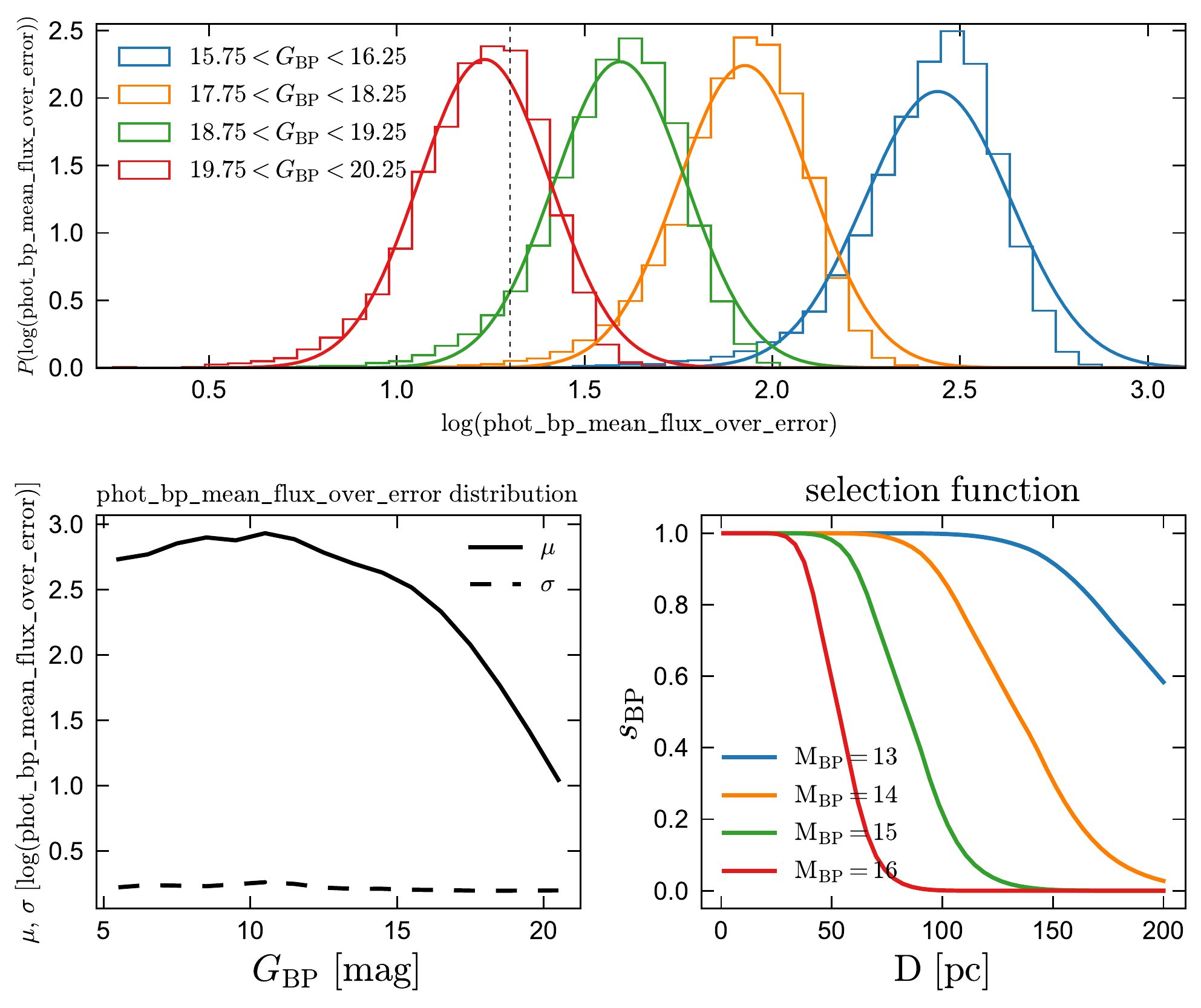}
    \caption{Top panel shows distributions of \texttt{phot\_bp\_mean\_flux\_over\_error} for nearby stars ($D<100$\,pc) in different 0.5 mag wide bins of $G_{\rm BP}$ apparent magnitude. Solid lines show lognormal fits; vertical dashed line shows the cut of \texttt{phot\_bp\_mean\_flux\_over\_error} > 20 that both components of a binary must satisfy to enter the catalog. Bottom left panel shows $\mu$ and $\sigma$ of lognormal fits to these distributions as a function of $G_{\rm BP}$. From these, we calculate the probability that a star at a given distance and absolute magnitude $M_{\rm BP}$ will have \texttt{phot\_bp\_mean\_flux\_over\_error} > 20 (bottom left panel).}
    \label{fig:selection_function_bp}
\end{figure*}

A similar strategy can be used to calculate the fraction of stars passing the photometric precision cuts. We find that the cut on \texttt{phot\_bp\_mean\_flux\_over\_error} is the most important: once it is applied and white dwarfs are excluded, all the sources that pass it also pass the other cuts in (ii). This quantity depends primarily on the apparent magnitude in the $G_{\rm BP}$ band. As Figure~\ref{fig:selection_function_bp} shows, the distribution of \texttt{phot\_bp\_mean\_flux\_over\_error} at a given $G_{\rm BP}$ is also roughly lognormal. We use the empirical dependence of the distributions of \texttt{phot\_bp\_mean\_flux\_over\_error} on $G_{\rm BP}$ from the bottom left panel of Figure~\ref{fig:selection_function_bp} to predict the distribution of \texttt{phot\_bp\_mean\_flux\_over\_error} for a hypothetical star with absolute BP-band magnitude at a given distance, and then calculate $s_{\rm BP}$ as the fraction of that distribution that exceeds the adopted threshold of 20:
\begin{align}
    \label{eq:s_bp}
    s_{{\rm BP}}\left(G_{{\rm BP}}\right)	&=\int_{20}^{\infty}P\left(\log\left(F_{{\rm BP}}/\sigma_{{\rm BP}}\right)\right)\,{\rm d}\log\left(F_{{\rm BP}}/\sigma_{{\rm BP}}\right)  \\
	&=\frac{1}{2}\left[{\rm 1}+{\rm erf}\left(\frac{\mu_{f}-\log20}{\sqrt{2}\sigma_{f}}\right)\right].
\end{align}
Here $F_{{\rm BP}}/\sigma_{{\rm BP}}$ represents \texttt{phot\_bp\_mean\_flux\_over\_error}, and $\mu_f$ and $\sigma_f$ represent the mean and dispersion of the log \texttt{phot\_bp\_mean\_flux\_over\_error} distributions at apparent BP-band magnitude $G_{\rm BP}$; i.e., the quantities plotted in the lower left panel of Figure~\ref{fig:selection_function_bp}.

\begin{figure*}
    \includegraphics[width=\textwidth]{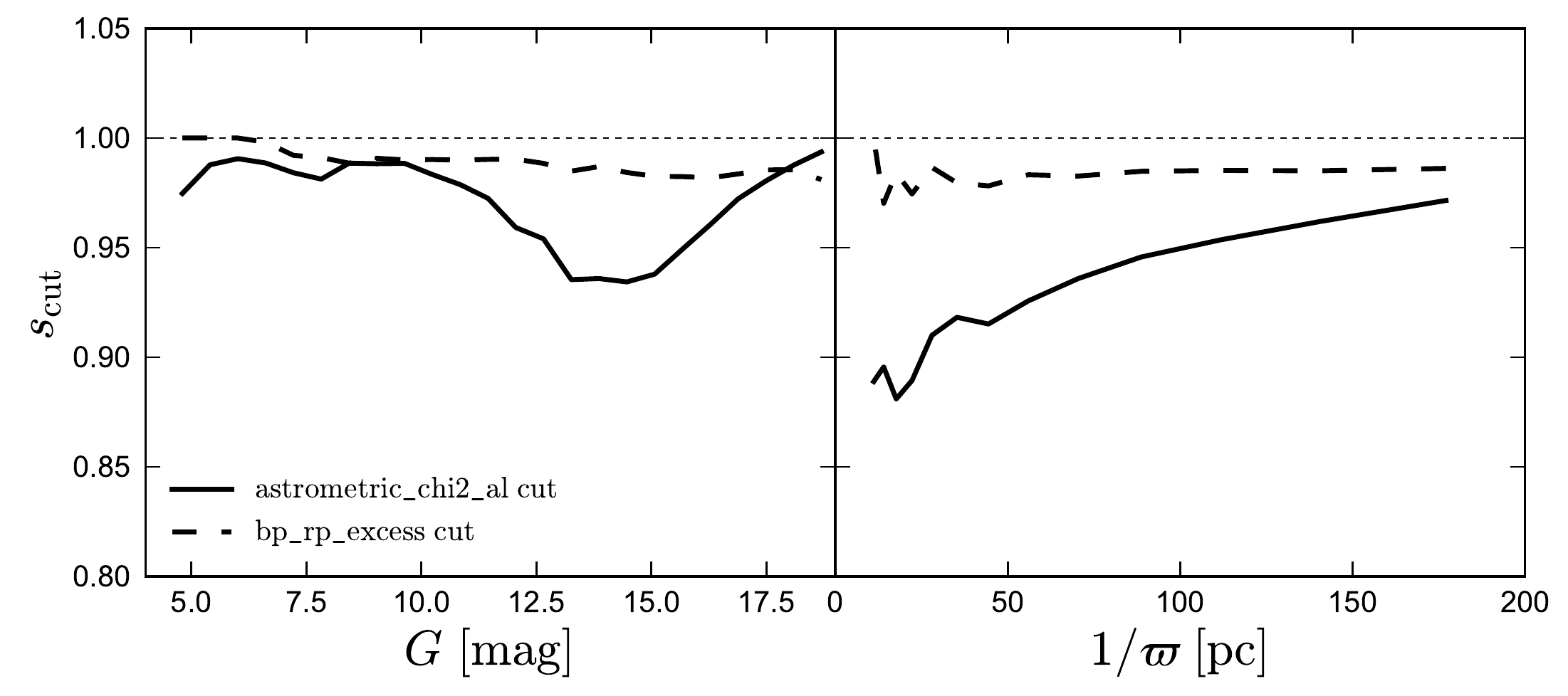}
    \caption{Fraction of suspected genuine sources within 200 pc that survive the cuts we impose on \texttt{astrometric\_chi2\_al} and \texttt{bp\_rp\_excess\_factor}, as a function of apparent magnitude (left) and distance (right). Both cuts remove only a small fraction of genuine sources. The cut on \texttt{bp\_rp\_excess\_factor} removes sources with contaminated photometry (section~\ref{sec:sensitivy}). The cut on \texttt{astrometric\_chi2\_al} preferentially removes nearby sources, perhaps because perturbations from a companion lead to the larger angular deviations from linear motion for nearby sources.}
    \label{fig:quality_cuts}
\end{figure*}

It is also worth considering whether the quality cuts (iii) and (iv) change the single-star selection function. We assess the effects of each quality cut as follows. We begin with the full sample of stars within 200\,pc satisfying $\varpi/\sigma_{\varpi}>10$. We then apply each quality cut and monitor the effects on the sample. For this nearby sample, it is straightforward to determine whether most of the sources removed by a particular cut are erroneous (generally distant faint stars with incorrect parallaxes) or real, because distant sources with incorrect parallaxes generally fall in a cloud below the main sequence (see \citealt{Lindegren_2018}). What we wish to quantify is the fraction of real sources that are removed by each cut, and whether these cuts introduce systematic biases.

Figure~\ref{fig:quality_cuts} shows the fraction of suspected genuine sources removed by cuts (iii) and (iv). We conclude that cut (iv) has a negligible effect on the single-star selection function. The effects of cut (iii) are also weak, but it does appear to preferentially remove nearby sources. This may be because unresolved astrometric binaries produce large deviations from linear motion of the light centroid at close distances. We tabulate $s_{\rm cut}$ due to cut (iv) as a function of parallax. The total single-star selection function is then
\begin{align}
    \label{eq:s1}
    s_{1}=s_{\varpi}(G)s_{\rm BP}(G_{{\rm BP}})s_{{\rm cut}}(\varpi),
\end{align}
where $s_{\varpi}(G)$ and $s_{\rm BP}(G_{{\rm BP}})$ are given by Equations~(\ref{eq:s_varpi}) and~(\ref{eq:s_bp}), respectively. 

\subsection{Contrast sensitivity for close pairs}
\label{sec:sensitivy}

\begin{figure}
    \includegraphics[width=\columnwidth]{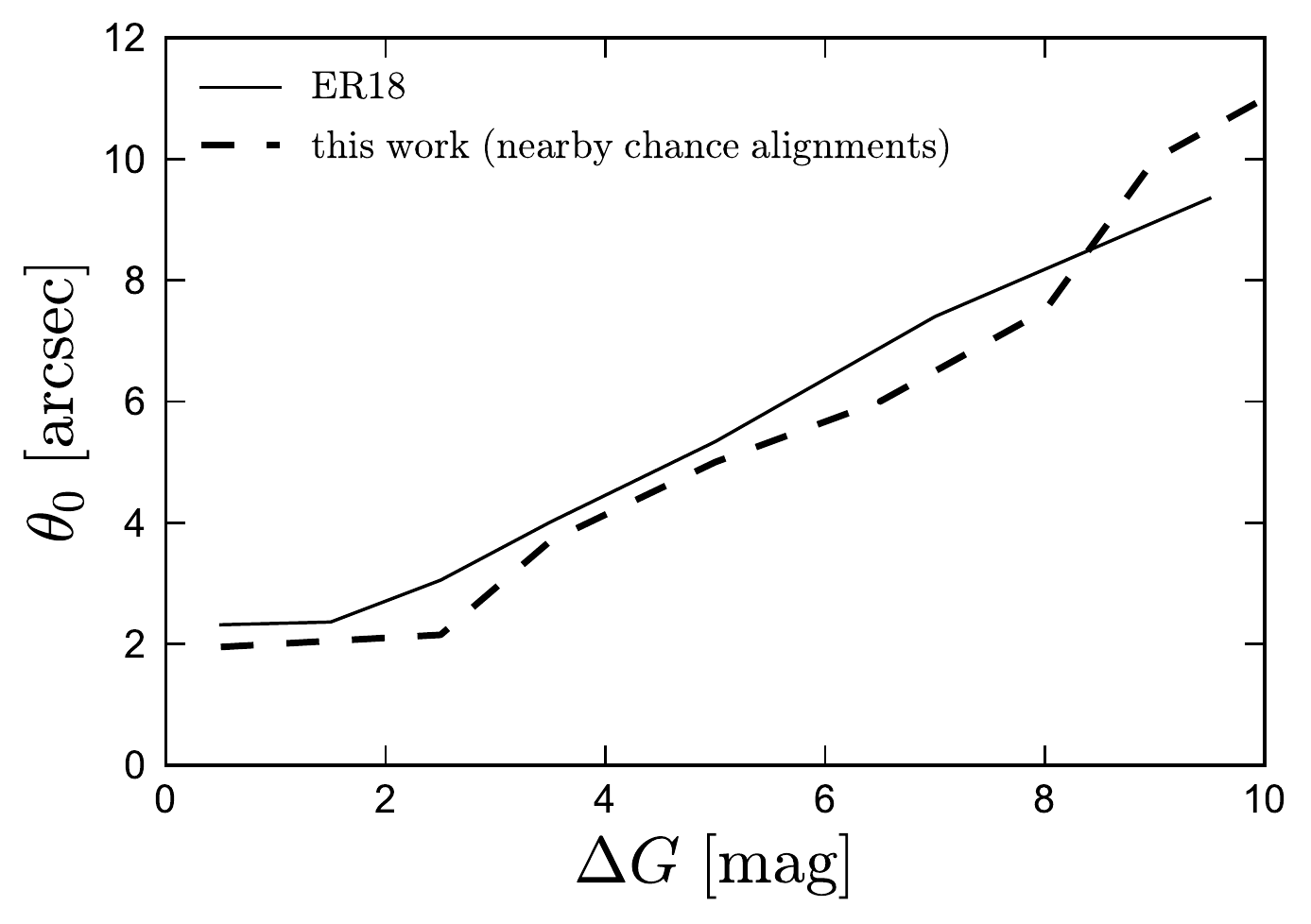}
    \caption{Contrast sensitivity given our quality cuts. $\theta_0$ is the angular separation below which the sensitivity to a companion with magnitude difference $\Delta G$ drops off rapidly (see Equation~\ref{eq:s_theta}). Companions can be detected at closer separations when the magnitude difference is small. We compare the values of $\theta_0$ found by \citetalias{ElBadry_2018} (for all stars passing our quality cuts in a dense field) to those derived from our all-sky catalog of nearby chance alignments with precise parallaxes.}
    \label{fig:contrast}
\end{figure}

A critical aspect of the selection function for binaries is the reduction in sensitivity to a companion at close angular separations. Whether a binary enters our catalog depends both on the angular separation of the two stars and on their flux ratio, as a secondary is more likely to be outshone or contaminated by light from the primary at fixed angular separation if the flux ratio is large than if it is small. 

The sensitivity to a companion can be measured by comparing the two point correlation function of chance alignments to what would be expected in the absence of crowding/blending effects \citep[e.g.][]{Arenou_2018, Brandeker_2019}. Of course, the sensitivity to companions in a particular catalog depends on the quality cuts imposed: cuts that remove objects with somewhat contaminated photometry will lower the sensitivity. 

We quantified the reduction in sensitivity to a companion as a function of angular separation $\theta$ and magnitude difference $\Delta G$ in \citetalias{ElBadry_2018}. There we found that, given our cuts on astrometric $\chi^2$ and photometric quality (primarily \texttt{bp\_rp\_excess\_factor}), the sensitivity to a companion goes to 0 at $\theta \ll \theta_0$, where $\theta_0\approx 2$\,arcsec for sources with similar magnitude, and $\theta_0$ increases with $\Delta G$. We quantified this dependence by fitting a function:
\begin{align}
    \label{eq:s_theta}
    s_{\Delta G}\left(\theta\right)=\frac{1}{1+\left(\theta/\theta_{0}\right)^{-\beta}},
\end{align}
where $\beta \approx 10$ and we fit for $\theta_0$ as a function of $\Delta G$. We note that the angular resolution of {\it Gaia} DR2 is actually significantly better than 2\,arcsec: most companions are detected down 1\,arcsec separations, and the detection fraction only drops to zero at $\theta < 0.5$\,arcsec \citep{Arenou_2018, Ziegler_2018}. The $\sim$2\,arcsec limit for our catalog is a result of our requirement that both stars have a measured \texttt{bp\_rp} color and the cut on \texttt{bp\_rp\_excess\_factor}.

The dependence of $\theta_0$ on $\Delta G$ calculated in \citetalias{ElBadry_2018} was derived from the source counts of all sources passing our quality cuts (iii) and (iv) in a dense field, most of which are fainter than the objects in the binary catalog. Here, we improve slightly on the \citetalias{ElBadry_2018} measurement by repeating their analysis, but using the sources from the chance alignment catalog described in Section~\ref{sec:data} instead of all sources in a dense field. The advantage of this approach is that the sources in the chance alignment catalog, being nearby and having precise parallaxes, are more representative of objects in the binary catalog. Unsurprisingly -- since it was verified in \citetalias{ElBadry_2018} that there are no strong trends in contrast sensitivity with apparent magnitude or source density; see also \citet{Brandeker_2019} -- the improved constraints are fairly similar to those derived in \citetalias{ElBadry_2018}. We compare them in Figure~\ref{fig:contrast}. On average, we find that the angular resolution at fixed separation is marginally better for the bright chance alignments than was found in \citetalias{ElBadry_2018}.

\section{Model validation}
\label{sec:model_validation}

To test whether the assumptions of the model we use to fit for the intrinsic mass ratio distribution are valid, we use it to predict properties of the population of single stars that pass the same quality cuts as the binaries in our catalog (Section~\ref{sec:single_star_validation}) and the separation distribution of chance alignments (Section~\ref{sec:chance_alignment_validation}).

\subsection{Single stars}
\label{sec:single_star_validation}

\begin{figure*}
    \includegraphics[width=\textwidth]{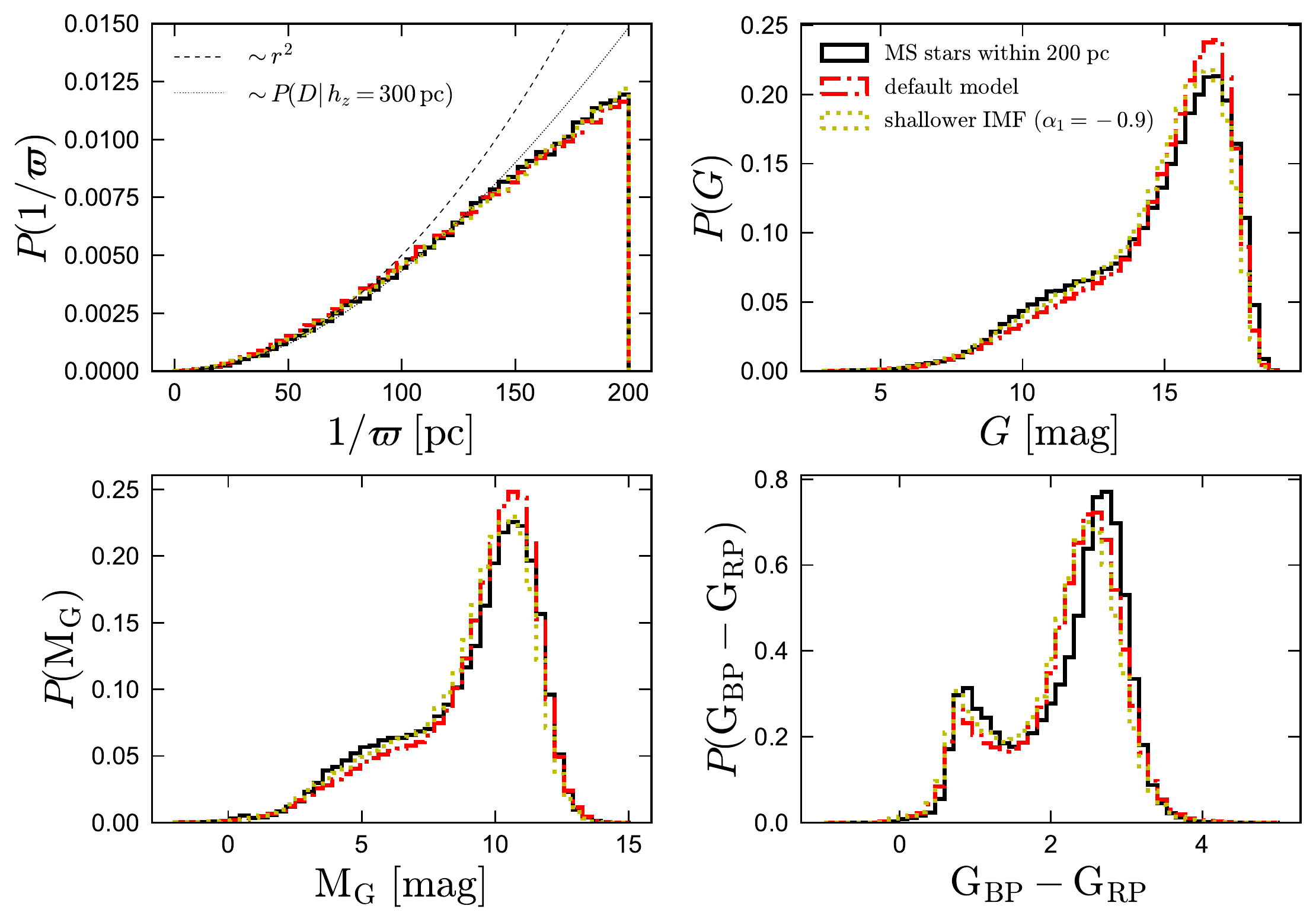}
    \caption{Validation of the single-star selection function and underlying Galactic model. Black histograms show distributions of distance, apparent magnitude, absolute magnitude, and color for all stars (not just binaries) within 200 pc that pass the quality cuts of the binary catalog. Black line show the distributions predicted by our model; i.e., assuming the same single-star selection function, IMF, Galactic scale height, star formation history, and metallicity distribution that we use when fitting the binary population. The reasonably good agreement with the observed distributions suggests that our empirical selection function and Galactic model are reasonable. Our fiducial model slightly overpredicts the number of faint stars. This tension can be resolved if a shallower IMF is assumed (with a logarithmic slope of -0.9 instead of -1.3 at $M< \rm 0.5M_{\odot}$; gold). Overall, we regard the agreement between the fiducial model and observations as quite satisfactory.}
    \label{fig:single_star_validation}
\end{figure*}

We query the {\it Gaia} catalog for all stars with $\varpi > 5$\,mas that pass the cuts on astrometric and photometric quality and precision that we require both members of the binary catalog to pass. Distributions of their distance, apparent magnitude, absolute magnitude, and color are shown in Figure~\ref{fig:single_star_validation} with a black histogram. To compare to the model predictions, we draw masses, ages, metallicities, distances from the fiducial distributions described in Section~\ref{sec:modeling}, compute synthetic photometry using PARSEC isochrones, and pass the observables through the single-star selection function (Appendix~\ref{sec:selection_function}). The resulting selection-function weighted distributions are compared to the data in Figure~\ref{fig:single_star_validation} (red histogram). In the upper left panel, we also show the distance distributions predicted for a uniform spatial distribution and for an exponential disk with scale height of 300 pc (comparable to the Milky Way; see \citealt{Juric_2008}), both assuming no incompleteness. The distance distribution of single stars is not that different from the exponential disk prediction, but it begins to deviate at $d\gtrsim 130$\,pc, which is where incompleteness effects become significant for $\rm M_G \approx 14$ (Figure~\ref{fig:selection_function_parallax}). 

The agreement between model and data is reasonably good. However, the fiducial model predicts slightly too many faint stars and too few bright stars.  The gold histograms shows that the agreement can be improved if a slightly shallower IMF is assumed, with a logarithmic slope of -0.9 (instead of -1.3, as assumed in the fiducial model) at $M<0.5\,M_{\odot}$.\footnote{This slope was found by fitting the IMF from the CMD, as described in Appendix C of \citet{ElBadry_2017}.}  This is consistent with the recent result from \citet{Sollima_2019}, who used the {\it Gaia} nearby star sample to measure the IMF. For the sake of this work, we are agnostic of whether Figure~\ref{fig:single_star_validation} indicates that the low-mass IMF in the solar neighborhood is slightly shallower than that assumed in our fiducial model or points towards a systematic in some other aspect of the model (such as the metallicity distribution function or stellar models). We note that changing the assumed IMF has very little effect on our inferred mass ratio distribution (Figure~\ref{fig:corner}).

We also note that the color distributions predicted by our model do not exactly match the observed distribution for either choice of IMF. This could be due to reddening, the adopted MDF (as the subset of stars with spectroscopic metallicity measurements is not guaranteed to be an unbiased sample), or imperfect stellar models. 

\subsection{Chance alignments}
\label{sec:chance_alignment_validation}

\begin{figure*}
    \includegraphics[width=\textwidth]{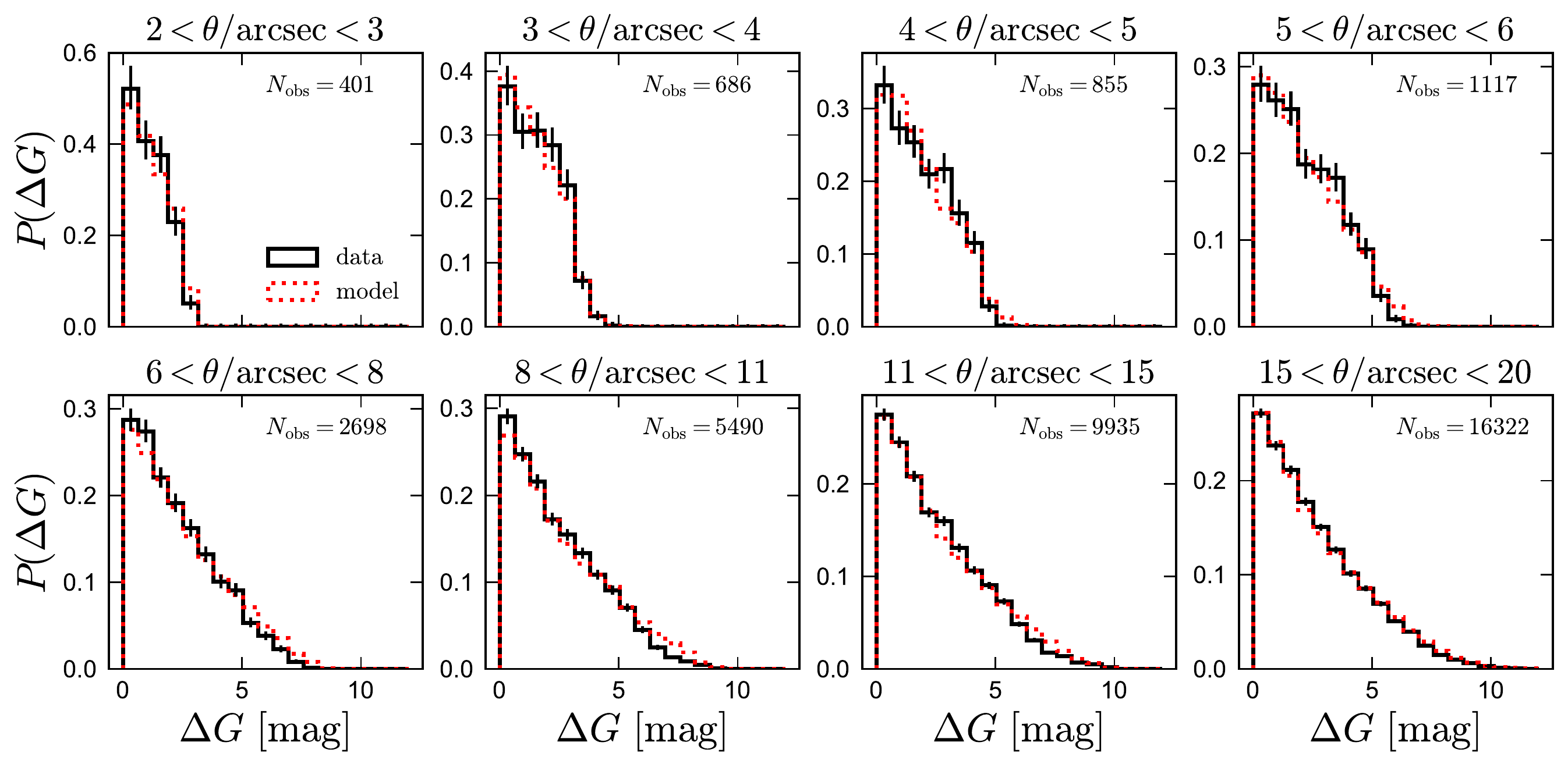}
    \caption{Distributions of apparent magnitude difference for our observed sample of chance alignments within 400 pc (black) and a simulated sample based on the same selection function and distributions of mass, age, metallicity, and distance assumed in modeling genuine binaries. Agreement between data and model validates the model assumptions, particularly the sensitivity to a companion as a function of angular separation and magnitude difference.}
    \label{fig:chance_alignments_prediction}
\end{figure*}

To validate our model for the contrast sensitivity as a function of angular separation, we predict the distributions of magnitude difference at fixed separation using our model and compare to the chance alignment catalog described in Section~\ref{sec:data}. This is accomplished as follows. We draw masses, ages, distances, and metallicities for both ``components'' of a chance alignment independently, and we draw angular separations assuming $P\left(\theta\right){\rm d}\theta \propto 2\pi\theta$. Just as for true binaries, we compute the selection function for each pair by multiplying the two single-star selection functions and the angular contrast sensitivity term. The results are shown in Figure~\ref{fig:chance_alignments_prediction} $\sigma_0$. The agreement with the real chance alignment catalog is good.

\section{RV variability}
\label{sec:rv_variability}

Although {\it Gaia} DR2 does not contain multi-epoch radial velocity measurements, the published radial velocity uncertainties contain information that can be exploited to detect a large fraction of RV-variable close binaries. The \texttt{radial\_velocity\_error} reported in the {\it Gaia} archive represents the uncertainty on the median of velocity measurements from several transits (see \citealt{Katz_2019}). It is calculated as 
\begin{align}
    \label{eq:err_median}
    \epsilon_{{\rm RV}}=\left[\frac{\pi}{2N_{{\rm obs}}}\sigma_{{\rm RV}}^{2}+\sigma_{0}^{2}\right]^{1/2},
\end{align}
where $\sigma_0=0.11\,\rm km\,s^{-1}$ is a constant term that represents the minimum achievable RV uncertainty due to calibration issues, $N_{\rm obs}$ is the number of radial velocity transits (\texttt{rv\_nb\_transits} in the {\it Gaia} archive), and $\sigma_{\rm RV}$ is the standard deviation of the RVs measured in individual transits. The standard deviation of the measured transit RVs can thus be reconstructed as
\begin{align}
    \label{eq:sigma_rv}
    \sigma_{{\rm RV}}=\left[2N_{{\rm obs}}\left(\epsilon_{{\rm RV}}^{2}-\sigma_{0}^{2}\right)/\pi\right]^{1/2}.
\end{align}
We expect $\sigma_{\rm RV}$ to be larger than usual if the variation in the true radial velocity of a star between transits is large compared to the observational RV precision. This suggests that close binaries could be identified as stars with unusually large $\sigma_{\rm RV}$ for their color and apparent magnitude.

\begin{figure}
    \includegraphics[width=\columnwidth]{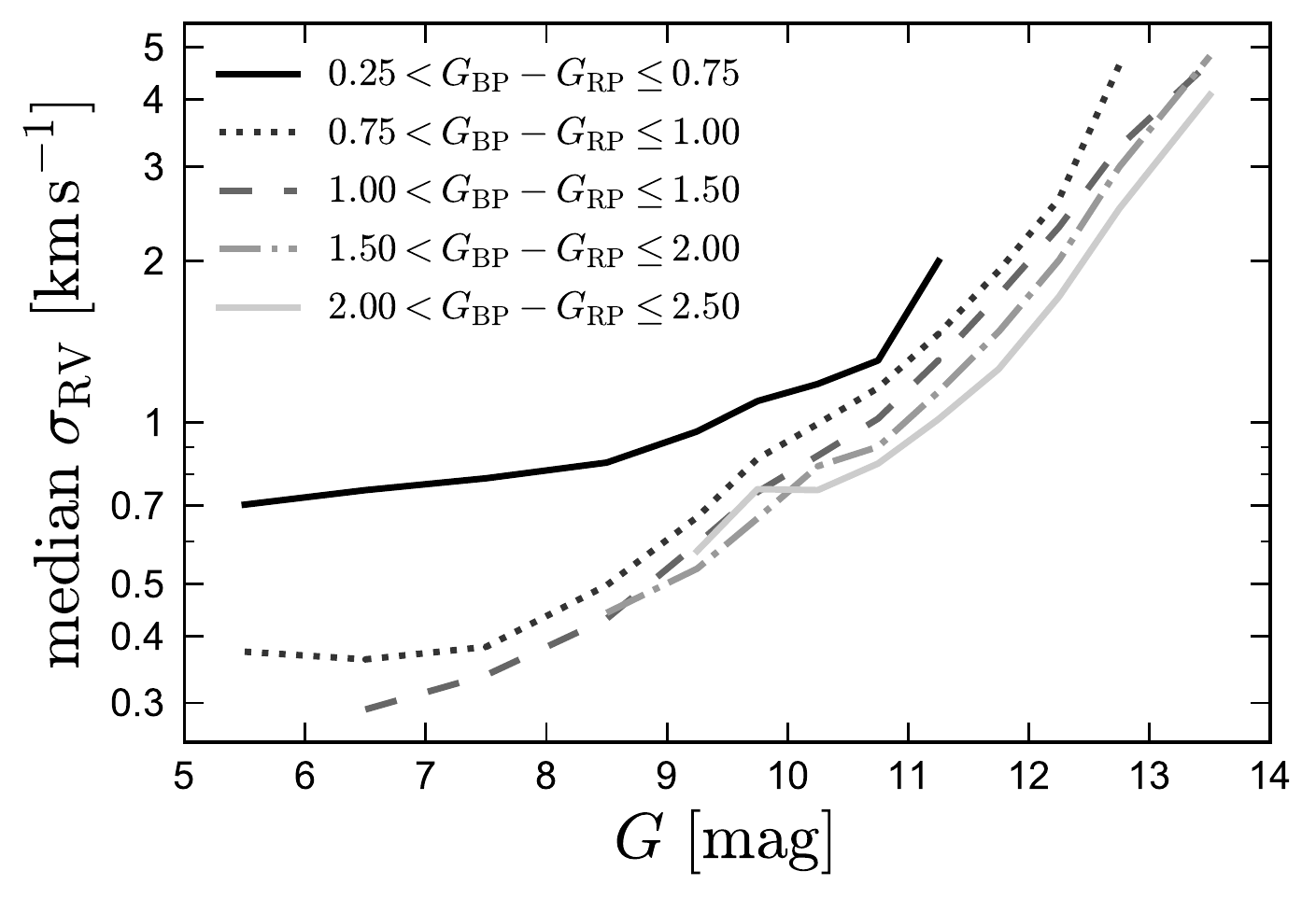}
    \caption{Median $\sigma_{\rm RV}$ (Equation~\ref{eq:sigma_rv}) for main-sequence stars within 200 pc that pass our quality cuts and \texttt{rv\_nb\_transits} > 2. Since a majority of stars are not intrinsically RV-variable, this represents the typical radial velocity precision for stars of a given brightness and spectral type. Most stars that have $\sigma_{\rm RV}$ substantially larger than the median for their color and magnitude are close binaries. }
    \label{fig:rv_err}
\end{figure}

To quantify this, we queried the {\it Gaia} archive for all stars within 200 pc that pass the quality cuts imposed on binary components and additionally have \texttt{rv\_nb\_transits} > 2. Selecting main-sequence stars in bins of \texttt{bp\_rp} color and $G$-band magnitude, we calculated the median $\sigma_{\rm RV}$ in each bin (Figure~\ref{fig:rv_err}). As expected, the typical RV error increases with increasing $G$ magnitude and is larger for bluer stars, which have weaker and broader absorption lines, at fixed magnitude. 

We designate sources that have $\sigma_{\rm RV}$ larger than 2.5 times the median for their color and apparent magnitude as likely close binaries. The factor of 2.5 is a practical choice to balance the number of false-positives and false-negatives. To assess the false-negative rate for this designation, we cross-matched  {\it Gaia} DR2 with the catalog of RV-variable main-sequence SB1s identified by \citet{Elbadry_2018a} using APOGEE spectra and calculated $\sigma_{\rm RV}$ for the subsample of that catalog that passes our quality cuts. Among SB1s for which \citet{Elbadry_2018a} found the radial velocity to vary by at least 5\,km\,s$^{-1}$ between visits, 62\% are correctly identified as binaries based on the {\it Gaia} $\sigma_{\rm RV}$. This fraction climbs to 82\% for sources whose APOGEE RVs varied by at least 15\,km\,s$^{-1}$. On the other hand, the false-positive rate is relatively low: only 2.6\% of the stars classified by \citet{Elbadry_2018a} as likely to be single are classified as binaries based on the {\it Gaia} $\sigma_{\rm RV}$. This means that, although some true binaries will be missed and there will be some false positives,  the {\it Gaia} $\sigma_{\rm RV}$ can be used to obtain an estimate of the close binary fraction in a population (where ``close'' means $a\lesssim {\rm few\,AU}$). 

To assess whether the components of wide twins are more likely to have an unresolved close companion than the components of non-twins, we use the $\sigma_{\rm RV}$ distributions to estimate the fraction of components of twin ($\Delta G < 0.25$) and non-twin ($0.25 \leq \Delta G < 1$) wide binaries with an unresolved close companion. We only consider binaries with $100 < s/{\rm AU} < 500$, since at wider separations, the number of ``excess'' twins is subdominant relative to the underlying population (Figure~\ref{fig:posterior_predictive}). We consider all components with masses in the range $0.5 < M/M_{\odot} < 1.2$ that have {\it Gaia} RVs with \texttt{rv\_nb\_transits} > 2 (i.e., we do not require both resolved components to have measured RVs). Among 546 eligible components of wide twins, 29 have $\sigma_{\rm RV}$ consistent with having unresolved close companion, implying a close companion fraction per wide binary component of
\begin{align}
    \label{eq:fclose_twins.}
    f_{{\rm close\,companion,\,twins}}=0.053\pm0.012.
\end{align}
Of the 780 eligible components of non-twins, 76 have $\sigma_{\rm RV}$ consistent with having unresolved close companion. This implies
\begin{align}
    \label{eq:fclose_twins.}
    f_{{\rm close\,companion,\,non-twins}}=0.097\pm0.013.
\end{align}
That is, the fraction of wide binary components that have a close unresolved companion is {\it higher} at the 2 sigma level for non-twins than for twins. This would seem to speak against a scenario in which the excess of wide twins is causally linked to hierarchical triples. Indeed, given the non-zero false positive rate of the $\sigma_{\rm RV}$-based close binary identification and the fact that ``excess'' twins represent only about half of the population with $\Delta G < 0.25$, the data are consistent with the excess twins having no close companions at all.

\section{Full fitting constraints}
\label{sec:full_constraints}

Constraints on fitting parameters for all bins of primary mass and separation are listed in Table~\ref{tab:full_fit}. Error bars are 2 sigma (middle 95.4\%).  

\begin{table*}
\centering
\caption{Marginalized 2 sigma constraints on fitting parameters for all bins of primary mass and separation. For  $0.1 < M_1/M_{\odot} < 0.4$, we fit a single power law, so $\gamma_{\rm smallq} = \gamma_{\rm largeq}$.}
\label{tab:full_fit}

\begin{tabular}{ |l|c|c|c|c|c|} 
 \hline
     & $0.1 < M_1/M_{\odot} < 0.4$ & $0.4 < M_1/M_{\odot} < 0.6$ & $0.6 < M_1/M_{\odot} < 0.8$ & $0.8 < M_1/M_{\odot} < 1.2$ & $1.2 < M_1/M_{\odot} < 2.5$ \\
      \hline \hline
$F_{\rm twin}$ & & & & & \\
    \hline 
$50<s/{\rm AU}<350$ & $0.189\pm_{0.049}^{0.042}$ & $0.112\pm_{0.038}^{0.034}$ & $0.059\pm_{0.021}^{0.019}$ & $0.101\pm_{0.029}^{0.027}$ & $0.086\pm_{0.075}^{0.058}$ \\ 
$350<s/{\rm AU}<600$ & $0.105\pm_{0.035}^{0.036}$ & $0.062\pm_{0.020}^{0.018}$ & $0.043\pm_{0.019}^{0.016}$ & $0.054\pm_{0.018}^{0.014}$ & $0.088\pm_{0.047}^{0.037}$ \\ 
$600<s/{\rm AU}<1,000$ & $0.049\pm_{0.039}^{0.032}$ & $0.054\pm_{0.020}^{0.017}$ & $0.032\pm_{0.019}^{0.016}$ & $0.035\pm_{0.016}^{0.015}$ & $0.015\pm_{0.034}^{0.024}$ \\ 
$1,000<s/{\rm AU}<2,500$ & $0.023\pm_{0.033}^{0.025}$ & $0.049\pm_{0.022}^{0.018}$ & $0.024\pm_{0.016}^{0.016}$ & $0.013\pm_{0.011}^{0.010}$ & $0.002\pm_{0.020}^{0.015}$ \\ 
$2,500<s/{\rm AU}<5,000$ & $-0.001\pm_{0.039}^{0.048}$ & $0.047\pm_{0.029}^{0.025}$ & $0.018\pm_{0.019}^{0.020}$ & $0.023\pm_{0.021}^{0.017}$ & $-0.003\pm_{0.023}^{0.023}$ \\ 
$5,000<s/{\rm AU}<15,000$ & $0.005\pm_{0.043}^{0.040}$ & $0.043\pm_{0.031}^{0.030}$ & $0.014\pm_{0.021}^{0.018}$ & $0.007\pm_{0.015}^{0.015}$ & $0.003\pm_{0.035}^{0.023}$ \\ 
$15,000<s/{\rm AU}<50,000$ & $0.024\pm_{0.120}^{0.094}$ & $-0.009\pm_{0.043}^{0.039}$ & $-0.012\pm_{0.032}^{0.028}$ & $0.009\pm_{0.032}^{0.024}$ & $0.001\pm_{0.057}^{0.037}$ \\ 
\hline
$q_{\rm twin}$ & & & & & \\
\hline 
$50<s/{\rm AU}<350$ & $0.953\pm_{0.006}^{0.010}$ & $0.959\pm_{0.014}^{0.023}$ & $0.969\pm_{0.005}^{0.019}$ & $0.962\pm_{0.009}^{0.014}$ & $0.950\pm_{0.029}^{0.018}$ \\ 
$350<s/{\rm AU}<600$ & $0.937\pm_{0.011}^{0.006}$ & $0.959\pm_{0.009}^{0.017}$ & $0.954\pm_{0.016}^{0.021}$ & $0.967\pm_{0.010}^{0.027}$ & $0.936\pm_{0.018}^{0.006}$ \\ 
$600<s/{\rm AU}<1,000$ & $0.954\pm_{0.025}^{0.022}$ & $0.963\pm_{0.012}^{0.025}$ & $0.939\pm_{0.018}^{0.008}$ & $0.957\pm_{0.028}^{0.023}$ & $0.954\pm_{0.040}^{0.022}$ \\ 
$1,000<s/{\rm AU}<2,500$ & $0.975\pm_{0.017}^{0.043}$ & $0.963\pm_{0.017}^{0.029}$ & $0.941\pm_{0.030}^{0.011}$ & $0.959\pm_{0.030}^{0.025}$ & $0.957\pm_{0.037}^{0.026}$ \\ 
$2,500<s/{\rm AU}<5,000$ & $0.955\pm_{0.039}^{0.025}$ & $0.953\pm_{0.019}^{0.022}$ & $0.950\pm_{0.029}^{0.019}$ & $0.945\pm_{0.039}^{0.014}$ & $0.958\pm_{0.035}^{0.027}$ \\ 
$5,000<s/{\rm AU}<15,000$ & $0.959\pm_{0.038}^{0.028}$ & $0.951\pm_{0.030}^{0.020}$ & $0.965\pm_{0.027}^{0.034}$ & $0.970\pm_{0.027}^{0.038}$ & $0.960\pm_{0.035}^{0.029}$ \\ 
$15,000<s/{\rm AU}<50,000$ & $0.961\pm_{0.031}^{0.030}$ & $0.961\pm_{0.033}^{0.029}$ & $0.961\pm_{0.032}^{0.029}$ & $0.957\pm_{0.038}^{0.026}$ & $0.957\pm_{0.039}^{0.026}$ \\ 
\hline 
$\gamma_{\rm largeq}$ & & & & & \\
\hline 
$50<s/{\rm AU}<350$ & $0.52\pm_{0.45}^{0.45}$ & $0.17\pm_{0.39}^{0.40}$ & $-1.43\pm_{0.51}^{0.48}$ & $-1.22\pm_{0.67}^{0.66}$ & $-0.89\pm_{1.00}^{0.90}$ \\ 
$350<s/{\rm AU}<600$ & $-0.16\pm_{0.43}^{0.43}$ & $-0.01\pm_{0.27}^{0.29}$ & $-0.89\pm_{0.33}^{0.35}$ & $-1.83\pm_{0.38}^{0.42}$ & $-1.88\pm_{0.55}^{0.54}$ \\ 
$600<s/{\rm AU}<1,000$ & $0.23\pm_{0.45}^{0.41}$ & $-0.44\pm_{0.27}^{0.28}$ & $-1.19\pm_{0.30}^{0.30}$ & $-1.43\pm_{0.30}^{0.30}$ & $-1.16\pm_{0.28}^{0.33}$ \\ 
$1,000<s/{\rm AU}<2,500$ & $0.39\pm_{0.39}^{0.41}$ & $-0.55\pm_{0.26}^{0.28}$ & $-0.90\pm_{0.25}^{0.26}$ & $-1.54\pm_{0.22}^{0.22}$ & $-1.55\pm_{0.19}^{0.21}$ \\ 
$2,500<s/{\rm AU}<5,000$ & $0.23\pm_{0.52}^{0.52}$ & $-0.43\pm_{0.38}^{0.40}$ & $-1.04\pm_{0.34}^{0.36}$ & $-1.52\pm_{0.33}^{0.32}$ & $-1.52\pm_{0.27}^{0.29}$ \\ 
$5,000<s/{\rm AU}<15,000$ & $0.22\pm_{0.59}^{0.51}$ & $-0.42\pm_{0.42}^{0.44}$ & $-0.86\pm_{0.35}^{0.37}$ & $-1.35\pm_{0.31}^{0.33}$ & $-1.31\pm_{0.25}^{0.28}$ \\ 
$15,000<s/{\rm AU}<50,000$ & $0.53\pm_{0.84}^{0.86}$ & $-0.43\pm_{0.67}^{0.67}$ & $-0.72\pm_{0.63}^{0.65}$ & $-1.39\pm_{0.58}^{0.55}$ & $-1.22\pm_{0.43}^{0.46}$ \\ 
\hline 
$\gamma_{\rm smallq}$ & & & & & \\
\hline 
$50<s/{\rm AU}<350$ &  & $0.33\pm_{0.56}^{0.60}$ & $0.20\pm_{0.58}^{0.51}$ & $0.13\pm_{0.69}^{0.64}$ & $0.13\pm_{1.00}^{0.98}$ \\ 
$350<s/{\rm AU}<600$ &  & $-0.37\pm_{0.47}^{0.44}$ & $0.24\pm_{0.47}^{0.41}$ & $0.40\pm_{0.43}^{0.41}$ & $-0.01\pm_{0.86}^{0.79}$ \\ 
$600<s/{\rm AU}<1,000$ &  & $0.17\pm_{0.46}^{0.44}$ & $0.24\pm_{0.36}^{0.33}$ & $0.12\pm_{0.26}^{0.26}$ & $0.15\pm_{0.72}^{0.55}$ \\ 
$1,000<s/{\rm AU}<2,500$ &  & $0.53\pm_{0.44}^{0.44}$ & $0.06\pm_{0.27}^{0.25}$ & $0.12\pm_{0.16}^{0.17}$ & $0.56\pm_{0.47}^{0.42}$ \\ 
$2,500<s/{\rm AU}<5,000$ &  & $0.02\pm_{0.58}^{0.56}$ & $-0.15\pm_{0.34}^{0.33}$ & $-0.14\pm_{0.21}^{0.21}$ & $0.09\pm_{0.43}^{0.36}$ \\ 
$5,000<s/{\rm AU}<15,000$ &  & $0.13\pm_{0.69}^{0.59}$ & $0.12\pm_{0.40}^{0.39}$ & $-0.20\pm_{0.22}^{0.21}$ & $-0.18\pm_{0.40}^{0.40}$ \\ 
$15,000<s/{\rm AU}<50,000$ &  & $-0.02\pm_{0.88}^{0.84}$ & $-0.13\pm_{0.60}^{0.69}$ & $-0.24\pm_{0.41}^{0.42}$ & $-0.09\pm_{0.66}^{0.64}$ \\ 
\hline 
$\gamma_{s}$ & & & & & \\
\hline 
$50<s/{\rm AU}<350$ & $-1.36\pm_{0.21}^{0.22}$ & $-1.55\pm_{0.20}^{0.18}$ & $-1.22\pm_{0.38}^{0.32}$ & $-1.02\pm_{0.46}^{0.40}$ & $-0.84\pm_{0.93}^{0.77}$ \\ 
$350<s/{\rm AU}<600$ & $-1.10\pm_{0.42}^{0.43}$ & $-1.20\pm_{0.27}^{0.29}$ & $-0.89\pm_{0.38}^{0.39}$ & $-0.80\pm_{0.43}^{0.47}$ & $-1.28\pm_{1.00}^{1.00}$ \\ 
$600<s/{\rm AU}<1,000$ & $-1.48\pm_{0.46}^{0.48}$ & $-1.16\pm_{0.29}^{0.31}$ & $-1.75\pm_{0.37}^{0.34}$ & $-1.13\pm_{0.36}^{0.36}$ & $-0.03\pm_{0.78}^{0.74}$ \\ 
$1,000<s/{\rm AU}<2,500$ & $-1.67\pm_{0.26}^{0.23}$ & $-1.56\pm_{0.14}^{0.14}$ & $-1.52\pm_{0.16}^{0.15}$ & $-1.47\pm_{0.13}^{0.13}$ & $-1.41\pm_{0.26}^{0.24}$ \\ 
$2,500<s/{\rm AU}<5,000$ & $-2.05\pm_{0.46}^{0.51}$ & $-1.71\pm_{0.30}^{0.30}$ & $-1.75\pm_{0.29}^{0.30}$ & $-1.67\pm_{0.26}^{0.23}$ & $-1.40\pm_{0.45}^{0.42}$ \\ 
$5,000<s/{\rm AU}<15,000$ & $-1.74\pm_{0.36}^{0.34}$ & $-1.74\pm_{0.21}^{0.21}$ & $-1.62\pm_{0.21}^{0.21}$ & $-1.84\pm_{0.17}^{0.16}$ & $-1.35\pm_{0.27}^{0.28}$ \\ 
$15,000<s/{\rm AU}<50,000$ & $-1.61\pm_{0.83}^{0.90}$ & $-1.68\pm_{0.49}^{0.47}$ & $-2.21\pm_{0.50}^{0.54}$ & $-2.00\pm_{0.40}^{0.37}$ & $-1.50\pm_{0.63}^{0.66}$ \\ 
\hline
\end{tabular}
\end{table*}

\bsp	
\label{lastpage}
\end{document}